\documentstyle[prb,aps,epsf]{revtex}
\tighten
\begin{document}
\draft 
 
\title{ Magnetotunneling through a quantum well in a tilted field I: 
Periodic orbit theory}    
\author{E.~E.~Narimanov   and  A.~Douglas ~Stone}   
\address{Applied  Physics,  Yale  University,  P.O.  Box  208284,  
New Haven CT06520-8284}  
\date{Preprint:  \today}  
\maketitle  
\begin{abstract}  
{ A semiclassical theory is developed  
and compared to experiments on the tunneling 
resonance spectrum for a quantum well in magnetic field tilted by 
an angle $\theta$ with respect to the tunneling direction.  As the tilt 
angle is increased from zero the classical mechanics of an electron  
trapped within the well undergoes a smooth transition from integrable 
to chaotic dynamics.  Perturbation theory is invalid for most of the 
regime of experimental interest, motivating a semiclassical treatment 
based on short periodic orbits within the well.  In this paper we 
present a unified theory of all the periodic orbits within the well 
which are of relevance to experiments and show that they are all related to 
bifurcations of the period-one traversing orbits.  An analytic theory 
is derived for the period and stability of these traversing orbits. 
An unusual feature of the classical mechanics of this system is 
the existence of certain important periodic orbits only in finite  
energy bands.  We calculate the widths of these bands 
and relate them to experimental data. 
In the following paper the results for these short periodic orbits 
are used in conjunction with a novel semiclassical tunneling formula 
to calculate the magnetotunneling current, which is then compared  
with experiments. }  
\end{abstract}    
\pacs{   
\hspace{1.9cm}   
}  
\vspace{1ex} 

\section{Introduction}
 
Most of our  intuition  about the properties of quantum  systems 
comes from the  consideration  of hamiltonians  with high symmetry, for 
which the classical motion is integrable and hence the Schr\"odinger 
equation is separable.  Symmetry-breaking terms are typically treated by 
perturbation theory and the physics is described in terms of transitions 
induced between stationary states of the symmetric problem. 
This approach fails when the  symmetry-breaking 
terms become too large and many levels of the unperturbed system are 
strongly mixed.  In this situation one approach is direct numerical  
solution of the non-separable Schro\"dinger equation using a large  
basis set and calculation 
of the expectation values of interest from the numerically-determined 
eigenstates.  For most problems of interest the computational effort 
involved is substantial, particularly if one wishes to explore a large 
parameter space of hamiltonians and not just a single fixed set of 
parameters.  Moreover, an exclusively numerical approach makes it 
very difficult to understand qualitatively the dependence of physical 
properties on the parameters of the problem and thus to generalize the 
results to other related systems.   
 
An alternative approach which can give greater physical insight  
is to use the semiclassical methods
developed for non-integrable systems during the past two decades by 
researchers studying ``quantum chaos'', i.e. the quantum manifestations 
of chaotic classical dynamics.  This approach has been used successfully 
in atomic physics during the past decade.  Of particular note is the 
theory of the spectra of Rydberg states in a high magnetic field 
(diamagnetic Kepler problem), \cite{Garton,Holle} 
where a qualitative {\it and} quantitative 
understanding has been obtained semiclassically in excellent agreement 
with experiments.  In that case the essential idea behind the theory 
is a relationship between the quantum density of states (DOS) and a sum over 
isolated unstable periodic classical orbits first derived by Gutzwiller 
(the ``Gutzwiller Trace Formula'')\cite{Gutzwiller}.  However this 
formula had to be extended to account for experimental spectra 
which depend on other factors in addition to the density of states
\cite{Delos}.   
 
Until recently there were no comparable applications of semi-classical 
theory to condensed matter systems.  Within the past few years however 
several such systems have been identified: ballistic microcavities  
\cite{Jalabert,Marcus}, two-dimensional anti-dot arrays \cite{Weiss,Gregor}, 
and the system which is the subject of this paper, resonant tunneling 
diodes (RTD) in a magnetic field tilted by an angle $\theta$ with 
respect to the tunneling direction.  It has become clear that of the 
three, this system allows the most detailed comparison between 
theory and experiment, because the microscopic hamiltonian is known so 
accurately and because several continuous experimental control parameters 
may be tuned {\it in situ} to map out a large parameter space.  
 
This system was first identified and studied by Fromhold et al.   
\cite{Fromhold}, who immediately understood the close analogy to the 
Garton-Tomkins \cite{Garton} spectral oscillations  
in the diamagnetic Kepler problem.  When the tilt angle $\theta$ 
is zero the experiment corresponds to a conventional resonant  
magnetotunneling geometry; there is resonant structure in the I-V 
characteristic (causing peaks in $d^2I/dV^2$) with each peak 
corresponding to the sub-band thresholds in the quantum well.   
The experiments were done at fixed 
magnetic field $B=11 {\rm T}$, for which the emitter state of the resonant 
tunneling device is primarily the $n=0$ Landau level, 
so that the observed peaks were only due to quantum well states with sub-band 
quantum number $p$ and Landau index $n=0$, as selection rules prohibit 
tunneling to other Landau levels.   Typically of order twenty such 
resonance peaks (sub-bands) were observed over the interval zero to one 
volt.  However, when the magnetic field was tilted by a substantial amount  
($\theta > 20^{\circ}$), 
Fromhold et al. \cite{Fromhold} found that in certain voltage intervals 
the number of tunneling resonances would abruptly increase indicating 
the presence of tunneling processes which could not be explained 
by the sub-band structure of the well at $\theta =0$. 
They interpreted these new peaks in terms of density of states oscillations 
associated semiclassically with the short periodic orbits of the 
well which collide with both the emitter and collector barriers.   
Numerical integration of the classical equations of motion revealed 
a number of relevant periodic orbits  
(we will discuss the different orbit types 
in detail below), and that in most of the voltage range at $B=11{\rm T}$ 
these orbits were unstable fixed points in an almost completely chaotic 
phase space.  It was found that  
the spacing of the new resonances in voltage were consistent with 
the period of the orbits identified, as was their appearance at  
particular values of the magnetic field.  In more recent work those 
authors \cite{scars} have emphasized that in many cases these  
oscillations should be interpreted as arising from individual  
electron eigenstates in the well which concentrate on the relevant 
classical periodic orbit (the ``scarred'' wavefunctions), and not 
by the level-clustering normally associated with the DOS oscillations 
given by Gutzwiller's trace formula.  All of this work was done 
at high magnetic field and large tilt angles such that the classical 
dynamics is almost completely chaotic.

Another important series of experiments \cite{Muller} 
looked at the I-V peaks in the entire 
(plane) parameter space of magnetic field and voltage, varying the tilt 
from $\theta=0$ to $\theta=45^{\circ}$ in small increments so that 
the resonance structure could be carefully analyzed in the transition  
regime between chaos and integrability.  They found a complicated  
pattern of peak-doubling and peak-tripling  
in various regions of the $B-V$ plane, which extended to much lower 
magnetic field than previously reported.  Such experiments are 
particularly interesting from the theoretical point of view because, 
as discussed below, classically the system is undergoing a
transition to chaos as a function of 
continuous parameters ($\theta,B,V$). In our view no quantum system of  
comparable controllability existed previously for the study of the 
quantum manifestations of the {\it transition} to chaos 
with its associated KAM (Kolmogorov-Arnold-Moser) behavior in phase space
\cite{caveat}. 
It is these experiments 
which we shall analyze in detail in the this paper and its companion 
work \cite{sc}.  As will be shown in the companion work, 
the non-linear conductance of the well is related to a weighted 
local density of states in the well, which takes into account the
coupling of well states to the emitter wavefunction.
Electrons  tunnelling from the emitter into the well at high voltages 
gain  kinetic 
energy as they  accelerate in the field and collide with the collector 
barrier.  Over several  collisions in the well the electron loses this 
energy by optic phonon  emission.  Therefore the tunneling  resonances 
are substantially broadened and only are sensitive to structure in the 
DOS on energy scales $ >  \hbar/T_{opt}  \sim 5 meV$.  The semiclassical
tunneling theory we will develop in the following paper \cite{sc}
relates the tunneling oscillations  in the spirit of
Gutzwiller's trace formula \cite{Gutzwiller} to
a sum of contributions from each periodic orbit (PO) :
\begin{eqnarray}
w_{\rm osc} & = & \sum_{\alpha, n} A_{\alpha,n} 
\exp\left( - n T_\alpha / \tau_{\rm opt} \right)
\cos\left( \frac{n S_\alpha}{\hbar} + \phi_\alpha \right)
\label{tun_rate}
\end{eqnarray}  
where $w_{\rm osc}$ is the oscillatory part of the tunneling
rate from the emitter to the well per unit time, the summation
is carrier out over various primitive periodic orbits in the well
reaching the emitter wall ($\alpha$) and their repetitions ($n$).
$S_\alpha$ is the action of a primitive orbit, the amplitude
\begin{eqnarray}
A_{\alpha,n} & = & \sum_{m=1}^M a^{(\alpha, n)}_m 
\end{eqnarray}
where the integer $M$ is number of collisions of a particular periodic
orbit with the emitter wall. The general expressions for the 
``coupling coefficients'' $a_m$ are quite complicated \cite{sc}
and include both the stability properties of the periodic orbits 
and the velocity distribution of the tunneling electrons (which is
related to the Wigner transform of the wavefunction of the {\it
isolated} emitter state.

The broadening of the energy levels in the well due to inevitable
emission of optical phonons, which accounts for the 
$\exp\left( - n T_\alpha / \tau_{\rm opt} \right)$ in the tunneling
formula (\ref{tun_rate}), implies that only the
shortest PO's (period one to four orbits) 
will give resolvable structure in the experiments we analyze.
In this paper we focus on the the classical mechanics of these short
periodic orbits relevant to experiment.

Although the work of Fromhold et al. had identified several important 
periodic orbits in the classical mechanics, they had not provided  
a model of the global phase-space structure as the system undergoes 
the transition to chaos.  Shepelyansky and Stone \cite{ss} developed 
such a model by reducing the dynamics to a two-dimensional effective 
map which, in the limit where the emitter state energy is negligible, 
is equivalent to the Chirikov standard map.  This limit amounts 
to replacing the double-barrier system with a single-barrier model 
since the injected electron does not have enough energy to climb 
the potential hill and collide with the emitter barrier. 
In this limit the dynamics is controlled by a single chaos parameter 
$\beta =2 v_0 B/E$ where $B,E$ are the magnetic, electric fields 
and $\epsilon_0=m^*v_0^2/2$ is the total injection energy of the electron. 
Since for much of the experimental parameter range $eV \approx \epsilon_0$, 
Shepelyansky and Stone argued that the classical mechanics  
should be approximately  
constant along parabolas $V = 8 e d^2 {m^*}^{-1} \beta^{-2} B^2$  
($d$ is the distance between the barriers) and estimated the  
value of $\beta$ at which global chaos occurs using the Chirikov  
resonance overlap criterion 
\cite{Chirikov}.  They pointed out that the first appearance of additional 
resonance peaks at $B \approx 5 T, \theta=11^{\circ}$ 
appeared to be due to the bifurcation of the main period-one orbit 
however they did not analyze these bifurcations further at the time. 
 
In this paper we provide a detailed analysis of the classical mechanics 
of these bifurcations both within the single-barrier model (SBM) and the 
more accurate double-barrier model (DBM).  The experimental data
\cite{Muller} shows that at tilt angles 
less than $24^{\circ}$ the  
peak-doubling is ``re-entrant'' as the magnetic field is increased.
This effect is related to non-linear resonances between the 
longitudinal and cyclotron frequencies and is correctly described by 
the SBM.  These resonances lead to bifurcations of the main period-one 
orbit, which we shall refer to as the ``traversing orbit'' (TO), since 
near resonance this orbit is not isolated and new orbits can be born 
without violating the Poincar\'e index conservation theorem \cite{index}.   
Therefore 
it is qualitatively correct, as conjectured by Shepelyansky and Stone 
\cite{ss} and Muller et al. \cite{Muller}, to associated peak-doubling 
and tripling with bifurcations of the traversing orbit.   
Below we derive an exact analytic expression for the period and  
stability of the traversing orbit in both the SBM and DBM which allows 
us to locate precisely the bifurcation points for all values of 
$B,V,\theta$.  The existence of such exact analytic formulas for 
non-trivial periodic orbits of a hamiltonian in the KAM regime is 
to our knowledge unique to this system and suggests its value as 
a textbook example of bifurcation theory and the  
approach to hamiltonian chaos. 
 
Using our analytic formulas, supplemented with numerical results  
for the double-barrier model 
we find a more complicated and interesting periodic orbit structure 
than in the SBM.   
We define a period-N orbit to be a periodic orbits which collides with 
the collector barrier N times before retracing itself.  In the DBM 
it is possible to classify period-N orbits further by the number of 
times they collide with the emitter barrier M, so that an $(M,N)$ orbit 
is a period-N orbit which collides with the emitter M times during 
one period.  In general, for the DBM, orbits with $M=0,1, \ldots N$ 
can and do occur, although $M > N$ is forbidden by energy conservation. 
Bifurcations of the traversing orbit must, by continuity, produce 
$(N,N)$ orbits, since the TO collides with both barriers by definition. 
A major finding of this work is that all relevant $(M,N)$ orbits are related 
to the $(N,N)$ orbits (and hence to the bifurcations of the traversing 
orbit) by subsequent sequences of tangent bifurcations which occur  
(for the experimental parameters) quite near the bifurcations of the TO. 
Thus we should consider the set of $(M,N)$ orbits as a ``family'' spawned 
by bifurcations of the TO.  However we also find, in agreement with  
other work \cite{Fromprb}, \cite{Fromcomment}, that often the $(N,N)$ 
orbits born 
at the N-fold bifurcation of the TO are not most important for the 
experimentally observed tunneling resonances.  This will be discussed in 
great detail in the companion paper to this one \cite{sc}. 
In this paper we will develop the classical theory of these 
families of short periodic orbits. 
 
First, we briefly discuss qualitatively the origin of classical chaos 
in this system, which we shall refer to as the ``tilted well''. 
At zero tilt angle ($\theta =0$) the acceleration along the  
electric field ${\bf E}=E \hat{\bf z}$ normal to the barriers 
and the transverse cyclotron motion decouple and are 
integrable.  Collisions  with the  barriers reverse the  
longitudinal component of momentum ($v_z \to  -v_z$) and do not 
transfer energy between the cyclotron and longitudinal  motion.  Once 
the B field is tilted, so that ${\bf B}= B \cos \theta  \hat{z} + \sin \theta 
\hat{y}$,  between collisions  
the electron executes cyclotron motion around the $\hat{{\bf B}}$ 
 direction, with a superimposed drift velocity ${ \bf v}_d = (E/B) 
\sin \theta \hat{\bf x}$, and accelerates {\it along} $\hat{{\bf B}}$ 
due to the 
component ${\bf E}  \cdot  {\hat  {\bf B}} = E \cos  (\theta)$.  This 
motion is still  integrable.  However now collisions with the barriers 
in  general  {\it  do} mix the  cyclotron  and  longitudinal  energies 
$\varepsilon_c,\varepsilon_L$ and make the total dynamics 
non-integrable.  (When $\theta \neq 0$ longitudinal will mean parallel 
to the magnetic field direction $\hat{{\bf B}}$, and transverse will 
refer to the plane perpendicular to $\hat{{\bf B}}$). 
The amount of energy exchange  $\Delta  \varepsilon = 
\varepsilon_L-\varepsilon_c$ depends sensitively on the {\it phase} of 
the cyclotron rotation at impact.  For example, we shall see below 
that when the phase is such that the velocity falls 
precisely in the $x-z$ plane there is no energy-exchange 
($\Delta  \varepsilon=0$), and periodic orbits with this property will 
be of great importance.  When degrees of freedom are non-linearly  
coupled so that the amount of energy exchange is determined by a  
rapidly varying phase, chaos is the inevitable result \cite{ss}. 
Since the rate of variation of the phase between collisions is  
$\omega_c=eB/m^*$,  we expect the degree of chaos to increase with 
increasing $B$.  Similarly, since the time between collisions decreases 
with increasing voltage, the rate of phase variation is a decreasing 
function of $V$ and we expect chaos to diminish as $V$ {\it increases}. 
This explains qualitatively the dependence of the chaos parameter 
$\beta \sim B/\sqrt{V}$ found by Shepelyansky and Stone \cite{ss}. 
To go beyond these qualitative considerations we need to perform a  
scaling analysis of the classical double-barrier hamiltonian, which 
we will describe in the next section. 
 
This paper is organized as follows.  In section II we introduce the 
scaled hamiltonian which is effectively two-dimensional and discuss 
the non-linear Poincar\'e map it generates, recovering the limiting 
behavior discussed by Shepelyansky and Stone, which is equivalent 
to the single-barrier model.  We introduce the crucial notion of 
non-mixing periodic orbits.  In Section III we discuss the periodic 
orbit structure of the SBM, deriving analytic expressions for the period and 
stability of all period-one orbits.  We consider the 
bifurcations of the traversing orbits in the SBM, enumerating the relevant 
period-two and period-three orbits.   In section IV we turn to the 
double-barrier model (DBM) and derive analytic formulas for the 
period-one orbits there.  The bifurcations of the TO in 
the DBM are discussed and the families of $(M,N)$ orbits are identified.  
Finally, we summarize the properties of the short periodic orbits 
and set the stage for their use to calculate the tunneling spectra 
semiclassically in the companion work \cite{sc}.

\section{Scaled dynamics and Poincar\'e map} 
 
\subsection{Scaled Hamiltonian} 
 
We now define the Hamiltonian we will use for analyzing the classical 
mechanics.  We neglect the coupling of the electrons to optic phonons 
within the well; we will take it into account in 
the semiclassical theory by introducing an appropriate level-broadening.   
The semiclassical tunneling theory expresses the tunneling current 
in terms of the emitter wavefunction, the tunneling rate through each 
barrier, and the periodic orbits of electrons trapped within the well. 
Therefore we are only concerned with the classical mechanics within 
the well and can represent the barriers by infinite hard walls 
separated by a distance $d$.  The z-axis will be chosen normal to the 
barriers (parallel to the electric field ${\bf E}$) and with an origin 
such that the collector barrier is at $z=0$ and the emitter barrier 
is at $z=d$.  The magnetic field is tilted in the $(y,z)$ plane, 
${\bf B} = \cos \theta {\bf {\hat z}} + \sin \theta {\bf {\hat y}}$ 
We choose a gauge where the 
vector potential ${\bf A} = (- B y \cos(\theta) + B z   
\sin(\theta)) {\bf {\hat x}}$. The Hamiltonian is 
 
\begin{eqnarray} 
H & = & \frac{(p_x - e B y \cos(\theta) + e B z \sin(\theta))^2}{2m} 
+ \frac{p_y^2}{2m} + \frac{p_z^2}{2m}  - eE z \nonumber \\  
&+& U(-z) + U(z - d) \label{h3d} \\
& = & \varepsilon \nonumber   
\end{eqnarray} 
where the function $U$ ( $U(z<0) = 0, \ U(z>0) = \infty $ ) represents  
the infinite hard walls at $z=0,-d$. 
 
The Hamiltonian (\ref{h3d}) involves four variable experimental parameters: 
$B,E,\theta$ and $d$.  It is of great convenience to  
rescale the variables in Eq. (\ref{h3d}) so as to express the 
dynamics generated by this Hamiltonian in terms of the minimum 
number of independent parameters.  This will simplify the analysis of 
the periodic orbits and also predict scaling relations relevant to 
the experimental data.  We present a rescaling below which is most 
useful for a periodic orbit theory of both the single-barrier and 
double-barrier models.  It is a natural extension of the simpler scaling 
introduced by Shepelyansky and Stone \cite{ss}.  An alternative scaling 
which applies to the DBM has been introduced by Monteiro {\it et al}  
\cite{Monteiro1,Monteiro2}. 
 
The natural unit of time for the problem is 
$\omega_c^{-1}$ where $\omega_c = e B/m$ is the cyclotron 
frequency.  The barrier spacing $d$ gives one length scale, and  
the only other energy independent length scale in the problem is 
$l_D = v_D \omega_c^{-1}$, where $v_D=E/B$ is the drift 
velocity for perpendicular 
electric and magnetic fields  ( the actual drift velocity 
when the fields cross at angle $\theta$ is $v_{d} \equiv  
v_D \sin \theta$). For electron total energies $\varepsilon < eV=eEd$ 
the emitter barrier is energetically inaccessible 
so the length scale $d$ is irrelevant. 
Since we wish to introduce a dimensionless hamiltonian related to 
Eq. (\ref{h3d}) 
by a canonical transformation, the scaling must be independent of 
energy and applicable to both the case $\varepsilon < eEd$ and  
$ \varepsilon > eEd$. Hence we must scale all lengths by $l_D$. 
 
In addition we want to exploit all symmetries of the Hamiltonian. 
The Hamiltonian (\ref{h3d}) is independent of the coordinate $x$ and 
therefore $p_x$ is conserved, so we can see immediately that 
the dynamics is two-dimensional for each value of $p_x$. However,  
there is an additional symmetry related to gauge invariance : the invariance of 
H under all transformations of $p_x$ and 
$y$, which keep the value of the difference $p_x - e B y \cos\theta$ 
unchanged. This implies that if a periodic orbit exists for one value 
of $p_x$, then an exact copy of this orbit exists for all $p_x$ 
translated by the distance $\Delta y= \Delta p_x \cos \theta / e B$. 
Combined with the translational invariance in the x-direction this  
means that any periodic orbits can be arbitrarily translated in 
the $x-y$ plane.  This is the classical analogue of the Landau-level 
degeneracy which is preserved in the Hamiltonian (\ref{h3d}). 
We want to rescale our Hamiltonian to eliminate this classical 
degeneracy in $p_x$ as well, so as to define a unique dynamics 
for each value of the total energy. 
This can be achieved by the following canonical transformation: 
 
\begin{eqnarray} 
\xi & = & \frac{x}{l_D} - \frac{\omega_c^{-1} p_y}{m l_D \cos\theta},
 \ \ \ \ \ \ \   
\eta = \frac{y}{l_D} - \frac{\omega_c^{-1} p_x}{m l_D \cos\theta}, 
\ \ \ \ \ \ \  
\zeta = z / l_D  \nonumber \\  
p_\xi & = & \frac{\omega_c^{-1}}{m l_D} p_x \ \ \ \ \ \ \  
p_\eta = \frac{\omega_c^{-1}}{m l_D} p_y \ \ \ \ \ \ \  
p_\zeta = \frac{\omega_c^{-1}}{m l_D} p_z \nonumber \\ 
\tau & = & \omega_c \frac{t} \nonumber \\ 
\end{eqnarray} 
 
which leads to the dimensionless Hamiltonian with two degrees of 
freedom : 
 
\begin{eqnarray}  
H_{\rm eff} & = & \frac{p_{\eta}^2 + p_{\zeta}^2}{2}  
+ \frac{1}{2} \left(\eta \cos\theta - \zeta \sin\theta \right)^2 + 
\zeta + U\left(-\zeta\right) + U\left(\zeta - \frac{d}{l_D} \right)  
\label{h2d} \\ 
& = & \frac{\varepsilon}{\varepsilon_D}   
\end{eqnarray} 
where rescaled energy is measured in units of the  
``drift energy'' $\varepsilon_D = m v_D^2/2$ and may be rewritten  
as  
$$\varepsilon/\varepsilon_D = \frac{v_0^2 B^2}{E^2} \equiv \beta^2/4.$$   
Note that both the coordinate 
$\xi$ {\it and} the momentum $p_{\xi}$ are absent in the scaled hamiltonian 
which is hence truly two-dimensional. 
 
\subsubsection{DBM vs. SBM : $\gamma$ parameter}

The only dependence on the barrier-spacing $d$ in the scaled hamiltonian 
is through the term $U(\zeta - d/l_D)$ representing the emitter barrier. 
As noted, when the total energy of the electron is less than the  
potential drop $e E d $ across the well, the electron can not reach  
the emitter barrier, and the term $U(\zeta - d/l_D)$ 
can be removed from the equation (\ref{h2d}).  In this case, for 
fixed $\theta$, the dynamics is uniquely defined by the value of the 
scaled energy, $\varepsilon/\varepsilon_D \equiv \beta^2/8$. 
This case corresponds to the single-barrier model studied by Shepelyansky 
and Stone \cite{ss}, who first showed that the dynamics of the SBM 
depends only on the parameter $\beta \equiv 2 v_0 B/E$. 
 
When $\varepsilon > e E d$, the electron can 
collide with the emitter barrier and the classical 
motion of the electron in such a case depends essentially on  
{\it both } $d/l_D$ and  $\beta$, 
leading to a more complicated and interesting dynamics. 
Since the crossover between these two regimes is determined by the 
condition $\gamma \equiv \varepsilon / eEd =1$, we re-express 
the parameter $d/l_D$ in Eq. (\ref{h2d}) in terms of the dimensionless 
parameters $\beta,\gamma$: $d/l_D= \beta^2/(8 \gamma)$, so  
that the dynamics 
in the DBM is determined by the values of $\beta,\gamma$. 
This is particularly convenient because in experiments the ratio of 
the emitter state energy to the applied voltage is approximately  
unchanged, so $\gamma$ is approximately constant over the $B-V$ 
parameter space.  Therefore both the dynamics of the SBM {\it and} the  
DBM can be analyzed 
fully by varying a single dimensionless parameter, $\beta$. 
This is how we will proceed in the remainder of this work. 
 
Before making any further analysis of the dynamics we note that there is  
one completely general prediction which follows from the 
scaled hamiltonian of Eq. (\ref{h2d}) if $\gamma$ is constant.  We can write 
\begin{equation} 
\beta^2= \frac{4 \gamma eV}{\varepsilon_D} =  
\frac{8\gamma e d^2}{m} \frac{B^2}{V}, 
\end{equation} 
which implies that for a given $\theta$  
{\it the classical mechanics is constant along parabolic 
boundaries in the $B-V$ plane: $V = (8 \gamma e d^2/m \beta^2) B^2$}. 
This is true of the exact dynamics of the double-barrier model as 
long as $\gamma$ is constant and the variation of effective mass 
with injection energy is negligible. 

\subsection{Poincar\'e Map} 
 
In order to analyze the two-dimensional hamiltonian dynamics 
of the canonical coordinates $(\eta,p_{\eta};\zeta,p_{\zeta})$ 
we use the Poincar\'e surface of section (SOS) method which is 
standard in non-linear dynamics \cite{Lichtenberg,Reichl,Gutzwiller}.   
For fixed values of $\beta$ and $\gamma$ the classical 
trajectories in this four-dimensional phase space lie on 
a 3-dimensional surface determined by energy conservation.  When 
$\theta \neq 0$ the system is non-integrable, 
there is no additional constant of 
motion other than the energy, and there exist chaotic trajectories 
which cover a finite fraction of this three-dimensional surface. 
To define the stability matrix for the periodic orbits and also 
to better visualize the phase-space structure we plot the behavior of 
a set of trajectories on a two-dimensional cross-section of this surface. 
The motion of an electron in the tilted well is bounded 
and all trajectories collide eventually with the collector barrier 
at $\zeta =0$.  Therefore it is convenient to choose the cross section 
to be the plane $(p_{\eta},\eta)$ when  
$\zeta = 0$ ($p_{\zeta}$ being 
then fixed by energy conservation).  If an initial condition is chosen on 
this plane then Hamilton's equations of motion can be used to obtain the values of 
$(\eta,p_\eta)$, when the trajectory again passes through the  
plane $\zeta = 0$.   This procedure defines a Poincar\'e map 
for the tilted well (other choices are possible, e.g.  
the emitter barrier map at $\zeta=d/l_D$  and may be used below). 
 
\begin{eqnarray}  
\eta_{n+1} = \Phi_q\left(\eta_n, (p_\eta)_n\right) \nonumber \\ 
(p_\eta)_{n+1} = \Phi_p\left(\eta_n, (p_\eta)_n \right)  
\label{Pmap} \end{eqnarray} 
 
Since every orbit reaches the collector barrier, {\it every} periodic 
orbit of the hamiltonian (\ref{h2d})  
corresponds to either a fixed point of the 
Poincar\'e map (period-1 orbits) or to a fixed point of the $N$-th iteration 
of the Poincare map (period-$N$ orbits). 
 
Note that the coordinates $\eta$ and momentum $p_\eta$ are  
proportional to the $x$- and $y$- components of the {\it velocity}  
of the electron in the original coordinate system : 
\begin{eqnarray} 
v_x & = & - \frac{ l_D \cos\theta}{T_c} \eta \nonumber \\ 
v_y & = & \frac{l_D}{T_c} p_\eta 
\label{vxvy_etapeta}
\end{eqnarray} 
 
This property allows us to relate the Poincar\'e map (\ref{Pmap}) in the 
coordinates $(\eta,p_\eta)$ to an equivalent Poincar\'e map in more familiar 
coordinates $(v_x/v_0,v_y/v_0) \equiv (\tilde{v}_x, \tilde{v}_y)$, which 
describes  the evolution of 
the velocity components of the electron in the plane perpendicular  
to the collector barrier:  
\begin{eqnarray} 
\left( \tilde{v}_x \right)_{n+1} & = &  
\Phi_x\left( 
\left( \tilde{v}_x \right)_n , 
\left( \tilde{v}_y \right)_n   
\right) \nonumber \\ 
\left( \tilde{v}_y \right)_{n+1} & = &  
\Phi_y\left( 
\left( \tilde{v}_x \right)_n , 
\left( \tilde{v}_y \right)_n  
\right) \label{PmapV} \\ 
\end{eqnarray} 
where the relations between $\Phi_x, \Phi_y$ and $\Phi_q, \Phi_p$
follow from Eqs. (\ref{vxvy_etapeta}) and (\ref{Pmap}).
 
Note that we have scaled the velocities by the maximum allowed  
velocity $v_0$ so that 
the values of this Poincar\'e map will be contained within the unit circle, 
independent of the energy (this would not be true of the variables $(\eta,p_{\eta})$ 
as the size of the energetically allowed region of the plane varies with the scaled 
energy $\beta^2/4$).  Although the variables $(\eta,p_{\eta})$ were  
most convenient for discussions  
of scaling, we will use the energy-scaled velocity map (\ref{PmapV})  
henceforth since it 
is easiest to interpret and compare for varying $\beta$ values. 
 
\begin{figure}[htbp]
\begin{center}
\leavevmode
\epsfbox{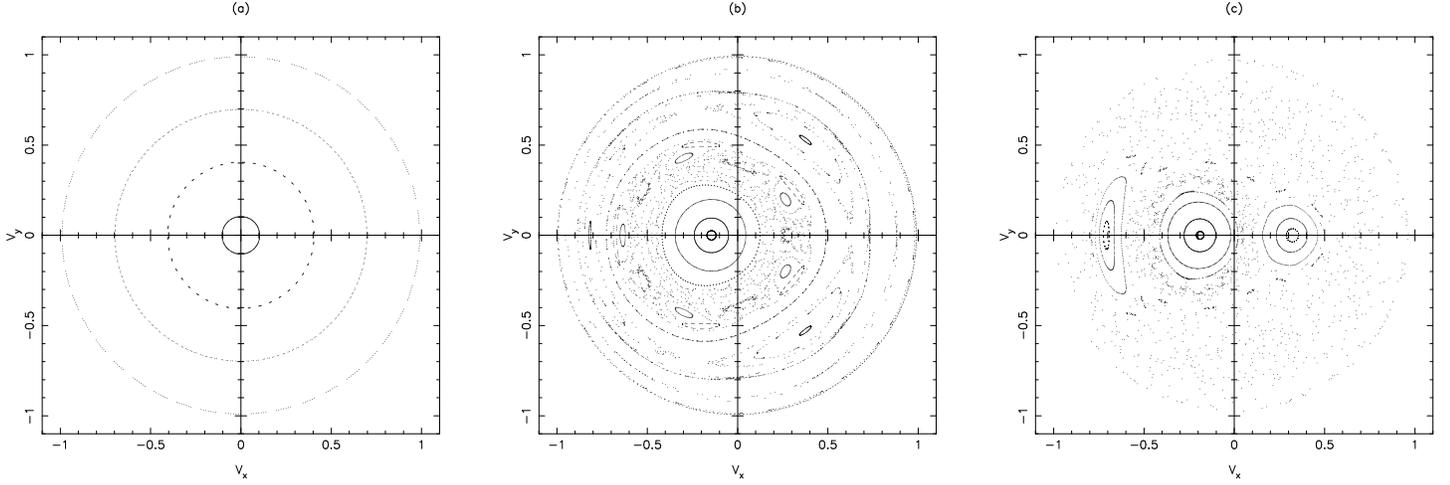}
\end{center}
\caption{
Three Poincar\'e surfaces of section for experimenatally relevant 
$\gamma = 1.17$ at : 
(a) $\theta = 0^\circ$, $\beta = 2$,
(b) $\theta = 20^\circ$, $\beta = 3.2$, (c) $\theta = 20^\circ$,
$\beta = 4.$
\label{fig_ps_dbm} 
}
\end{figure}

A plot of the Poincar\'e map (\ref{PmapV}), which is called {\it 
Surface of Section} (SOS) is generated by choosing a grid of initial 
conditions in the plane $(v_x/v_0, v_y/v_0)$  
corresponding to a particular value of $\beta$ and iterating the map 
many times for each initial condition.  Period-$N$ stable orbits 
appear as ``chains'' of $N$ ``islands''; whereas period-$N$ unstable orbits 
will be imbedded in the chaotic layers between the islands \cite{Reichl} 
and are not evident to the (untrained) eye.  In Fig. 
\ref{fig_ps_dbm} we show several 
examples of the collector barrier SOS as $\beta$ is increased for fixed 
$\gamma = 1.15$ (which corresponds to the approximate value in the 
relevant experiments \cite{Muller}). 
 
When $\theta=0$ the squared distance of a point in the SOS from the origin 
is proportional to the cyclotron energy, which is conserved, so each 
trajectory must lie on a circle (see Fig. \ref{fig_ps_dbm}a).  When 
$\theta \neq 0$ (Fig. \ref{fig_ps_dbm}b) 
we immediately see the appearance of stable islands and chaotic 
layers, coexisting with slightly distorted circular curves which represent 
the unbroken tori according to the standard KAM scenario \cite{Lichtenberg}. 
For larger $\beta$ (Fig. \ref{fig_ps_dbm}c) no KAM curves survive  
and the entire SOS is chaotic 
except for a few surviving stable islands, which however typically 
represent the 
features of most importance for the experimental tunneling oscillations. 
 
We now undertake a more explicit determination of the properties of the 
Poincar\'e map for the tilted well. 
To calculate the functions $\Phi_p$ and $\Phi_q$ of the Poincar\'e map, one 
has first to analyze the motion of the electron between collisions.  
This motion is integrable and is most easily represented  
in a frame of reference (denoted by $(x',y',z')$),  
rotated by the tilt angle $\theta$ around the $x$ axis, so  that  
$z'$ is parallel to the direction of the magnetic field : 
 
\begin{eqnarray} 
x' & = & x \nonumber \\  
y' & = & y \cos\theta - z \sin\theta \nonumber \\ 
z' & = & y \sin\theta + z \cos\theta \nonumber  
\end{eqnarray} 
 
In this frame of reference the motion of the electron in the $(x',y')$ 
plane between collisions is a superposition of the cyclotron rotation 
with the frequency $\omega_c \equiv 2\pi /T_c$ and a uniform drift along $x'$ 
with the velocity $v_{d} = E \sin\theta/B \equiv v_D
\sin\theta$, while the longitudinal motion is a uniform acceleration :
\begin{eqnarray} 
v_{x'}(\tau) & = & v_c \cos(\phi^0 + \tau)  
- v_{d}  \nonumber \\ 
v_{y'}(\tau) & = & v_c \sin(\phi^0 + \tau)  
\nonumber \\ 
v_{z'}    & = & v_{z'}^0 - \frac{E \cos\theta }{m} t  
 =  v_{z'}^0 - \frac{l_D \cos\theta }{T_c} \tau  
\label{vzprime}
\end{eqnarray}   
where $v_c$ is the cyclotron velocity (which remains constant 
between collisions) and $\phi^0$ is the initial phase of the  
cyclotron rotation.  
 
The energies associated with the transverse (cyclotron)  
and longitudinal motion are separately conserved between collisions. 
For $\theta \neq 0$ the cyclotron and longitudinal motions get mixed by the 
collisions with the barriers \cite{ss}: 
 
\begin{eqnarray}  
\bar{v}_{z'} & = & - \cos(2\theta) v_{z'} + \sin(2\theta) v_{y'} 
\nonumber \\ 
\bar{v}_{y'} & = & \sin(2\theta) v_{z'} + \cos(2\theta) v_{y'} 
\nonumber \\ 
\bar{v}_{x'} & = & v_{x'} \label{eqns_reflect} 
\end{eqnarray} 
 
where ${\bf v}$ and ${\bf \bar{v}}$ are the velocities immediately 
before and after collision respectively.  This transformation is 
equivalent to a clockwise rotation of the velocity vector by $2 \theta$  
in the $(y'-z')$ plane, followed by a reflection $v_{z'} \to -v_{z'}$; 
hence it leaves no vector in this plane invariant (for $\theta \neq 0$).  
Therefore generically there {\it is} exchange of kinetic energy between 
the longitudinal and cyclotron motion at each collision  
 
\begin{eqnarray} 
\delta \varepsilon_{L \leftrightarrow c} = \frac{m}{2}  
\left( v_{z'} \cos\theta - v_{y'} \sin\theta \right)^2 \equiv  
\frac{m}{2} v_y^2 
\label{energy_exchange}
\end{eqnarray} 
and the dynamics is non-integrable. 
 
Note that it is {\it possible} to have zero energy exchange upon collision  
for $\theta \neq 0$. The condition for this is simply that $v_y=0$ at collision, 
i.e. the cyclotron phase is such that the instantaneous motion is in the $x-z$ plane. 
The reason that no energy is exchanged in this case is that the  
impulse at collision is purely in the 
z-direction and reverses this component of velocity leaving $v_x$ and 
$v_y$ unchanged.  If $v_y=0$ at the time of collision then  
$v_{z'} = v_z \cos \theta \to \bar{v}_{z'}=-v_z \cos \theta = -v_{z'}$ 
and the longitudinal kinetic energy is conserved.  
Stable period-one orbits with $v_y=0$ ($p_{\eta}=0$) are 
visible in both Figs. \ref{fig_ps_dbm}b,\ref{fig_ps_dbm}c.  
We refer to these as {\it non-mixing} 
orbits since they involve no energy exchange; they will play a fundamental 
role in the periodic orbit theory developed below.  
 
The transformation equations for ${\bf v'}$ due to collisions at the emitter 
barrier are identical to (\ref{eqns_reflect}).  As we shall see below, 
it is useful to consider 
the dynamics in yet a third frame of reference which is parallel to the primed 
frame, but moving with the drift velocity $v_{d}$ in the $x'$ direction. 
In this moving frame the transverse motion is pure cyclotron rotation and each 
iteration of the Poincar\'e map is just a pair of non-commuting orthogonal 
transformations of the velocity: first the continuous cyclotron 
rotation around the $z'$ axis, followed by the instantaneous 
rotation/reflection around the $x'$ axis.  Since the latter is known explicitly 
(Eq. (\ref{eqns_reflect}), to get an explicit formula for the 
Poincar\'e map what is needed is 
an expression for the increment in the cyclotron phase between collisions. 
However, there is no simple general formula for this phase increment 
for $\gamma > 1$ because after a collision with the collector barrier 
an orbit may or may not have enough longitudinal energy to collide 
with the emitter barrier before its next collision with the collector. 
Since $v_{y'}$ changes discontinuously in a 
collision, the cylotron phase increment will change discontinuously 
due to the emitter collision.  If one varies the initial conditions of a trajectory 
so that it ceases colliding with emitter barrier in the next iteration of the map, one 
can show that the phase jump goes to zero as the impulse at the emitter 
goes to zero (i.e. as $v_z$ 
at collision goes to zero), but its derivative is discontinuous. 
Hence, in general the 
Poincar\'e map for $\gamma > 1$ does not have continuous derivatives  
everywhere on the surface of section.  As a consequence the stability matrix of periodic  
orbits for the exact map for $\gamma >1$ is not always defined. 
This has significant and novel consequences for 
the behavior of periodic orbits in the DBM: these can vanish without  
reaching marginal stability in a new kind of bifurcation we will refer 
to as a {\it cusp bifurcation}.  We shall return to this in detail below.  
 
As a result of this discontinuous behavior we can only present  
a simple explicit form of the Poincar\'e map in certain limiting 
cases. The simplest of these, previously analyzed by Shepelyansky and  
Stone \cite{ss}, is when $\gamma < 1 (\varepsilon < eV)$, 
in which case no orbit reaches the emitter barrier and classically the 
problem is equivalent to the motion of an electron in an infinite  
triangular well in a tilted $B$ field.  We now briefly review this limit. 
 
\subsection {The Single-Barrier Model (SBM)} 
 
When $\gamma \leq 1$, the cyclotron phase increment between collisions 
with the collector barrier is $\omega_c t_0$, where $t_0$ is the time it  
takes the electron launched ``upwards'' after the collision 
in the effective electric field, ${\bf E} \cos \theta$, to fall back  
down and hit the collector. The resulting 
Poincar\'e map takes the form : 
\begin{eqnarray} 
\Phi_x(\tilde{v}_x,\tilde{v}_y) & = & 
{\cal V}_x\left(\tilde{v}_x, \tilde{v}_y, 
\tilde{v}_z;\omega_c t_0 \right)
\nonumber \\
\Phi_y(\tilde{v}_x,\tilde{v}_y) & = & 
{\cal V}_y\left(\tilde{v}_x, \tilde{v}_y,
\tilde{v}_z; \omega_c t_0 \right)
\label{PmapVSBM}
\end{eqnarray}
where 
\begin{eqnarray}
{\cal V}_x\left(\tilde{v}_x, \tilde{v}_y,
\tilde{v}_z; \tau \right)
& = & 
\tilde{v}_x \cos(\tau) 
- \tilde{v}_y \cos\theta \sin(\tau) 
\nonumber \\  
& + & \tilde{v}_z \sin\theta 
\sin(\tau) - (2/\beta) \sin\theta (1 - \cos(\tau)) 
\nonumber \\ 
{\cal V}_y\left(\tilde{v}_x, \tilde{v}_y,
\tilde{v}_z; \tau \right)
& = & 
\tilde{v}_x \cos\theta \sin(\tau) 
+ \tilde{v}_y  
\left(  
\cos^2\theta \cos(\tau) + \sin^2\theta  
\right) \nonumber \\  
& + & \tilde{v}_z \sin\theta \cos\theta 
\left(1 - \cos(\tau) \right) \nonumber \\ 
& + & (2/\beta) \sin\theta \cos\theta  
\left(\sin(\tau) - \tau\right),  \label{define_Vfunc} 
\end{eqnarray} 
the scaled velocity ${\bf \tilde{v}} \equiv {\bf v} / v_0 $ 
(with $\tilde{v}_z(\tilde{v}_x,\tilde{v}_y) 
\equiv  \sqrt{1 -  \tilde{v}_x^2 -  \tilde{v}_y^2}
> 0$) 
and the time interval $t_0(\tilde{v}_x, \tilde{v}_y)$ 
between successive collisions  
of the electron with the collector barrier is the  
first positive root of the equation : 
\begin{eqnarray}
0 & = & z(t_0) \equiv \frac{v_0 {\cal Z}\left(\tilde{v}_x,
\tilde{v}_y, \tilde{v}_z; \omega_c t_0 \right)}{\omega_c}
\label{dtSBM}
\end{eqnarray}
where the function ${\cal Z}(\tilde{v}_x, \tilde{v}_y, \tilde{v}_z;
\tau)$ is defined as 
\begin{eqnarray} 
 {\cal Z}\left(\tilde{v}_x,
\tilde{v}_y, \tilde{v}_z; \tau\right) 
& = &    
- \tilde{v}_x \sin\theta \left( 1 - \cos(\tau) \right)  
+ \tilde{v_y} \sin\theta \cos\theta  
\left( \tau - \sin( \tau) \right) 
\nonumber \\ 
& + & \tilde{v}_z
\left( \tau \cos^2 \theta + \sin^2 \theta \sin(\tau ) 
\right) \nonumber \\ 
& - & (2/\beta) \left( \sin^2 \theta (1 - \cos(\tau )) +  
\cos^2\theta \frac{\tau^2 }{2} \right)    
\label{define_Zfunc} 
\end{eqnarray} 
 
If $\omega_c T \gg 1$, an approximate root is found easily, 
\begin{eqnarray} 
T = \frac{\beta \tilde{v}_{z'}}{\cos\theta}. 
\end{eqnarray} 
In this approximation the map when transformed to the $(x',y',z')$ 
coordinates becomes identical \cite{ss} 
to the kicked-top map introduced by Haake \cite{Haake},
\cite{explain_kicked-top}. 
 
As is indicated by the numerical analysis of both the kicked-top 
map and of the exact mapping  
(\ref{PmapVSBM}),  the KAM transition to chaos takes place 
when $\theta \beta \sim 1$.  We therefore take the limit 
$\beta \gg 1$ and $\theta \ll 1$. In this case both 
the kicked-top map and  
the exact map (\ref{PmapVSBM}) in the vicinity  
of a particular value of $ {\tilde{v}_{z'}} = \tilde{v}'$  
can be expressed precisely in the 
form of a local standard map (kicked rotor) \cite{Chirikov},  
\cite{Lichtenberg}  
\begin{eqnarray} 
I_{n+1} & = & I_n + K \sin\phi_{n+1} \nonumber \\ 
\phi_{n+1} & = & \phi_n + I_n  
\label{standard_map} 
\end{eqnarray} 
where 
\begin{eqnarray} 
I_n & = & \beta  {\tilde{v}_{z'}} \nonumber \\ 
K & = & 2 \theta \beta  \sqrt{1 - (\tilde{v}')^2}  
\end{eqnarray} 
and $\phi$ is the phase of the cyclotron rotation. 
 
The map is called local because the kick strength varies with $v_{z'}$, 
so that the chaos boundary, given by the condition\cite{Chirikov} 
$K \approx 1$  varies with $v_{z'}$.  The resulting  
condition for chaos as an explicit  
function of all system parameters is \cite{ss}: 
\begin{eqnarray} 
B^2 > \frac{m E \varepsilon}{32 e \theta^2 \varepsilon_c } 
\label{chaos_boundary} 
\end{eqnarray} 
where $\varepsilon_c \equiv \varepsilon (1 - (\tilde{v}')^2)$ is the  
instantaneous energy of the cyclotron motion. 
 
Athough the estimate $(\ref{chaos_boundary})$ was obtained only in the 
limiting case $\theta \ll 1$ and $\beta \gg 1$, it does predict the  
correct behavior 
 of the exact mapping (\ref{PmapVSBM}) for the SBM. 
Qualitatively it predicts that chaos increases with increasing magnetic field  
and energy and with decreasing electric field and quantitatively the 
condition given by Eq. (\ref{chaos_boundary}) is in good agreement with 
the onset of complete energy exchange between the cyclotron and longitudinal 
motion as determined from simulations of the exact map \cite{ss}.

\subsection{The Double-Barrier Model (DBM) } 
 
When $\gamma = \varepsilon/eV > 1$, the electron can
 retain enough longitudinal  
energy on collision with the collector barrier to reach the  
emitter wall, although it need not do so.  If we regards the coordinates 
$(\tilde{v}_x, \tilde{v}_y)$ in the SOS as initial conditions for the next 
segment of the trajectory, we may partition the SOS into inner and outer 
regions. 
Initial conditions $(\tilde{v}_x, \tilde{v}_y)$ in the inner region 
will define 
all trajectories which collide with the emitter before their next collision 
with the collector.  For such initial conditions the equation  
\begin{eqnarray} 
z\left( t \right) \equiv \frac{v_0 {\cal Z}\left(\tilde{v}_x,
\tilde{v}_y, \tilde{v}_z; \omega_c t \right)}{\omega_c}
& = & d \equiv \frac{v_0}{\omega_c} \frac{\beta }{4 \gamma} 
\label{dtUpDBM} 
\end{eqnarray} 
where the function ${\cal Z}$ was defined in (\ref{define_Zfunc}), 
must have a positive root  $t = t^{\uparrow}$, which corresponds 
to the time interval to the next collision with the emitter barrier. 
 
For initial conditions in the outer region 
Eq. (\ref{dtUpDBM}) has no positive roots, the electron 
does not reach the emitter barrier before the next collision with the 
collector barrier, and it's trajectory is exactly the same as in the  
SBM for this iteration of the map. Hence the Poincar\'e map is still
given by the expression (\ref{PmapVSBM}).

\begin{figure}[htbp]
\begin{center}
\leavevmode
\epsfysize=1.8in 
\epsfbox{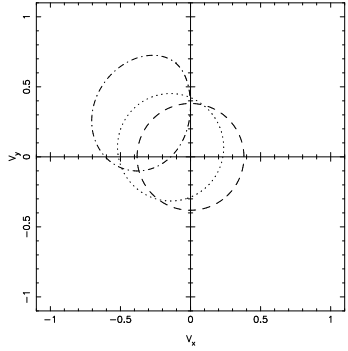}
\end{center}
\caption{
The critical boundary, separating initial conditions such that
the electron will reach the emitter barrier before the next 
collision with the collector wall (region enclosed by the
critical boundary) from those when the electron returns to
the collector wall without striking the emitter barrier
(the region outside the critical boundary).  $\gamma = 1.17$,
and (a) $\theta = 0^\circ$ (dashed line), (b) 
$\theta = 15^\circ$, $\beta = 3.$ (dotted line),
(c) $\theta = 30^\circ$, $\beta = 5$ (dashed-dotted line).
\label{fig_critical_boundary}
}
\end{figure}

The ``critical boundary'' betweeen the two regions is the curve 
$(\tilde{v}_x^{(c)},\tilde{v}_y^{(c)})$,  such that the electron
lunched from the collector barrier with the velocity 
$ {\bf v} = v_0 (\tilde{v}_x^c, \tilde{v}_y^c, \tilde{v}_z^c)$, 
will reach the emitter wall emitter wall with component of the total 
velocity perpendicular to the plane of the barrier equal to zero. 
For $\theta = 0$  
the critical boundary is a circle given 
by the equation : 
\begin{eqnarray} 
\tilde{v}_x^2 + \tilde{v}_y^2 = 1 - 1/\gamma \equiv  
\frac{\varepsilon - eV}{\varepsilon} 
\end{eqnarray} 
In Fig. \ref{fig_critical_boundary} we show a few examples 
of the  ``critical boundary'' for  
different values of $\beta$ and $\gamma$.  It is important to realize 
that in general trajectories can cross the critical boundary and indeed 
for large chaos parameter almost all trajectories do. However knowledge 
of the critical boundary is useful for formulating the Poincar\'e map 
of the DBM. 
 
\begin{figure}[htbp]
\begin{center}
\leavevmode
\epsfysize=2.5in 
\epsfbox{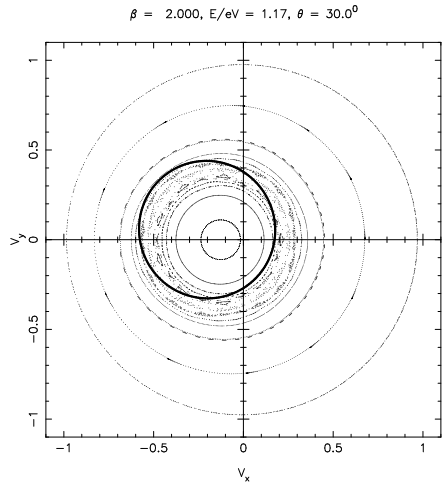}
\end{center}
\caption{
The Poincar\'e surface of section for $\gamma = 1.17$, and
$\beta = 2$, $\theta = 30^\circ$. The chaotic region near 
the critical boundary (thick solid line) is the 
``chaotic halo'', created by the nonanaliticity of the map.
\label{fig_halo}
}
\end{figure}

For $(\tilde{v}_x,\tilde{v}_y)$ outside the  
critical boundary, the next iteration of the Poincare map does not  
involve the collision with the emitter barrier,  
and the Poincare map is therefore still given by ({\ref{PmapVSBM}), as  
in the single barrier model. 
 
When  $(\tilde{v}_x,\tilde{v}_y)$ is inside the 
critical boundary, then the Poincar\'e map is given by : 
\begin{eqnarray} 
\Phi_x(\tilde{v}_x,\tilde{v}_y) & = &  
{\cal V}_x\left(\tilde{v}_x^e, \tilde{v}_y^e, 
\tilde{v}_z^e ; \omega_c t^{\downarrow} \right)
\nonumber \\
\Phi_y(\tilde{v}_x,\tilde{v}_y) & = &  
{\cal V}_y\left(\tilde{v}_x^e, \tilde{v}_y^e, 
\tilde{v}_z^e ; \omega_c t^{\downarrow} \right)
\label{PmapVDBM}
\end{eqnarray} 
where $\tilde{\bf v}_e$ is the scaled velocity immediately after 
collision with the emitter barrier and can be obtained as 
\begin{eqnarray} 
\tilde{v}_x^e & = & 
{\cal V}_x\left(\tilde{v}_x, \tilde{v}_y, 
\tilde{v}_z ; \omega_c t^{\uparrow} \right)
\nonumber \\
\tilde{v}_y^e & = & 
{\cal V}_x\left(\tilde{v}_x, \tilde{v}_y, 
\tilde{v}_z ; \omega_c t^{\uparrow} \right)
\nonumber \\
\tilde{v}_z^e & = & - \sqrt{1 - \tilde{v}_x^2 - \tilde{v}_y^2} 
\end{eqnarray}
$t^{\uparrow}$ is defined as the time interval until the 
next collision with the emitter barrier and is given by the first 
positive root of the equation (\ref{dtUpDBM}), and the parameter  
$t^{\downarrow}$ represents the time interval between the  
collision with the emitter barrier and the next collision with  
the collector map. The value of $t^{\downarrow}$ can be obtained 
from the equation   
\begin{eqnarray} 
d + \frac{v_0}{\omega_c} 
{\cal Z} \left( \tilde{v}_x^e, \tilde{v}_y^e,\tilde{v}_z^e ;
\omega_c t^\downarrow \right) = 0 
\end{eqnarray}
 
As noted above, an important property of the 
Poincar\'e map  (\ref{PmapVDBM}) is  
that it has a discontinuous derivative as the initial conditions  
$(\tilde{v}_x, \tilde{v}_y)$ are varied across the critical boundary. 
Therefore the conditions for the global validity of the KAM theorem are 
not satisfied by this map and the transition to chaos can be discontinuous 
here as in the stadium billiard \cite{stadium}.  However unlike the stadium 
billiard not all trajectories are affected by the discontinuity of the map 
for arbitrarily small chaos parameter.  Away from the critical boundary the 
map satisfies all the conditions for the existence of KAM tori and, for small 
chaos parameter, in the inner and outer regions there will exist an outermost 
and innermost KAM torus.  These two tori  
will define a set of trajectories which either 
always hit the emitter barrier (lie within the outermost KAM curve of the inner region) or 
always miss the barrier (lie outside the innermost KAM curve of the outer region). 
Between these two tori the non-analyticity of the map is felt by the 
trajectories and the numerics demonstrates clearly that there are no remaining 
KAM curves in an annular region bounded approximately by the maximum and 
minimum cyclotron energies of points on the critical boundary.  In this region 
the chaos does not appear to be associated with the separatrices corresponding 
to the hyperbolic fixed points as it would be for small chaos parameter in 
a KAM system.  The practical consequence is that one observes an anomalously large 
``chaotic halo'' around the critical boundary (see Fig. \ref{fig_halo}).  
In this region the effective map description fails 
badly and only analysis of the exact map can be used.  
In fact, as we shall see 
below, many of the important short periodic orbits first appear 
{\it at} the critical boundary 
at a finite value of $\beta$ and emerge from the chaotic halo region 
with increasing $\beta$. 
We will be able to develop an analytic theory of the simplest such orbits 
from the exact map. 
 
Although the effective map based on the SBM fails in the ``halo'' region, 
for small chaos parameter and small $\theta$ it should work just as
well in the 
outer region of the SOS as it does in the SBM, since here the trajectories  
are prevented by the innermost KAM curve from reaching the emitter and the 
DBM Poincar\'e map is {\it identical} to the SBM.  Since the local
chaos parameter 
in the effective map description of the SBM is  
$K = 2 \beta \theta \sqrt{1 - (\tilde{v}')^2}$ the chaos 
is weakest at the innermost KAM curve of the outer region (since the cyclotron 
energy is the smallest there) and this curve is the last in the outer 
region to break.  The quantitative prediction for the breaking of this curve  
from the local standard map approximation (Eq. (\ref{standard_map})) 
is in a good agreement with the exact behavior.
 
One may try to extend similar reasoning to the inner region of trajectories 
which always reach the emitter barrier.  Here the effective map is clearly 
somewhat different because of the additional  
energy exchange (``kick'') at the emitter barrier. 
It {\it is} possible to obtain an effective area-preserving map for 
small tilt angles which is similar to a standard map with two 
unequal kicks per period.  However the SOS generated by this approximation 
has little similarity to the exact map.  This is because when the energy  
is almost completely 
longitudinal (as it is in this region of phase space)  
the kick strength goes to zero at leading order in the 
tilt angle and the effective map description fails.  
Note that it is precisely the periodic orbits in the inner region 
(which reach the emitter) which are measured in the tunneling spectrum. 
Thus we are particularly interested in obtaining a good description of this region 
of phase space and must work with the exact map described 
by Eqs. (\ref{PmapVDBM}).   
 
Fortunately, as we show below, it is possible to obtain 
a good theoretical understanding of the short periodic orbits in 
the entire phase space, including the crucial central region of the SOS,  
based on analysis of the exact map.  In fact we are able to obtain 
analytic expressions for the period and stability of  
an infinite class of important periodic orbits for arbitrarily large 
values of the chaos parameter.   
} 

\section{Periodic Orbit Theory (Single-Barrier Model)}
 
\subsection{Integrable Behavior}

Eq. (\ref{tun_rate}) of Section I gives a quantitative semiclassical formula
for the tunneling current through the tilted well in terms of 
the contributions of different periodic orbits which connect emitter
and collector barriers.  Clearly these  
orbits can be fully described only within the framework of the double 
barrier model.  Nevertheless, the behavior of the periodic orbits 
in the DBM as a function of tilt angle and $\beta$ is exceedingly 
complex and has not been understood systematically up to this point.
In order to develop such a systematic understanding it is very helpful
to consider the SBM, which has a similar but simpler periodic orbit
structure. The similarity between the two models is easily seen
by considering the limit of zero tilt angle.
 
When $\theta = 0$, both systems are integrable and all of the periodic 
orbits can be divided into two groups. 
A {\it single} traversing orbit (TO) bouncing perpendicular to 
the barrier(s) with zero cyclotron energy and infinite families
of helical orbits (HO) with periods equal to an integer multiple of the
cyclotron period, $2\pi/\omega_c$.  The traversing orbit
corresponds to the fixed point of the Poincar\'e map in the centre  
$(0,0)$ of the surface of section - see Fig. \ref{fig_ps_dbm}; its period
is given by 
\begin{eqnarray} 
T_{\rm TO} & = & \frac{\beta}{\omega_c}  
\ \ \ \ \ \ \ \ \ \ ({\rm SBM})  
\label{period_to_sbm} \\ 
T_{\rm TO} & = & \frac{\beta}{\omega_c}  
\left( 1 - \sqrt{1 - \frac{1}{\gamma}} \right)  
\ \ \ ({\rm DBM}) \label{period_to_dbm}
\end{eqnarray} 
Unlike all other one-bounce orbits, the TO exists for arbitrarily
small energy, since its frequency need not be in resonance with the
cyclotron frequency.  Since it has zero cyclotron energy
its semiclassical quantization yields the states of the well with
Landau index equal to zero, and hence the TO 
determines the sub-band energy spacings of  
the triangular (SBM) or trapezoidal (DBM) well by the semiclassical 
rule for integrable systems: $\Delta \varepsilon = \hbar / T_{\rm TO}$. 
 
Due to the rotational invariance of the system at zero tilt angle all  
other periodic orbits in the well (in both the SBM  
and DBM) exist in degenerate families related by rotation around the
z-axis.   The union of all trajectories in a family defines a
torus in phase-space, known as a ``resonant'' torus 
in the nonlinear dynamics literature \cite{Reichl} because the
periodic motion of the two degrees of freedom are commensurate:
\begin{eqnarray} 
n \omega_c & = & k \omega_L 
\label{resonance} 
\end{eqnarray}  
where $n$ and $k$ are integers (which do not have a common divisor)  
and $\omega_L$ is the frequency of  
the periodic motion in the longitudinal direction).
Since longitudinal and transverse motion decouple, $\omega_L$ 
 is the frequency of  
the periodic motion of the uniformly accelerated electron bouncing 
normal to the barriers, and  it's value is : 
\begin{eqnarray} 
\omega_L & = & \frac{2 \pi \omega_c}{\beta \sqrt{\tilde{ \varepsilon}_L}}  
\ \ \ \ \ \ \ \ \ \  \ \ \ \ \ \ \ \ \ \  
({\rm SBM}) 
\label{wl_sbm} \\ 
\omega_L & = & 
 \frac{2 \pi \omega_c}{\beta \sqrt{ \tilde{\varepsilon}_L } }  
\times 
\left\{ 
\begin{array}{ll} 
1  &  \gamma \tilde{\varepsilon}_L < 1  \\ 
\left( 1 - \sqrt{1 - \frac{1}{\gamma \tilde{\varepsilon}_L} } \right)^{-1} 
&\gamma \tilde{\varepsilon}_L \geq 1  \\ 
\end{array} 
\right. 
\  \ \ \ \ \ \ \ ({\rm DBM})  
\label{wl_dbm} 
\end{eqnarray} 
where $\tilde{\varepsilon}_L \equiv \tilde{v}_z^2$ is the 
scaled longitudinal energy.

The resonance condition (\ref{resonance}) means that any periodic orbit  
of a family labelled by the integers $n$ and $k$ collides   
with the collector barrier $n$ times while making $k$ full
cyclotron rotations before retracing itself.
Therefore all such orbits in real-space trace out rational 
fractions of a helix (hence the term helical orbits) 
between successive collisions and have periods given by 
\begin{eqnarray} 
T_{\rm HO} & = & \frac{2 \pi k}{\omega_c} \label{t_ho} 
\end{eqnarray} 
for both the SBM and DBM.

A simplifying feature of these systems is that one of the oscillation
periods, the cyclotron period, is independent of energy
and voltage.  The longitudinal period varies with both energy
and voltage, going to zero as longitudinal energy tends to zero.  
If a family of helical orbits $(n,k)$ exists at a given energy, a family 
of the same type can be generated at a lower energy by simply
removing cyclotron energy (hence reducing the cyclotron radius)
until the radius of the helix shrinks to zero, at which point this
``family'' has become degenerate with the TO and ceases to exist.
These degeneracy points occur then, whenever 
the period of the traversing orbit $T_{\rm TO}$  
passes through the value  $k T_c / n $, for both
the SBM and DBM. 
 
When the magnetic field is tilted 
the rotational symmetry around the
field direction which was the 
origin of continuous families of helical orbits in 
the well is broken and {\it all} the resonant tori are destroyed. According  
to the Poincar\'e - Birkhoff theorem \cite{Lichtenberg} 
each of them is replaced by an integer number of pairs 
of stable and unstable orbits (normally just a single pair).
The degeneracy points of the untilted system, at which an $(n,k)$ 
resonant torus collapsed, evolve into n-fold bifurcations
of the TO. 

The reason that the periodic orbit theory of the DBM is more complicated
than that of the SBM stems from two facts. 1) In the unperturbed
DBM there are two distinct families of orbits for each pair $(n,k)$
(one which reaches the emitter and one which doesn't), 
whereas there is only one such family in the SBM.  2) These families
can collapse at the critical boundary and not just by reaching
degeneracy with the TO.  However in all the other respects mentioned
above the two models are similar, and in particular, the bifurcations
near the TO, which are crucial for explaining the experimental data
of Muller et al.\cite{Muller}, are very similar in the two models.  
We thus begin with the simpler case of the SBM \cite{SBMrealizable}.
 
\subsection{Periodic Orbits at $\theta = 0$} 
 
As just noted, the periodic orbits at $\theta=0$ are of two types:
the (usually) isolated traversing orbit and the families of helical
orbits.  The TO, with no cyclotron energy has a period which is 
independent of magnetic field and monotonically increasing from zero 
with increasing energy:
\begin{eqnarray} 
T_{\rm TO} = \frac{2 \sqrt{2 m^* \varepsilon}}{eE} \equiv  
\frac{\beta}{\omega_c} 
\label{t_to_sbm}
\end{eqnarray} 
For all HO's the period is finite and an integer multiple of 
$T_c=2\pi/\omega_c$.  Thus a given family of HO's
labelled by $(n,k)$ can only exist above the energy at which
$n T_{TO}=k T_c / n$.  These threshholds
are the degeneracy points discussed above. At the threshhold
all energy longitudinal ($\tilde{varepsilon}_L = 1$); together
with  (\ref{resonance}),(\ref{wl_sbm}) this yieds : 
\begin{equation}
\beta_{(n,k)} = \frac{2\pi k}{n}.
\end{equation}
Since $0 \leq \tilde{varepsilon}_L \leq 1$, 
for values of $\beta > \beta_{(n,k)}$ there always exists exactly
one root of the equation
\begin{eqnarray} 
\tilde{\varepsilon}_L(n,k) = \left(\frac{2 \pi k}{\beta n} \right)^2   
\end{eqnarray} 
where $\tilde{\varepsilon}_L = \tilde{v}_z^2$ is the {\it scaled}  
longitudinal energy. The scaled cyclotron energy for this family  
(resonant torus) is just $\tilde{v}_c^2 = 1 - \tilde{\varepsilon}_L$.
As the value of $\beta$ is increased, the existing helical  
orbits gain more cyclotron energy and move away from the traversing  
orbit, allowing for the creation of new families of HO near the TO.
We will now analyze what happens to the shorter
periodic orbits as the magnetic field is tilted, beginning with the
one-bounce orbits.

\subsection{One-bounce orbits} 

\subsubsection{Continuity argument} 
 
\begin{figure}[htbp]
\begin{center}
\leavevmode
\epsfysize=4in 
\epsfbox{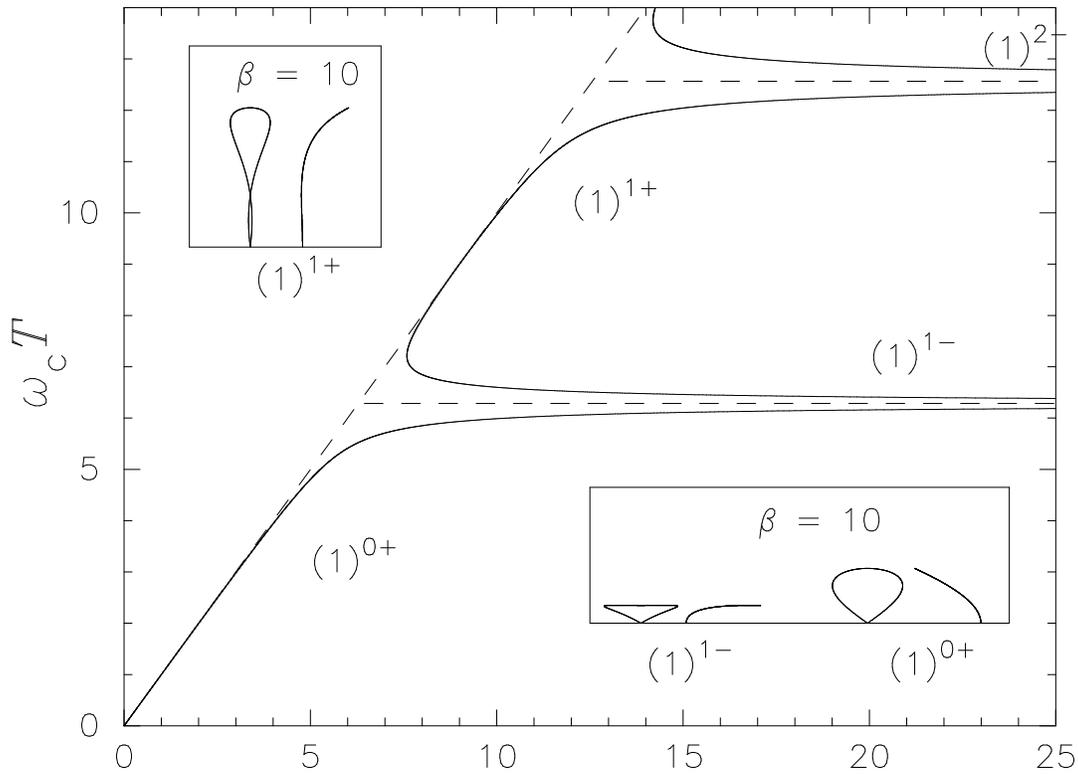}
\end{center}
\caption{
Periods of one-bounce orbits as functions of $\beta$ for 
the tilt angle $\theta = 11^\circ$. The dashed lines 
corresponds to the periods of one bounce orbits at zero 
tilt angle. The insets show the $y-z$
projections of the three existing one-bounce orbits 
at $\beta = 10$.
\label{fig_t_po1_sbm}
}
\end{figure}

One-bounce orbits are periodic orbits which have retraced themselves
between each bounce off the single barrier, i.e. they are fixed points of
the first iteration of the Poincar\'e map.  Note that different
one-bounce orbits may have widely differing periods, and may for
instance have periods longer than two or three bounce orbits.  
For $\theta=0$ the existing
one-bounce orbits consist of the TO and all HO families with $n=1$
which are above threshold, i.e. with $k< \beta /2\pi$.  The bahavior of 
the periods of these orbits 
is indicated by the dashed lines in Fig. \ref{fig_t_po1_sbm}.  
Since the periods $T$ of the HO families are fixed to be integer multiples
of $T_c$ they are independent of $\beta$ when we plot $\omega_c T$.
 
When the magnetic field is infinitesimally tilted, all helical families
(resonant tori) 
are immediately destroyed and replaced by pairs of stable  
and unstable periodic orbits. These surviving one-bounce orbits are
only infinitesimally distorted from their analogs at $\theta=0$ and
by continuity the periods of these orbits are also only infinitesimally
altered.  For our system it is clear which orbits from each infinite
family survive.  For each helical family there are exactly two orbits
which collide with the barrier with $v_y=0$, the condition for zero
energy exchange according to Eq. (\ref{energy_exchange}).  
It is these two orbits from
each family which survive.  This is easily seen by recalling that
longitudinal and transverse energy are separately conserved between
collisions even in the tilted system, so any {\it one-bounce} periodic
orbit for arbitrary tilt angle must also conserve these quantities
during the collision.  But the condition for this is just
$v_y=0$, which is satisfied for the two one-bounce helical orbits from each
family which hit with $v_x=\pm v_c$. By continuity these two orbits
must evolve into the two surviving isolated fixed points of the map
under tilting of the field. However this tilt spoils the $y \to -y$
symmetry of the system, so these two orbits are no longer symmetry-related
and their periods differ, one becoming longer than $kT_c$ and the
other becoming shorter.  As a result each of the horizontal lines
in Fig.  \ref{fig_t_po1_sbm}a, which there represent the one-bounce 
HO families, splits
into an upper and lower branch representing these two orbits.
Moreover for infinitesimal tilt angle one of these 
branches must be stable and one unstable (the lower branch is the
stable one as we shall see below).  Finally, there is no longer
a qualitative difference between the TO and the HO's once the
field is tilted.  For $\theta \neq 0$ 
the TO is required to have non-zero transverse energy
in order to satisfy the $v_y=0$ condition and since it was
degenerate with the $(1,k)$ family of HOs at $\beta= 2 \pi k$ it
must be continuously deformable into one of the HOs near these points.

To label the single-bounce orbits, 
it is convenient to introduce the following notation : 
\begin{eqnarray} 
(1)^{\pm (k)} \nonumber 
\end{eqnarray} 
which means, that it is a single - bounce periodic orbit (``$1$'') 
with the period $T$ such that $k T_c < T < (k+1) T_c$.
To distinguish the two orbits, which for $k \geq 1$ can satisfy
this inequality,  we introduce an additional index $\pm$, such that  
the sign ``$-$'' corresponds to the periodic orbit with a smaller period 
( we use this notation in fig. \ref{fig_t_po1_sbm}) 

The qualitative behavior of the complete set of one-bounce orbits of the SBM
follows from these continuity arguments and is shown in Fig.
\ref{fig_t_po1_sbm},  where
for definiteness we have plotted the exact analytical results of the
next subsection.  Note that for $\beta \neq 2 \pi k$ there is always one orbit
with a nearly linear variation of its period with $\beta$.  This 
is the $(1)^{+k}$ orbit and {\it it} is 
the analog of the TO of the untilted system.  However near 
$\beta= 2 \pi k$ the period of each of the $(1)^{+k}$ orbit saturates to
$k T_c $ as it becomes primarily helical, while a new
pair of orbits is born at a tangent bifurcation near $\beta = 2 \pi k$.
One of these, the $(1)^{+(k+1)}$ takes over the role of the TO 
while the other, the $(1)^{-(k+1)}$ becomes the
unstable partner of the helical orbit generated by the $(1)^{+k}$ orbit.
Thus, qualitatively speaking, the system repeats itself every
time $\beta$ is increased by $2 \pi$.  Quantitative scaling relations
between the behavior in each interval are discussed in Appendix B.
Note finally that the continuity argument suggests that in 
the tilted system the period
$k T_c$ is forbidden for one-bounce orbits since 
the two surviving HO's from each resonant torus are shifted away
from this value and the period of the ``TO'' can no longer cross
that of the HOs as $\beta$ varies; we shall prove this
statement rigorously shortly.

\subsubsection{Quantitative theory} 

We now derive exactly the periods of all one-bounce orbits
for arbitrary tilt angle. We also prove that there can exist no
one-bounce orbit not identified by the continuity argument given
above.
As just noted, it is trivial to see that all one-bounce orbits
must be non-mixing (i.e. bounce with $v_y=0$) for any tilt angle.
Therefore we can impose this condition in order to find all one-bounce
orbits and their periods.  The derivation is most 
easily performed in the coordinate system $(x'',y'',z'')$, which  
{\it moves} in the 
direction perpendicular to $B$ and $E$ with the drift velocity  
$v_d = E \sin(\theta)/B$ : 
\begin{eqnarray} 
x'' & = & x' - v_d t \nonumber \\ 
y'' & = & y' 
\nonumber \\ z'' & = & z'  
\label{drift_frame_define} 
\end{eqnarray} 
 
\begin{figure}[htbp]
\begin{center}
\leavevmode
\epsfbox{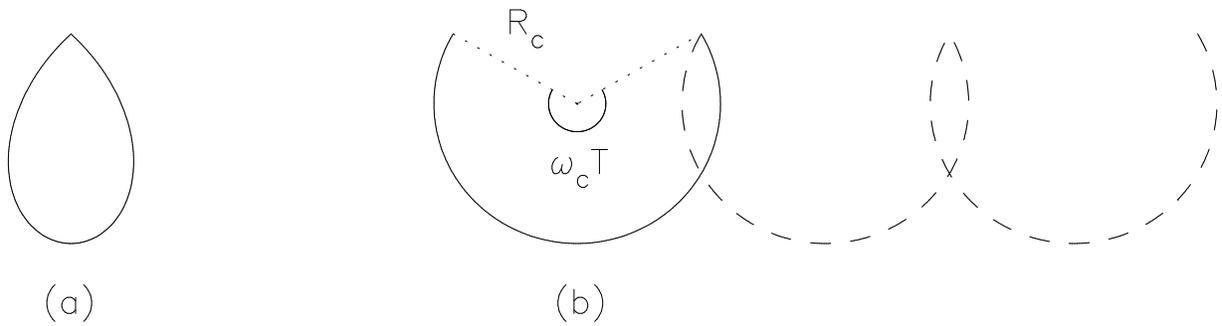}
\end{center}
\caption{
A single-bounce orbit projected onto the $(x',y')$ plane (a) and
$(x'',y'')$ plane of the ``drifting'' frame of reference (b).
\label{fig_po1_driftXY_sbm}
}
\end{figure}

Projected on the plane $(x'',y'')$, the trajectory of the electron 
between successive collisions is a portion of a circle of radius 
$v_c/\omega_c$ with an angular size $\omega_c T$, where $v_c$ is 
the cyclotron velocity and $T$ is the time interval between 
collisions (period of the 1-bounce orbit).  For $T > 2 \pi/\omega_c$  
the trajectory retraces the 
circle several times (see Fig. \ref{fig_po1_driftXY_sbm}).  
Any orbit which is periodic in the
lab frame will not be so in the drift 
frame, instead the initial and the final points of the trajectory between 
successive collisions must be separated by the distance $ \delta x'' = 
v_{d} T$ ( where $T$ is the period of the orbit) and have the same value 
of $y''$.  On the other hand, for one-bounce periodic orbits the 
distance $ \delta x''$ can be expressed as 
(see Fig. \ref{fig_po1_driftXY_sbm}) 
\begin{eqnarray} 
\delta x'' = 2 v_c/\omega_c \sin(\alpha) = 2 
v_c/\omega_c \sin(\omega_c T/2) \nonumber, 
\end{eqnarray}  
so that 
\begin{eqnarray} 
v_c = v_{d} \frac{\omega_c T/2}{\sin(\omega_c T/2)} 
\nonumber 
\end{eqnarray}  
and at the point of collision therefore 
\begin{eqnarray}  
v_{x''}|_{z=0} & = & v_d \left(\omega_c T/2 \right)  
\cot \left( \omega_c T/2 \right) \nonumber, \\  
v_{y''}|_{z=0} & = & v_d \left( \omega_c T/2 \right)  
\label{vxy1} 
\end{eqnarray} 
 
Since the motion along the direction of the magnetic field $\hat{z}'' = 
\hat{B}$ is a uniform acceleration under the force $eE\cos(\theta)/m$, 
at the point of collision 
\begin{eqnarray} 
v_{z''} = \frac{eE \cos(\theta)}{m \omega_c} 
\frac{\omega_c T}{2} 
\end{eqnarray}  
Note, that at the point of collision $v_y = v_{y''} \cos(\theta) - 
v_{z''} \sin(\theta) = 0$, as expected. 
 
Substituting $v''$ into the equation of energy conservation $ 
\varepsilon = m ({\bf v}'' - {\bf v}_{d})^2/2$ at the barrier,  
we finally obtain : 
\begin{eqnarray}  
{ {(\beta/2)^2 - (\omega_c T/2)^2} \over { \left[ 1 - (\omega_c 
T/2)   \cot(\omega_c   T/2) \right]^2}}   =   \sin^2(\theta).  
\label{t1sbm}  
\end{eqnarray} 
This is the basic equation determining the periods $T(\beta,\theta)$ for 
all one-bounce orbits.  As $\beta \to 0$ the only solutions which exist
require $T \to 0$ also, and it is easily seen by expanding the left-hand
side that there is in fact only one solution for any value of $\theta$,
and this solution has $\beta = \omega_c T$ as for the TO in the 
unperturbed system.  For any $\beta$
there are no solutions with $\omega_c T = 2 \pi k$ (as argued above) due
to the divergence of the denominator in the left-hand side at these values.
If there were solutions with this value of the period, then viewed 
in the drift frame the orbit would be an integer number of full circles, 
which is one can see intuitively is impossible due to the 
collision (see. Fig.  \ref{fig_po1_driftXY_sbm}). 

For $\beta \gg 2 \pi k$ there are many solutions as can be 
easily shown graphically by plotting the single-valued function
\begin{eqnarray}
\beta & = & {\cal F}\left(\sin\theta, \frac{\omega_c T}{2} \right)
\label{beta_f_po1_sbm}
\end{eqnarray}
where
\begin{eqnarray}
{\cal F}\left(x,y\right) & = & 2 \sqrt{y^2 + x^2 
\left(1 - y \cot y\right)^2}
\label{define_f} 
\end{eqnarray} 
as is done in Fig. \ref{fig_t_po1_sbm}.

\begin{figure}[htbp]
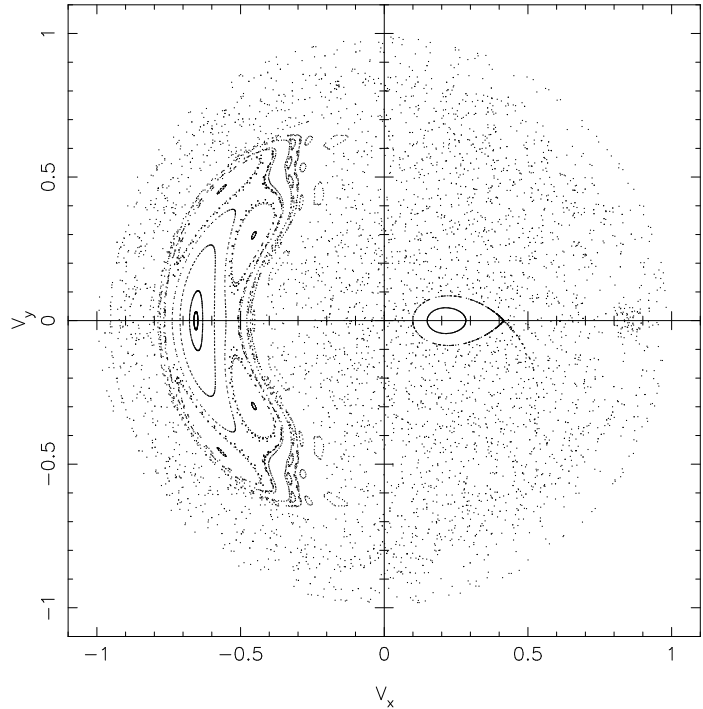

\begin{center}
\leavevmode
\epsfysize=4in 
\epsfbox{fig06a.epsi}
\end{center}
\begin{center}
\leavevmode
\epsfysize=4in 
\epsfbox{fig06b.epsi}
\end{center}
\caption{
Poincar\'e surface of section for the single-barrier model for 
$\theta = 11^\circ$ and (a) $\beta = 5$ (as in the unperturbed system, 
the single-bounce orbit ($(1)^{+(0)}$) is still surrounded by a large
stable island, but has nonzero $x$-component of the total velocity
at the collision with the collector barrier),
and (b) $\beta = 7.7$ (the $(1)^{+0}$ orbit is still stable, but
moved to the periphery of the surface of section; a tangent
bifurcation has just prodiced two new single-bounce orbits :
stable $(1)^{+(1)}$ near the origin, which now takes the role of the 
TO, and unstable $(1)^{-(1)}$, which produces an elongated flow pattern
near the stable island of $(1)^{+(1)}$).
\label{fig_ps_po1_sbm}
}
\end{figure}

The single solution at $\beta < 2\pi k$ corresponds to the 
$(1)^{+0}$ which
is a slightly deformed version of the TO; it is visible as
the central island in the SOS of Fig. \ref{fig_ps_po1_sbm}a 
with $v_y =0$ (as is required, cf. above discussion), 
but with now some small value of $\tilde{v}_x$.  
As $\beta$ is increased, this orbit gains
cyclotron energy, and the corresponding fixed point moves away 
from the center to the left side of the surface of section.  
As discussed above, for  $\beta > 2 \pi$ the period of 
the orbit $(1)^{+(0)}$ approaches asymptotically
$T_c$ as the majority of its energy is fed into transverse motion and 
it becomes a recognizable deformation of a $k=1$ helical orbit of
the untilted system 
(see Figs. \ref{fig_t_po1_sbm}, \ref{fig_ps_po1_sbm}b ).
 
The two new orbits $(1)^{\pm k}$ which must arise by continuity 
in each interval appear in tangent bifurcations at 
thresholds given by $\beta = \beta_k^{\rm tb}$, where  
\begin{eqnarray} 
\beta_k^{\rm tb} = {\cal F}\left( \sin\theta,\varrho_k \right) 
\label{bt_tb_sbm}  
\end{eqnarray} 
and $\varrho_k$ is the $k$-th positive root of the equation 
\begin{eqnarray}  
\frac{{\varrho} \tan{\varrho}} 
{\left( 1 - {\varrho} \cot{\varrho}\right) 
\left(1 + 2 {\varrho} \csc{\varrho}\right)} 
& = &  
\sin^2\theta. 
\nonumber 
\end{eqnarray} 

This is clearly seen in the SOS of Fig.  \ref{fig_ps_po1_sbm}b, 
the fixed point of the stable periodic orbit 
$(1)^{+(1)}$ is at the center of the stable 
island near the origin, whereas its unstable partner 
is (less obviously) visible as the elongated flow pattern at
$v_y=0$ and slightly larger values of $v_x$. 
The evolution of these orbits above threshold is precisely as
predicted by the continuity argument above: the $(1)^{+k}$
initially has a period close to that of the TO before saturating
to $T \approx (k+1)T_c$; whereas the $(1)^{-k}$ orbit immediately becomes
helical with $T \approx kT_c$.
We must emphasize that Eq. (\ref{t1sbm})
uniquely identifies all one-bounce orbits for arbitrary $\theta$. Thus
there are no one-bounce orbits for any $\theta$ which cannot be related
to one-bounce orbits of the untilted system (this is not the case for
period-two and higher orbits).  Hence we have a qualitative and 
quantitative
understanding of the periods and topology of all one-bounce orbits.
The next issue to address is their stability properties.

\subsubsection{Stability}

We define the stability of a periodic orbit in the standard manner
\cite{Reichl,Lichtenberg}.  The non-linear Poincar\'e 
velocity map (Eq. (\ref{PmapVSBM}) ) is
linearized for small deviations of the initial velocity from the values
corresponding to the periodic orbit (fixed point of the map).
This linear map is represented by a $2 \times 2$ {\it monodromy} matrix
$M_1$
which has determinant one due to conservation of phase-space volume in
the hamiltonian flow.  The PO is unstable if one of the eigenvalues
of $M_1$ has modulus larger than one (the other being necessarily less
than one), so that an initial deviation along the associated eigenvector
grows exponentially.  The PO is stable if the eigenvalues are 
$e^{i\phi}, \phi \neq \pi,2\pi$, implying that any initial deviation
will simply rotate around the fixed point.  The points of marginal
stability are when the eigenvalues are $\pm 1$; and by the
continuity of the map, $M_1$ must pass through marginal stability
in order for the orbit to go unstable.  Equivalently,
if $|Tr [M_1]|$ is less than two the orbit is stable, if 
greater than two it is unstable, and when $|Tr [M_1]|=2$ 
it is marginally stable.  There are additional general constraints.
As already noted, new orbits must appear in stable-unstable pairs in
what are called {\it tangent bifurcations} (TB).  Exactly at the point
of TB the orbits are marginally stable with $Tr [M_1] = 2$, before 
the stable one moves to $Tr [M_1] < 2$ and the unstable one moves to
$Tr [M_1] > 2$.  Conversely, the other value for marginal stability,   
$Tr [M_1] = - 2$ corresponds to forward or backwards
period-doubling bifurcations of the PO.  These will be of 
great interest below as they are closely-related to the peak-doubling
transitions seen in the magnetotunneling experiments. 
 
We can obtain the monodromy (stability) matrix for all one-bounce
orbits analytically, but again will first extract its qualitative features
by continuity arguments.  As just noted, for infinitesimal tilt angle the TO
is deformed into the $(1)^{+k}$ orbit in the interval $ 2\pi k < \beta <
2 \pi (k+1)$.  Therefore the stability properties of the $(1)^{+k}$
orbits must be continuous with those of the TO in these intervals.
For the case of the TO of the untilted system the monodromy matrix is
trivial.  The TO has $v_x=v_y=0$, therefore a
small increment of velocity in the $x-y$ plane leaves the time
interval between collisions unchanged to linear order in $\delta v$.
Thus each iteration of the monodromy matrix is just rotation of this
deviation vector by the angle $\omega_c T$, leading to
$Tr [M_1] = 2 \cos \omega_c T$.  Therefore the TO is stable 
at all values of $\beta$ except such that
$\omega_c T = m \pi$; $m = 1,2,3,\ldots$.  It follows by continuity that 
the orbits $(1)^{+k}$ will be stable everywhere in the interval 
$ 2 \pi k < \beta < 2 \pi (k+1)$ except in infinitesimal intervals around
these values.  

The lowest value at which instability can occur is
$\beta = 2 \pi k$, but this is precisely the point of tangent bifurcation
where the $(1)^{+k}$ and $(1)^{-k}$ orbits are born. 
Since $(1)^{+k}$ must evolve immediately into the
analog of the (stable) TO above threshold, it must become the stable
member of the pair immediately after the TB; whereas the
$(1)^{-k}$ orbit must then be unstable.  This is  
allowed by continuity since the $(1)^{-k}$ immediately evolves into the analog 
of the HOs, which are marginally stable for all $\beta$ and can hence
become unstable under infinitesimal perturbation.

Near the midway points of the relevant interval, $\beta=2 \pi (k + 1/2)$,
the $(1)^{+k}$ orbit can again go unstable, but it must immediately
restabilize by continuity for higher values of $\beta$ in this interval.
We find that in fact all $(1)^{+k}$ do go unstable by period-doubling
bifurcation (PDB) near this value, and for sufficiently 
small tilt angles they all
restabilize by inverse PDB at slightly higher $\beta$.

As $\beta$ increases past the value $2 \pi (k +1)$ the $(1)^{+k}$
orbit ceases to play the role of the TO (which is taken over by the
$(1)^{+(k+1)}$ orbit) and continuity alone does not determine its stability.
However from the effective map arguments of subsection 
II.B we know that 
at $\beta \geq 1/\theta$ the system 
undergoes the KAM transition to global chaos, and we  
therefore expect all existing periodic orbits to finally go unstable  
for sufficiently high values of $\beta$.  In other words, for any
non-zero $\theta$ the continuity argument will fail for sufficiently
high $\beta \sim 1/\theta$ and new orbits can appear which have no
analog in the untilted system.  In fact this second destabilization
of the $(1)^{+k}$ orbit occurs by a PDB which creates a period-two
orbit with no analog in the untilted system, as we shall see below.
 
As $\theta$ becomes of order unity, 
the $\beta$ value as which global chaos sets in becomes also of order
unity and we do not expect any of the $(1)^{+k}$ orbits to remain stable
over a large interval.  As already shown above, however, we can 
prove from Eq. (\ref{t1sbm}) that a $(1)^{k\pm}$ pair is born by 
tangent bifurcation
in each interval. Thus the $(1)^{+k}$ must be stable over some small
interval for arbitrarily large $\beta$, but it need not
restabilize after its first PDB.  (Note that the effective map
argument only predicts global chaos in the sense of no remaining KAM
tori for large $\beta$; it does not prove that no stable periodic orbits
can exist, and indeed we have proved the converse: stable one-bounce 
orbits do exist above any finite value of $\beta$). 
To interpolate continuously between the limits of infinitesimal and
large $\theta$ the second PDB moves continuously to lower $\beta$ values until
it eliminates the inverse PDB and hence eliminates the 
restabilization of the $(1)^{+k}$ PO.
 
To make all of these features explicit and quantitative
we have derived the 
monodromy matrix for all single-bounce orbits.  The straightforward but
tedious calculation is sketched in 
Appendix A. We find:
 
\begin{eqnarray}  
{\rm Tr}(M_1) & = &  4\cos^4(\theta) [  \tan^2(\theta) + 
(\omega_c   T/2) \cot(\omega_c T/2) ] \nonumber    
\\ &  \times & 
\{\tan^2(\theta)  + \sin(\omega_c  T)/(\omega_c T) \} - 2 
\label{trm1sbm} \end{eqnarray} 
 
\begin{figure}[htbp]
\begin{center}
\leavevmode
\epsfysize=4.5in 
\epsfbox{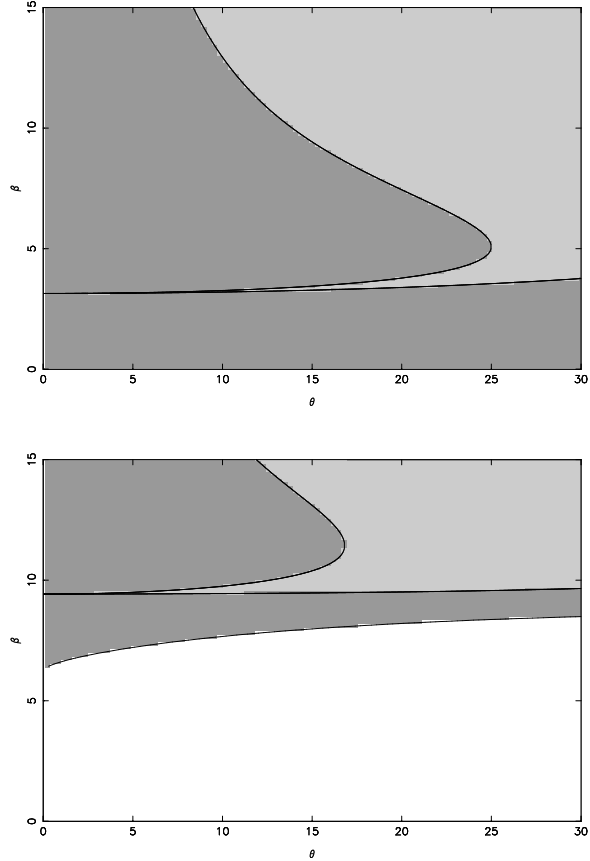}
\end{center}
\caption{
Regions of existance (shaded areas) of one-bounce orbits 
$(1)^{(0)+}$ (a) and $(1)^{(1)+}$ (b) in the $(\theta,\beta)$
plane. Dark and light shading correspond to stable and unstable
regions respectively. 
\label{fig_exist_intervals_po1_sbm}
}
\end{figure}

This equation describes precisely the stability properties of the
one-bounce orbits sketched above.  First, every time a new pair of 
roots of Eq. (\ref{t1sbm}) appear with 
increasing $\beta$, ${\rm Tr}(M_1)=+2$ corresponding
to a tangent bifurcation, as discussed.  
As $\beta$ increases from this threshold
one root (describing the $(1)^{-(k)}$ PO) becomes increasingly unstable with
${\rm Tr}(M_1) \to + \infty$.  In contrast, the other root
corresponding to the $(1)^{+k}$ orbit initially becomes stable
(${\rm Tr}(M_1) < 2$) and remains so for a finite interval before
going unstable at ${\rm Tr}(M_1)=-2$ by PDB.  For sufficiently small
$\theta$, ${\rm Tr}(M_1)$ will pass through the value $-2$ twice more
before tending to $-\infty$, corresponding to the restabilization and
subsequent destabilization of the $(1)^{+k}$ predicted by the continuity
arguments above.  As $\theta$ increases for any fixed interval $k$
eventually a critical angle is reached at which this restabilization
ceases, just as predicted.  The behavior of the ${\rm Tr}(M_1)$ for
$(1)^{\pm k}$ orbits with $k=0,1,2$ is shown in Fig.
\ref{fig_trm_po1_sbm}. 
Since increasing $k$ corresponds
to larger $\beta$, the critical angle becomes smaller as $k$ increases. 
The intervals of restabilization of the $(1)^{+k}$ orbits are shown
in Fig. \ref{fig_exist_intervals_po1_sbm}  terminating 
at the critical angles $\theta_k^\dag$.
The most experimentally relevant interval is $k=0$, for
which Eq. (\ref{trm1sbm}) predicts a critical angle of 
$\theta_0 \simeq 25^{\circ}$, very
close to the value at which the peak-doubling regions merge in
the data of ref. \cite{Muller}.  We will later show how the occurence
of this critical angle relates to the size and evolution 
of the peak-doubling regions in the data. 

Quantitative results for the $\beta$ values at which the PDBs occur and
for the critical angle are easily obtained from Eq. (\ref{trm1sbm}) 
for the monodromy matrix.  Equation (\ref{trm1sbm}) can be written as 
\begin{eqnarray}
{\rm Tr}(M_1) + 2 & = & R (\theta, \omega_c T)  \nonumber \\
& = & 4\cos^4(\theta) R_1 (\theta, \omega_c T) R_2 (\theta, \omega_c T)
\end{eqnarray}
where the zeros of the function $R (\theta, \omega_c T)$ (known as 
the {\it residue}) give the parameter values for all PDBs.
It is easily seen from Eq.  (\ref{trm1sbm}) that factor 
$R_1$ has exactly one root
in each interval $k$, whereas the factor $R_2$ has either two or
zero roots in each interval, corresponding to the presence or absence
of the restabilization.  The set of transcendental equations which
determine the roots of $R_1,R_2$ and hence the bifurcations points and
critical angles are summarized in Appendix B.

The existence and stability properties of the one-bounce orbits
as predicted by Eqs. (\ref{t1sbm}),(\ref{trm1sbm}) are confirmed 
by the numerically-generated
SOS and indeed reveal the underlying pattern to the complex behavior
seen in the SOS.  The period-doubling bifurcations of the one-bounce
orbits are of particular interest because they are closely-related
to the peak-doubling phenomena observed experimentally.  We will 
elucidate this behavior in the next section on period-two orbits.

\subsection{Two-bounce orbits} 

\subsubsection{Qualitative description, $\beta \theta \ll 1$}

For $\theta = 0$ all two-bounce periodic orbits occur in helical 
families satisfying the resonance condition : 
\begin{eqnarray} 
(2k + 1) \omega_L & = & 2 \omega_c \;\;\;\;\;\; k=0,1,2,\ldots  
\label{res2_sbm}
\end{eqnarray} 
Only odd integers appear in the resonance condition since even
integers yield orbits equivalent to the period-one helical family.  
As follows from Eqs. (\ref{t_ho}) and (\ref{res2_sbm}), the periods of the 
two-bounce helical orbits are  given by 
\begin{eqnarray}  
T & = & (2 k + 1) \frac{2 \pi}{\omega_c} 
\label{t_ho2_sbm} .
\end{eqnarray}

Therefore, just as for the one-bounce helical orbits, the resonant tori 
corresponding to the two-bounce orbits
can only appear above a threshold value of $\beta$ at which the 
longitudinal period becomes long enough to satisfy Eq. (\ref{t_ho2_sbm}).
At this threshold the two-bounce orbits are indistinguishable from
the second repetition of the traversing orbit.  Thus the thresholds
$\beta_c^{(2)}$ are given by the condition 
$ 2 T_{\rm TO} =  (2k + 1) T_c$, which gives 
\begin{eqnarray} 
\beta_c^{(2)} & = &  \pi (2k + 1).
\end{eqnarray} 
Once emerged, the period-2 resonant tori remain in the phase space of 
the system for arbitrary large value of $\beta$, simply moving towards
the periphery of the surface of section as $\beta$ increases. 

Again, as for the helical one-bounce periodic orbits, when the
magnetic field is tilted, the resonant tori of the two-bounce 
orbits are destroyed and replaced by an integer number of pairs of 
stable and unstable two-bounce periodic orbits.  By continuity, these
orbits must appear in the vicinity of the $(1)^{+k}$ traversing 
orbits (which are
now playing the role of the TO) and near the values $\beta \approx \pi (2k+1)$
at which the two-bounce tori appear.  Our previous analysis for small 
tilt angles has already identified one direct and one 
inverse period-doubling bifurcation of the $(1)^{+k}$ 
near these values of $\beta$ (see  Fig. \ref{fig_trm_po1_sbm}). 
In a direct PDB a stable
one-bounce PO becomes unstable while generating a stable two-bounce
PO in its neighborhood; in an inverse PDB an unstable one-bounce
PO becomes stable while creating an unstable two-bounce PO in its
neighborhood.  Hence for consistency we
conclude that exactly one pair of two-bounce PO's is created from
each two-bounce family for infinitesimal tilt angle.  Furthermore,
one of these arises from the direct PDB and is therefore stable, 
whereas the other arises from the inverse PDB and is unstable. 
(For infinitesimal tilt angle the interval $\Delta \beta$ between
these two PDBs is also infinitesimal and they are created at the
same ``time'' in agreement with the Poincar\'e-Birkhoff theorem;
for any finite angle they are separated by some finite interval
in $\beta$).

\begin{figure}[htbp]
\begin{center}
\leavevmode
\epsfysize=2.5in 
\epsfbox{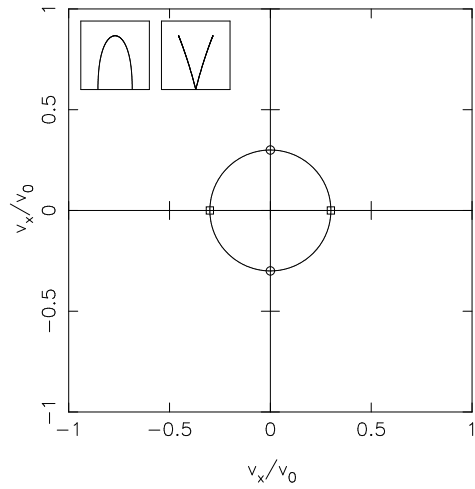}
\end{center}
\caption{
Torus of two-bounce orbits in the Surface of Section. Marked are
the only ``self-retracing'' (in the $y-z$ plane) two-bounce orbits :
(a) the orbit with $v_y = 0$ at collisions,which evolves into
the non-mixing two-bounce orbit $(2)^-$, and (b) the orbit with 
$v_x = 0$ 
at collisions - which becomes the self-retracing mixing orbit 
$(2)^+$. Insets show the $y-z$ projections of these orbits.
\label{fig_torus_po2_sbm}
}
\end{figure}

It follows that there must be exactly two orbits from each helical family
which are continuously deformed into the stable and unstable 
two-bounce POs created at these two PDBs.  It is easy to identify
one of the two in analogy to our earlier reasoning.  There is 
only one two-bounce PO in each helical family for which both of
its two collisions with the barrier occur with $v_y=0$ 
(see Fig. \ref{fig_torus_po2_sbm}).
This orbit can be continuously deformed into a non-mixing 
two-bounce orbit which will become degenerate with the non-mixing
$(1)^{+k}$ at the PDB - see  Fig. \ref{fig_po2_pic_YZ_sbm}a.  
However, unlike the case for one-bounce
HOs, there is no second orbit with fixed points at $v_y=0$ which
can evolve into the second two-bounce orbit which we know must
be created.  Hence this second orbit at $\theta \neq 0$ 
must be mixing; i.e. it must generate fixed points with non-zero $v_y$.  
Thus it must be obtained by a deformation of one of the two-bounce
orbits in the helical torus with finite values of $v_y$ at collision.

To identify which orbit this is we must consider the general properties
of mixing two-bounce orbits in this system.  We have noted above that
due to time-reversal symmetry the SOS has to be symmetric under the
transformation $v_y \to -v_y$.  It is obvious that a two-bounce
orbit with the same value of $v_x$ at each collision will generate
two fixed points in the SOS which satisfy this reflection symmetry.
Note that since $v_x \propto y$, such a mixing period-two orbit
strikes the barrier at the same value of $y$ 
in each collision.  We will refer to such
orbits as self-retracing since they retrace themselves in $y-z$ projection.
All self-retracing two-bounce orbits are mixing.
However there exist non-self-retracing two-bounce mixing orbits.
These must collide with different values of $v_x$ at each collision,
but still satisfy the required reflection symmetry of the SOS in a 
more subtle manner.  In such an orbit
the values of $v_x$ at collision differ {\it for any one sense of traversal},
but traversing the orbit in the opposite sense generates
two additional fixed points which restore
the $v_y \to -v_y$ symmetry of the SOS which has four fixed points for
such orbits.  Such an orbit
is shown in Fig.  \ref{fig_po2_pic_YZ_sbm}c, and analogous orbits 
exist for higher-bounce POs as well.  We will discuss their origin later.  

However,these non-self-retracing two-bounce orbits cannot
be created at a PDB of a one-bounce orbit (period-one fixed point)
since such a PDB cannot create more than two new fixed points 
\cite{mao,meyer}.
Therefore the second, mixing orbit we seek for $\theta \neq 0$
must be a self-retracing orbit, i.e. it must have the same value of
$v_x$ at each of its two collisions with non-zero $v_y$
 - see  Fig. \ref{fig_po2_pic_YZ_sbm}b.
The only orbit in the $\theta=0$ helical family with this property 
is the one which collides with
the barrier with $v_x=0$ at each collision 
(see Fig. \ref{fig_torus_po2_sbm}).  Hence 
by continuity it is this orbit which must be continuously deformed
to give the mixing orbit which must, by the Poincar\'e-Birkhoff 
theorem, exist for infinitesimal tilt angle.
Intuitively, the PDB of the $(1)^{+k}$ orbit to the
non-mixing two-bounce orbit corresponds to splitting the $(1)^{+k}$
at the point of collision, whereas the PDB corresponding to the
mixing one corresponds to splitting the $(1)^{+k}$ at the
point furthest away from the collision (see Table I).

Since lack of mixing at collision should enhance the stability 
of an orbit for given $\beta,\theta$,
we may expect that the non-mixing two-bounce orbit is born stable in
the direct PDB and the mixing one is born unstable at the inverse
PDB which occurs at a slightly higher value of $\beta$.  This conjecture
is confirmed by our analytic calculations below.  In accord with
our earlier notation we will label this pair of two-bounce orbits,
which must exist in each interval by continuity, as:
\begin{eqnarray} 
(2)^{\pm k} 
\end{eqnarray} 
where the sign ``$+$'' corresponds to the orbit with the longer period
as before.  Note that the stable non-mixing orbit
then will be the $(2)^{-k}$ and the unstable mixing orbit will be the
$(2)^{+k}$ (one should not then interpret $+,-$ as stable,unstable).
For simplicity we drop the interval index $k$ below.  The same 
scenario occurs in each interval, just at smaller $\theta$ as $k$
is increased.
 
\subsubsection{Qualitative description, $\beta \theta \sim 1$}

\begin{figure}[htbp]
\begin{center}
\leavevmode
\epsfysize=2.5in 
\epsfbox{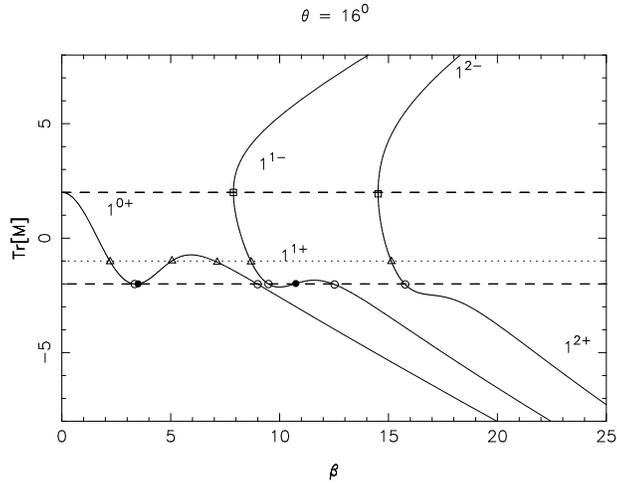}
\end{center}
\caption{
Trace of the monodromy matrix for single-bounce orbits $(1)^{(0)+}$,
$(1)^{(1)+}$, $(1)^{(1)-}$, $(1)^{(2)+}$, $(1)^{(2)-}$ for 
$\theta = 16^\circ$.  The dotted
line represents the condition for the $1:3$ resonance,
the dashed lines show the boundaries of the stability region
${\rm Tr}[M] \leq 2$. Open circles show the locations of the 
direct PDBs, the solid circles correspond to inverse PDBs, 
open triangles represent $1:3$ resonances, squares represent
tangent bifurcations.  
\label{fig_trm_po1_sbm}
}
\end{figure}

Up to now we have focused on the limit of small $\beta \theta$ where
each orbit must by continuity have an analog for $\theta =0$.
Unlike single-bounce orbits in the tilted well,
there will exist orbits with two or more bounces which have
no analogs in the integrable case.
In fact we have already shown above (see Fig. 
 Figs. \ref{fig_trm_po1_sbm}, \ref{fig_exist_intervals_po1_sbm}) 
that after restabilizing 
by inverse PDB the $(1)^{+}$ orbit must eventually go unstable
by a third PDB which must give rise to a stable two-bounce orbit
with no analog in the untilted system.  We denote these new orbits
as $(2)^{*}$; one such orbit must exist for each $(1)^{+}$ orbit although
for small tilt angle they will not appear until values of 
$\beta \sim 1/\theta$. 

Will the $(2)^{*}$ orbits be mixing or non-mixing?  One can also
decide this by reference to our stability analysis of the $(1)^{+}$ 
orbit (see Fig. \ref{fig_trm_po1_sbm}
 above).   As we showed, for each $(1)^{+}$ orbit,
as $\theta$ is increased to a critical value, the second and third
PDBs move closer together and finally merge, after which no restabilization
of the $(1)^{+}$ orbit occurs.  But the second PDB is associated with
the mixing $(2)^{+}$ orbit; if it merges with the $(2)^{*}$ orbit
when the second and third PDB coincide, then $(2)^{*}$ orbits must also
be of the same symmetry, i.e. mixing.  

\begin{figure}[htbp]
\begin{center}
\leavevmode
\epsfysize=5.5in 
\epsfbox{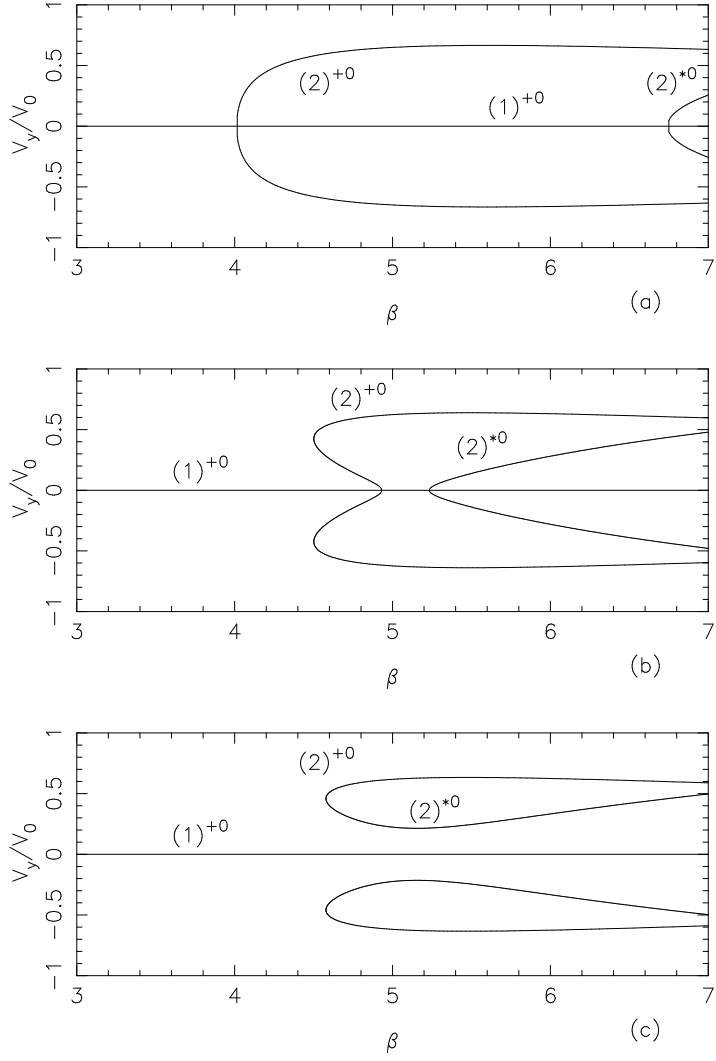}
\end{center}
\caption{
Bifurcation diagrams in the coordinates $(\beta, \tilde{v}_y)$ for the
period-$2$ mixing orbits, related to the bifurcations of the
single - bounce orbits.
The two branches with non-zero $\tilde{v}_y$ correspond 
to the two-bounce mixing orbits $(2)^{+(0)}$ and $(2)^{*(0)}$, while 
the horizontal line represents the single-bounce orbit $(1)^{+(0)}$.
The non-mixing period-2 orbit $(2)^{-(0)}$ has $v_y = 
0$ at each of the points of collision and cannot be seen in this diagram. 
For a small tilt angle the period-$2$ orbits are born in
period-doubling bifurcations - see panel (a). When $\theta > 
\theta_k$ the mixing period-2 orbits are born in a tangent bifurcation
- see panel (c). 
The transformation from the two types of behavior cannot  
happen in a single step. If it were possible, then at the 
critical angle {\it two} new mixing two-bounce orbits were  
created at the location of  the single-bounce orbit, which can {\it 
not} happen in a generic conservative 2D system.  
The alternative is provided by the following two-step process. 
First, at some critical angle $\theta_k^0 < \theta_k$ the behavior of 
the first to appear mixing orbit $(2)^{-(k)}$ is changed, as is shown  
in the bifurcation diagram at the panel (b). 
When $\theta_k^0 < \theta < 
\theta_k$, the unstable orbit  $(2)^{+(k)}$ appears in a tangent  
bifurcation with a new self-retracing mixing stable period-2 orbit, 
which is 
soon to be absorbed by the single-bounce orbit in an {\it  
inverted} period-doubling bifurcation, while the 
qualitative  behavior of the stable $(2)^{*(k)}$ orbit remains unchanged.
As the tilt angle is 
increased, the interval of stability of the single-bounce orbit 
shrinks, while the interval of existence of the auxiliary mixing orbit  
increases. At the critical tilt angle the inverted and
standard period-doubling bifurcations merge and annihilate each other,
so that at greater values of the tilt angle the mixing period-2  
orbit are no longer directly related to the single-bounce orbit - see
panel (c).
\label{fig_po2_bifdiag_m_sbm}
}
\end{figure}

What happens to the $(2)^{+},(2)^{*}$ orbits for tilt angles above
$\theta_k$?  On the one hand above $\theta_k$ they cannot
be created by PDBs of the $(1)^{+}$ 
orbit, since we have shown that it never restabilizes.  On the other 
hand, these two periodic orbits cannot cease to exist
suddenly, since they exist for an infinite interval above the threshold
for PDB and the orbit far from threshold is negligibly perturbed by
a small increase in tilt angle.  The resolution of this apparent
paradox is that above $\theta_k$ the two orbits 
are created by a tangent bifurcation
in a region of the SOS and at a value of $\beta$ very close to that
at which the PDBs occur below $\theta_k$.  The detailed 
description of the transition from the 
PDB scenario to the TB scenario is sketched in
Fig. \ref{fig_po2_bifdiag_m_sbm} and described in the caption.  
In contrast, nothing qualitatively new happens 
to the behavior of the initially stable $(2)^{-}$ as
$\theta$ is increased beyond $\theta_k$; its interval of stability
just shrinks continuously. 

So for {\it all} $\theta$ we are able to locate all
two-bounce orbits which are related originally to the one-bounce
$(1)^{+k}$ orbit, and to describe their evolution
qualitatively.  There are exactly three such orbits associated with
each $(1)^{+}$ orbit: the $(2)^{-}$ which is initially stable and
non-mixing, the $(2)^{+}$ which is initially unstable and mixing,
and the $(2)^{*}$ which is initially stable and mixing.  

The last point to understand is the evolution of these orbits 
with increasing $\beta$ once they are created.  Since these orbits
exist for all $\beta$ above threshold at $\theta =0$, we expect
the same behavior for nonzero $\theta$.  However, as
both the $(2)^{-}$ and $(2)^{*}$ orbits are initially stable, we 
expect them both to become unstable as $\beta \to \infty$.
It turns out that the $(2)^{-}$ orbit goes unstable as the second 
stage of an infinite period-doubling transition to chaos.
The $(2)^{*}$ on the
other hand follows a more complex route to its final unstable
form.  As the parameter $\beta$ is increased, the orbit $(2)^*$
goes unstable via a period-doubling bifurcation, but soon
restabilizes and finally goes unstable via a {\it pitchfork}
bifurcation. In such bifurcation 
a new stable (mixing) orbit is created with a period identical
to that of the orbit which has gone unstable.  In this case the new
orbit is precisely of the non-self-retracing type shown in 
Fig. ( \ref{fig_po2_pic_YZ_sbm}c)
and described above.  Thus this one new two-bounce orbit creates
four fixed points in the SOS and satisfies the required conservation
of the Poincar\'e index.  From the generic properties of 2D conservative
maps it can be shown that such orbits can {\it only} be created in
these pitchfork bifurcations.  Although it is interesting to 
note the origin of the non-self-retracing two-bounce orbits, they
are of a little importance for the description of the
experimental tunneling spectra, since generally the pitchfork 
bifurcations appear at relatively high values of $\beta$, as we will
show in the quantitative description of 
the two-bounce orbits in the next subsection.

In principle completely new two-bounce orbits can also arise by tangent
bifurcations at sufficiently large tilt angles and values of $\beta$,
in fact no visible islands due to such orbits are seen in the 
SOS for any tilt angles of interest in the range of $\beta$ values
which are accessible experimentally.  Thus for understanding the 
experimentally observed peak-doubling regions only the 
the three two-bounce orbits $(2)^{-},(2)^{+},(2)^{*}$ for the intervals 
$k=0,1$ are most relevant.  Their properties are summarized in table I.
These orbits, once their generalization to the double-barrier model is
understood, will be sufficient to explain the peak-doubling data of
references \cite{Fromhold,Muller}.

We now give an analytical description of the periods and stability 
of the two-bounce orbits identified above.

\subsubsection{Quantitative theory : Non-mixing two-bounce orbits} 
 
The derivation of the periods of the non-mixing two-bounce orbits can 
be performed using the same technique developed in the analysis of
the single-bounce orbits. In the drift frame introduced in 
section III.B.2
the orbit consists of two identical and overlapping arcs of a circle 
of angular size $\omega_c T > \pi$ with their endpoints 
displaced by $v_D T/2$.  Imposing the non-mixing condition at the two
collisions determines $T$.  Conservation of energy is not required to
fix the period and this leads to the striking result that the period
is independent of energy (this is the only relevant orbit with this
property).  This calculation, the details of
which are given in the Appendix D, yields :
\begin{eqnarray} 
\frac{\omega_c T}{ 4 } \cot\frac{\omega_c T}{4} & = & - \tan^2\theta 
\label{t_po2_sbm} 
\end{eqnarray} 
The $k$-th positive root of this equation gives the value of the 
period of the $(2)^{+(k)}$ orbit. Note that the solutions $T$ do not
depend on $\beta$. This is the only orbit with this property.

We have also calculated the stability properties of these orbits
by evaluating the trace of the corresponding monodromy matrix
using the general expressions developed in the Appendix C. This
straightforward but tedious derivation is given in Appendix E.
In fig. \ref{pic_trM_po2_nm_sbm} we plot 
${\rm Tr}(M)$. In agreement with our qualitative analysis, ${\rm Tr}(M)$
is a monotonically decreasing function 
of $\beta$, so that the initially stable two-bounce non-mixing orbit 
destabilizes by a period-doubling bifurcation and then remains unstable 
for all $\beta$.  The 4-bounce periodic orbit, 
which is born in this bifurcation, will in turn bifurcate, 
producing an infinite series of period-doubling bifurcations 
of the same type as the period-doubling sequence in the 
quadratic DeVogelaere  map \cite{qdV},\cite{Reichl}. However, since   
the periodic orbits of this sequence have long periods and relatively
large cyclotron energy, they are of a little importance for the 
description of the tunneling spectra in the tilted well, and will not 
be discussed in the present paper.

\subsubsection{Mixing period-2 orbits}

Due to nonzero energy exchange at the points of collision the
analytical description of a general mixing two-bounce periodic orbit
will be very complicated.  However, as we pointed out 
before, the most important two-bounce mixing orbits are 
self-retracing (in $y-z$ projection) leading to the symmetry property
that $v_x$ is the same a both collisions.  Imposing this condition
simplifies the analytical treatment. For each of
these orbits, the electron collides with the barrier twice at the same
point with exactly the same {\it absolute values} of the velocity components 
$v_x,v_y,v_z$. Using this property,
one can show (see Appendix F), that the periods $T$ of the two-bounce
self-retracing orbits must satisfy the following system of 
coupled transcendental equations :
\begin{eqnarray}
\label{t_po2_m_sbm}
\left\{
\begin{array}{cclr}
\frac
{\textstyle \sin\left(\frac{ \textstyle \omega_c T}{2}\right)}
{\frac{\textstyle \omega_c T}{ \textstyle 2}}
& = & - \tan^2\theta
\frac
{\textstyle \sin\left(\frac{\textstyle \omega_c \delta T}{2}\right)}
{\frac{\textstyle \omega_c \delta T}{ \textstyle 2}} 
& ({\protect\ref{t_po2_m_sbm}\rm a})
\\
\left(\frac{\beta}{2}\right)^2
& = & 
\sin^2\theta \left( 1 - 
\frac{
\frac{\omega_c T}{2} \left( \cos\left(\frac{\omega_c T}{2}\right)
+  \cos\left(\frac{\omega_c \delta T}{2}\right) \right)
}
{
2 \sin\left(\frac{\omega_c T}{2}\right)
}
\right)^2 
+
\left( \frac{\omega_c T}{4}\right)^2
+ 
\cot^2\theta \left( \frac{\omega_c \delta T}{4}\right)^2
& ({\protect\ref{t_po2_m_sbm}\rm b})
\end{array}
\right.
\nonumber
\end{eqnarray}
where $\delta T < T$ is the difference of the time intervals between 
successive collisions $t_1$ and $t_2$ (see Appendix F). This system 
of two equations determines the periods of all of the self-retracing
two-bounce orbits as functions of $\beta$ and the tilt angle.

Although the equations (\ref{t_po2_m_sbm}a), (\ref{t_po2_m_sbm}b) 
look quite complicated,
they allow a further analysis. Assume at least one solution exists
for some fixed value
of $T$ and find the corresponding value(s) of the time difference
$\delta T$ from equation (\ref{t_po2_m_sbm}a) which 
depend explicitly only on $T,\theta$
(but only implicitly on $\beta$). As an equation for $\delta
T$ at fixed $T$ and $\theta$, this relation can have multiple
solutions $\delta T = \delta T_n$ :
\begin{eqnarray}
\delta T_n & = & 2 \wp_n\left( - \cot^2\theta 
\frac{\sin\left(\frac{\omega_c T}{2}\right)}
{\frac{\omega_c T}{2}} \right), \ \ \ \delta T_n < T
\label{dt_t_po2_m_sbm} 
\end{eqnarray}
where the function $\wp_n(x)$ was defined in Appendix B (see Eq. 
(\ref{eq_sin}) )
and the maximal value $n$ depends on the value of $T,\theta$.
If $T$ is not a solution of the system for any $\beta$, Eq. (8) will
have no roots with $\delta T < T$.  One knows (from the calculation
of the stability matrix for the single-bounce orbits) the exact
values of $T$ at which the $(2)^{+k},(2)^{*k}$ orbit are born by
PDB and inverse PDB of the $(1)^{+k}$. Hence we can find the starting value of
$T$ for each $(2)^{+},(2)^{*k}$ orbit and follow it continuously as
$\beta$ increases. 
Each root $\delta T_n$ when inserted into Eq. (\ref{t_po2_m_sbm}b) 
yields a solution ``branch'' $\beta_n (T)$ for a two-bounce orbit.

\begin{figure}[htbp]
\begin{center}
\leavevmode
\epsfysize=3.5in 
\epsfbox{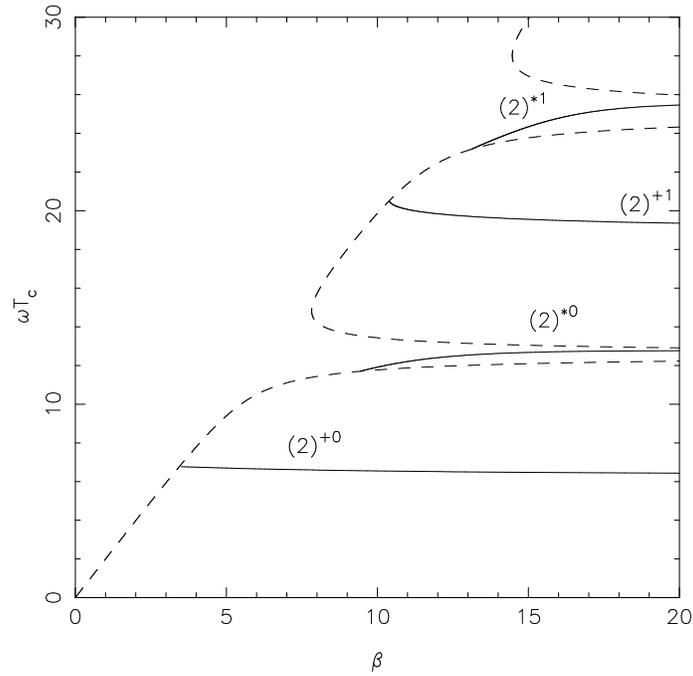}
\end{center}
\caption{ 
Periods of the self-retracing mixing two-bounce orbits $(2)^{+(0)}$,
$(2)^{*(0)}$, $(2)^{+(1)}$, and
$(2)^{*(1)}$, related to the bifurcations of the single-bounce 
periodic orbits as functions of $\beta$. 
The tilt angle is $\theta = 15^\circ$. The dashed lines
show the (scaled) time intervals of two repetitions of 
single-bounce orbits (i.e. twice the period of single-bounce orbits).  
\label{fig_po2_periods_m_sbm}
}
\end{figure}

There does not however need to be exactly one self-retracing two-bounce
orbit for each solution branch $\beta_n (T)$.  If the period of such
an orbit is a non-monotonic function of $\beta$ then the same orbit
will give rise to multiple solution branches which must merge at
the extrema of $T(\beta)$.  One can show that there can be no more
than one extremum at finite $\beta$ for $T(\beta)$, thus each orbit will
be described by either one or two such branches.
Conversely, one solution $\beta_n(T)$ can be non-monotonic in $T$,
hence it must describe two different two-bounce 
orbits with different periods at the same
value of $\beta$.  With care, 
{\it any} two-bounce self-retracing orbit can be obtained by this approach.
This procedure yields the plots of the periods for the $(2)^{+0},
(2)^{*0} $ orbits shown in Fig.  \ref{fig_po2_periods_m_sbm}.
  Note that unlike the non-mixing
$(2)^{-k}$ orbits, the periods of the mixing orbits depend on $\beta$.

In fact for small tilt angles the period of the $(2)^{+k}$ orbit is 
a monotonically decreasing function of $\beta$ and there is only
the $n=1$ solution branch to consider.  In this case we can expand
Equations (7),(8) for $\beta \theta \ll 1$ and obtain an explicit
formula for the periods of these orbits:
\begin{eqnarray}
\omega_c T & = & \pi \left(1 + 2 k\right)
\left( 1 + \theta^2 + \frac{10 + \pi^2 - 6 \beta^2}{6}\theta^2
\right)  +  O\left(\theta^6\right).
\label{t_smalltilt_po2_m_sbm}
\end{eqnarray}

Although the $(2)^{*}$ orbits have the same topology as the 
$(2)^{+}$ (and at large $\theta$ they are born together in a 
tangent bifurcation), they have no analogs in the untilted system
so their periods cannot be obtained from such an expansion.
The quantitative analysis of Eqs. (\ref{t_po2_m_sbm}a),
(\ref{t_po2_m_sbm}b) confirms the transition
scenario between PDB and TB for the $(2)^{+},(2)^{*}$ 
for large tilt angles described in Fig. \ref{fig_po2_bifdiag_m_sbm}.
 
Once the values of $T$ and $\delta T$ are known from the Eqs.
(\ref{t_po2_m_sbm}a),(\ref{t_po2_m_sbm}b), the components 
of the velocity at the
points of collisions can be obtained from (\ref{v_inplane_po2_m_sbm}),
and one can calculate the monodromy matrix for each such orbit using
(\ref{m_mk_sbm}) and (\ref{mk_analyt_sbm}). In Fig.
\ref{pic_trM_po2_m_sbm} 
we show the behavior of the trace of the
monodromy matrix for $(2)^{+}$ and $(2)^{*}$ orbits.
As argued above, one finds that the $(2)^{+}$ orbits are
unstable for all $\beta$, whereas the $(2)^{*}$ orbits which are
born stable (since they arise from a direct PDB of the $(1)^{+}$
orbit), and go unstable in the complicated sequence ending with a
pitchfork bifurcation which we have described above and in the 
caption to Fig. \ref{pic_trM_po2_m_sbm}. 

In Table I we summarize the relevant period-$1$ and period-$2$ 
orbits.

\subsection{Three-bounce periodic orbits} 
 
The scenario for the three-bounce periodic orbits is similar in
many ways to that for the two-bounce orbits just described.
When the magnetic field is not tilted all three-bounce periodic
orbits belong to resonant tori and correspond to the resonances
\begin{eqnarray}
k \omega_L & = & 3 \omega_c 
\label{resonance3}
\end{eqnarray}
where the integer $k$ is not a multiple of $3$. Thus as $\beta$
increases from zero in the first interval there are two thresholds
for the birth of resonant tori.  When $\beta = 2\pi/3$ the family
of helical orbits which perform $1/3$ of a cyclotron
rotation per collision with the barrier appears, and at $\beta= 4\pi/3$
the family which makes $2/3$ of a rotation per collision appears.
As for the two-bounce orbits, the analogous orbits in the higher 
intervals behave in exactly the same manner qualitatively, and
so we focus here on those in the first interval.  

When the magnetic field is tilted, the period-$3$ resonant tori are 
destroyed and replaced by pairs of stable and unstable three-bounce
orbits. Here some important differences from the two-bounce orbits 
enter.  First, we cannot have a {\it single} three-bounce orbit created
at some value of $\beta$ since there is no analog of a period-doubling
bifurcation for creating three-bounce orbits.  At the threshold
for creation of the three-bounce helical families, when they are 
degenerate with the third repetition of the traversing orbit, the
$Tr M_1 = -1$ and its stability cannot change.  Therefore
period-three orbits must always be created in stable-unstable pairs by
tangent bifurcation.  Moreover there is generically no constraint
that such a tangent bifurcation occur at the fixed point corresponding
to a period-one orbit\cite{mao}.  In this sense there are no trifurcations in 
a generic system.  When $\theta=0$ the rotational symmetry of the system 
does constrain the entire family of three-bounce orbits to appear 
degenerate with the third repetition of the traversing orbit, but
as soon as $\theta \neq 0$ the pair of three-bounce orbits which
survive are created away from the period-one fixed point.  However,
by continuity the tangent bifurcation (TB) which creates this pair must
occur near this fixed point and at approximately the same value of 
$\beta$.  We infer that for small tilt angles there are at least 
two TB's in
the first interval, each of which creates a stable-unstable pair of
three-bounce orbits, at $\beta_1 \approx 2 \pi/3, \beta_2 \approx 4 \pi/3$.
Extending our earlier notation, we will denote these four orbits
by $(3)^{\pm}_{1},(3)^{\pm}_{2}$.  

Which orbits of the resonant tori survive?
In this case there is no orbit in the helical family which has
all of its collisions with $v_y=0$; therefore by continuity there
can be no three-bounce non-mixing orbits for small tilt angles (and
one can easily show that this result holds for any $\theta$).
However there are two orbits in each torus which collide with
$(v_y)_1=0,(v_y)_2=-(v_y)_3$ corresponding to two possible orientations
of the appropriate equilateral triangle along the $v_x$ axis.
These two orbits satisfy the required symmetry of the SOS upon tilting,
while no others in the torus do.  Therefore it is these orbits
which survive (slightly distorted due to the tilt, of course).

\begin{figure}[htbp]
\begin{center}
\leavevmode
\epsfysize=3.in 
\epsfbox{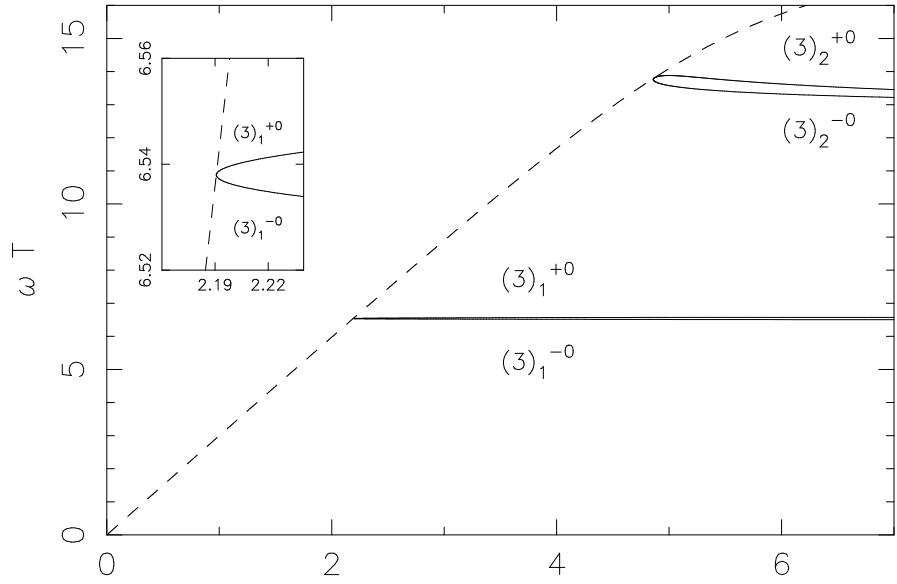}
\end{center}
\caption{
The periods of the three-bounce orbits $(3)_1^{\pm 0}$ and 
$(3)_2^{\pm 0}$
vs. $\beta$ for tilt angle for $\theta = 11^\circ$ (solid lines).
The dashed line represents the period of single-bounce orbit
$(1)^{+(0)}$, multiplied by $3$.
\label{fig_period_po3_type1_sbm}
}
\end{figure}

This conclusion, while correct, must be reconciled with our earlier
statement that the two orbits must appear at a tangent bifurcation.
At a TB the two orbits are identical, yet the two orbits we have 
identified correspond to opposite orientations of the equilateral triangle
and would not coincide for any finite size of the triangle defining
the three fixed points (see Fig. 
\ref{fig_touch_and_go_po3_sbm}).  In order to coincide at the TB
the unstable member of the pair must actually pass through the single-bounce
fixed point at the center of the triangle in what is known as a 
``touch-and-go'' bifurcation\cite{mao}.  
At this point the unstable three-bounce orbit coincides with
the third repetition of the $(1)^+$ orbit, which is no longer
isolated and $Tr M_1^3=2$ (or equivalently $Tr M_1 =-1$).  
So as $\beta$ is reduced to the 
threshold for the TB, first the unstable three-bounce orbit shrinks
to a point coinciding with the period-one fixed point, and then at
even lower $\beta$ reappears on the other side with the approriate 
symmetry to disappear by TB with the stable member of the pair.
In fig. \ref{fig_touch_and_go_po3_sbm} we show the surfaces of section 
just before (a) and soon after (b) the ``touch-and-go'' 
bifurcation of the orbits $(3)_1^{-}$ and $(1)^{+(0)}$.
This ``touch-and-go'' (TAG) bifurcation of the three-bounce orbits occurs
over such a small $\beta$ interval for small tilt angles that it
is hard to distinguish from a trifurcation of the $(1)^+$ orbit
without careful magnification of the transition, but it is required
by continuity and the generic principles of 2D conservative maps.
In Fig. \ref{fig_period_po3_type1_sbm} we plot the periods of these 
four three-bounce orbits, $(3)_1^{\pm},(3)_2^{\pm}$, 
which are related to the resonant tori of the untilted system.

As in the case of the two-bounce orbits, our knowledge of the
behavior of the $(1)^+$ orbit allows us to predict that in the
first interval their must exist a further (pair) of three-bounce
orbits which have no analog in the untilted system.  The reason is
the following.  From Fig. \ref{fig_trm_po1_sbm},
 for small tilt angle, we know that
the $Tr M_1$ for the $(1)^+$ orbit passes through $-1$ three times
before the $(1)^+$ orbit becomes permanently unstable.  Each time
$Tr M_1=-1$ there must be a TAG bifurcation, so there must
be three such bifurcations.  Two of them are associated with
the $(3)_1^{-},(3)_2^{-}$ orbits we have already identified and occur
near $\beta=2\pi/3, 4\pi/3$; the
third TAG bifurcation must be associated with a third pair of orbits
born by TB at large $\beta \sim 1/\theta$.  This pair plays a similar
role for the three-bounce orbits as does the $(2)^*$ orbit for the
two-bounce orbits in each interval, hence we denote them by $(3)_*$

\begin{figure}[htbp]
\begin{center}
\leavevmode
\epsfysize=4.5in 
\epsfbox{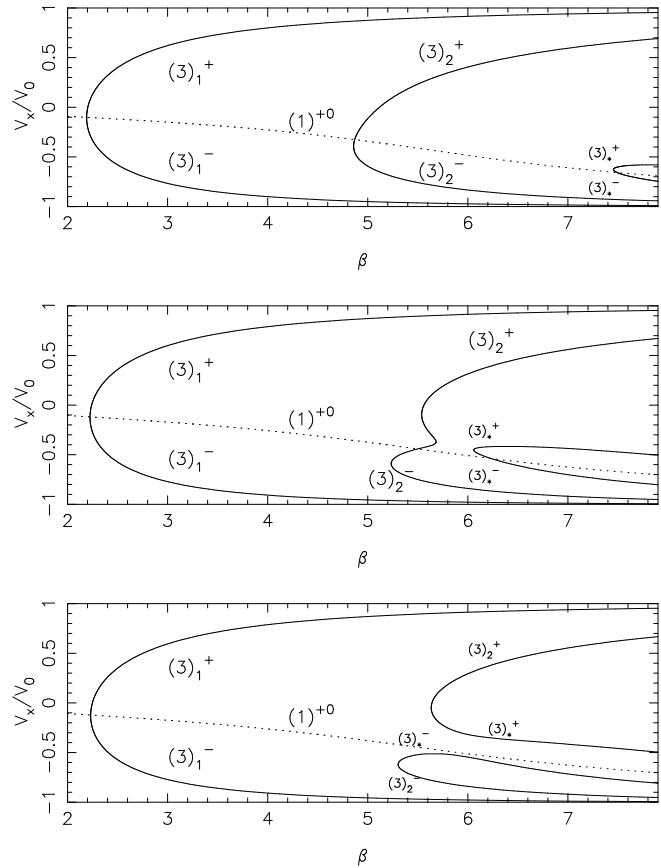}
\end{center}
\caption{
The bifurcation diagrams of the self-retracing three-bounce orbits 
in three 
different regimes (see text). The vertical axis represents the $x$
component of the scaled velocity of the electron at the point of
collision with $\tilde{v}_y = 0$. The dotted line represents the 
single-bounce orbit. Note the exchnage of partners bifurcation
between (b) and (c).
\label{fig_bifdiag_po3_sbm}
}
\end{figure}

As $\theta$ is increased to order unity, the
TAG bifurcation of the $(3)_*$ orbits moves to lower $\beta$ till
it eventually coincides with the TAG bifurcation of the $(3)_2^-$
orbit and the two bifurcations ``annihilate''.  
We know this must occur since $Tr M_1$ ceases
passing through $-1$ the second and third times 
(see Fig. \ref{fig_trm_po1_sbm}).
The TAG resonances relating the orbits to
the resonances of the $(1)^+$ orbit no longer exist for higher $\theta$
(just as the PDBs of the $2^+,2^*$ no longer exist above some critical
angle), but the orbits do not disappear.  Instead, they demonstrate an
``exchange of partners'' bifurcation , which for higher tilt angles 
allows them to exist without ever evolving into TAG resonances of the 
$(1)^+$ - see Fig. \ref{fig_bifdiag_po3_sbm}.
Again, just like for the two-bounce orbits, the transformation from 
the small tilt angle to large tilt angle behavior requires the 
appearance of auxiliary three-bounce orbits in additional tangent
bifurcations to provide a smooth evolution.
This scenario is illustrated by the bifurcation diagrams in Fig.
\ref{fig_bifdiag_po3_sbm}.

In principle an analytic theory of the periods and stability of
these three-bounce orbits is possible, but the system of three
coupled transcendental equations which define the period are not
easily analyzed.  Since we already know the qualitative scenario,
we have simply used the symmetry properties of these three-bounce  
orbits to locate numerically the fixed points and hence find the period
and time interval between collisions.  These quantities are all we
need to use the general formalism for the monodromy matrix developed
in Appendix C.  

In fig. \ref{fig_trM_po3_sbm} we show the behavior of the trace of 
the monodromy matrix for three-bounce orbits $(3)_1^{0\pm}$, 
$(3)_2^{0\pm}$ and $(3)_*^{0\pm}$.
The stability properties of the three-bounce orbits show a clear
analogy with the behavior of two-bounce orbits. The
$(3)_1^{\pm},(3)_2^{\pm}$ orbits related to the resonant tori, 
are either always unstable, or go
unstable via period-doubling bifurcations and never regain stability.   
Whereas the behavior of the new $(3)_*$ is different.  
As follows from Fig. \ref{fig_trM_po3_sbm}, the initially 
unstable $(3)_*$ restabilizes via pitchford bifurcation after its TAG
bifurcation with the $(1)^+$ orbit, before 
eventually going unstable in a period-doubling 
bifurcation at higher value of $\beta$. The initially stable
$(3)_*$ orbit has a monotonically decreasing monodromy 
and goes unstable via a period-doubling bifurcation.
All of these orbits are self-retracing in the sense defined above.
At the pitchfork bifurcation of the $(3)_*$ orbit just described,
a new three-bounce orbit appears which is non-self-retracing.  
Thus, as for the two-bounce orbits, orbits of this type only appear
after the creation of the self-retracing orbits and hence arise at
relatively high $\beta$ values.  Hence they have little effect on the 
experimental observations and will be disregarded below.

\subsection{Many-bounce orbits}

The analysis of period-$n$ ($n > 3$) orbits can be conducted in a
similar framework. First, one can identify the periodic orbits,
which survived from  the resonant tori of the untilted system, and 
then relate these orbits to the $1:n$ resonances of the single-bounce
orbits $(1)^+$. Since for small tilt angles $Tr M_1$ is non-monotonic
with $\beta$ and crosses the stability region three times, the third
crossing will always give rise to new orbits which are born at 
$\beta \sim 1/\theta$ and which have no analogs in the untilted 
system.  As $\theta$ is increased these resonances will move to
lower $\beta$ and annihilate with earlier resonances leading to 
new tangent bifurcations and the ``exchange of partners'' already 
understood and observed for the two-bounce and three-bounce orbits.
Additional new orbits can be formed both by pitchfork bifurcations of
self-retracing orbits and by completely new tangent bifurcations,
however such orbits appear to play no role in the first and second
interval for experimentally relevant values of $\beta$.
Moreover, even the n-bounce orbits which {\it are} related to the 
resonant tori of the unperturbed system generally have too long 
periods and/or too much cyclotron energy to be observed in the 
tunneling spectra. As they introduce no essentially new physics 
we will not present any detailed treatment of these orbits.

\section{Periodic Orbits in the DBM}

We now analyze the periodic orbit structure of the double-barrier model
(DBM).  This model will provide a description of periodic orbits 
relevant to the experiments of refs. \cite{Fromhold,Muller}.  A
crucial point discussed in section IIA above is that in general 
for a fixed tilt angle the
classical dynamics of the DBM depends on two dimensionless parameters:
the parameter $\beta = 2v_0B/E$ already used in analyzing the SBM, and
the parameter $\gamma = \epsilon_0/eV$ measuring the ratio of the 
injection energy to the voltage drop.  Fortunately, in the experiments
this second parameter is roughly constant \cite{Muller,Monteiro2}, 
$\gamma \approx 1.15 - 1.17$.
Therefore the periodic orbit theory (and ultimately the semiclassical 
tunneling theory) need only be done varying $\beta$ with $\gamma$ 
fixed to the experimental value.  We will focus on this case henceforth.
In interpreting the results of this section however, it must be borne
in mind that $\beta$ no longer is the product of three independent
variables; $v_0$ and $E$ are related by the condition of constant 
$\gamma$.  The magnetic field however is still an independent variable
and thus it is easiest to think of increasing $\beta$ as increasing
the magnetic field. 

Many of the periodic orbits we will discuss below have been previously
identified by Fromhold et al. \cite{Fromhold} or Monteiro and Dando
\cite{Monteiro1}.  What has not been done is to systematize all the
experimentally-relevant orbits and find their intervals of existence
and stability.  This we attempt to do below.

As previously noted, the periodic orbit theory of the DBM is in many 
respects similar to that of the SBM, but there are three significant
differences.  First, orbits can be born or disappear in a manner which
violates the generic bifurcation principles for conservative systems
since the Poincar\'e map for the DBM is nonanalytic on the critical
boundary of the SOS (the curve separating initial conditions which will reach
the emitter barrier from those which will not, cf. section IIC).  
The novel bifurcations
which result (which we call cusp bifurcations) play a crucial role
in the behavior of the short periodic orbits in the system.
Second, the unperturbed system has a more complicated structure as
there can exist two distinct resonant tori corresponding to the
same resonance condition $n \omega_c = k \omega_L$, one corresponding
to helical orbits which do reach the emitter, and the other corresponding
to helical orbits which do not.  
Third, once the field is tilted, orbits which are periodic after $n$
bounces with the collector may collide with the emitter any number
of times from zero to $n$.  As a function of $\beta$ such orbits
can change their connectivity with the emitter.  
In fact it can be shown that any orbit which does reach the emitter
can only exist for a finite interval of $\beta$.
We will now explain these important points in detail.

\subsection{Periodic orbits at $\theta = 0$}

First let us assume there exists an $(n,k)$ resonant torus of the 
unperturbed system which does not make any collisions with the
emitter barrier for a given value of $\beta$.  At $\theta=0$ longitudinal
and cyclotron energy decouple and, as the emitter barrier plays no role,
the frequency of the longitudinal motion must be given by Eq. (\ref{wl_sbm})
for the SBM.  Using this formula for $\omega_L$, 
the resonance condition $n \omega_c = k \omega_L$
leads to a condition on $\beta$:
\begin{equation}
\beta = 2 \pi \frac{k}{n} \sqrt{\varepsilon_0/\varepsilon_L}
\label{bt_hel_col}
\end{equation}

Exactly as for the SBM, if such an orbit exists for one value of 
the longitudinal energy $\varepsilon_L$, another such family will 
exist at the same total energy but with smaller longitudinal energy, 
since adding to the cyclotron energy does not change $\omega_c$. 
From Eq. (\ref{bt_hel_col}) the new family with smaller $\varepsilon_L$
will exist at higher $\beta$ as the magnetic field will have to 
be increased to keep it in resonance.  As $\beta$ increases for such 
families the orbits will just move further away from the emitter
but will always exist above the threshold value defined by the maximum
value of $\varepsilon_L$.
Unlike the SBM however, the maximum allowed value is not $\varepsilon_0$, 
since
before all the energy is put into longitudinal motion the orbit begins
to hit the emitter barrier; this happens of course when $\varepsilon_L
=eV \equiv \varepsilon_0/\gamma$.  We will call orbits which don't reach the
emitter ``collector'' orbits and those which do ``emitter'' orbits.
Our argument implies that
there exist families of $(n,k)$ helical collector orbits for all $\beta$
{\it above} the threshold $\beta_c = 2 \pi (k/n) \sqrt{\gamma}$.
These orbits are identical to those in the SBM and the only change
introduced by the emitter barrier is that the threshold for their
creation has been raised by a factor $\sqrt{\gamma} = 
\sqrt{\varepsilon_0 /eV}$.  

\begin{figure}[htbp]
\begin{center}
\leavevmode
\epsfysize=1.8in 
\epsfbox{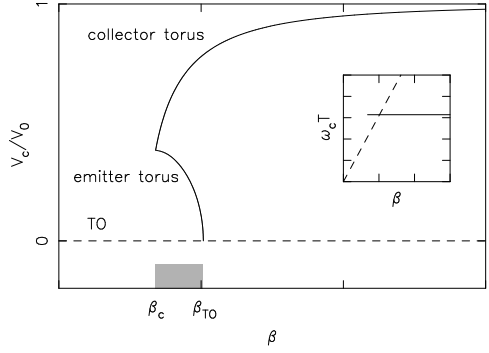}
\end{center}
\caption{
The scaled cyclotron velocity for the resonant tori 
($n = 1$, $k = 1$) as function of
$\beta$ at zero tilt angle; $\gamma = 1.2$, number of cyclotron rotations per
period $m = 1$. The horisontal line $\tilde{v}_c = 0$  corresponds to the
travesing orbit.
\label{fig_tori_dbm} 
}
\end{figure}

Now assume there exists an $(n,k)$ family for a given value of $\beta$
which {\it does} reach the emitter barrier.  The longitudinal frequency of 
any such orbit is easily calculated to be:
\begin{eqnarray}
\omega_L & = &
 \frac{2 \pi \omega_c}{\beta} \sqrt{\frac{\varepsilon_0}{ \varepsilon_L}} 
\left( 1 - \sqrt{1 - \frac{eV}{\varepsilon_L} } \right)^{-1}.
\label{wl_hel_em}
\end{eqnarray}
Note the crucial difference here from Eq. (\ref{wl_sbm}); for the 
emitter orbits $\omega_L$ is an {\it increasing} function of 
$\varepsilon_L$. Imposing the resonance condition then leads to the relation:
\begin{equation}
\beta = 2 \pi \frac{k}{n} \sqrt{\frac{\varepsilon_0}{ \varepsilon_L}} 
\left( 1 - 
\sqrt{1 - \frac{\varepsilon_0}{\gamma \varepsilon_L} } 
\right)^{-1}, 
\label{bt_hel_em}
\end{equation}
which implies that $\beta$ is also an increasing function of $\varepsilon_L$
in the interval of interest.
For emitter orbits the {\it smallest} value that $\varepsilon_L$ can take
is $eV$, otherwise they will cease to reach the emitter, and for this
value $\beta = \beta_c$. 
Therefore, like the collector families, the emitter $(n,k)$ families 
also do not exist below $\beta_c$.  They  
are born when $\beta$ increases through $\beta_c$
at the critical boundary simultaneously with the collector family 
corresponding to the same values of $(n,k)$ (see Fig. 
\ref{fig_tori_dbm}).

When created, the emitter families have 
non-zero cyclotron energy
(see Fig. \ref{fig_tori_dbm}) and can be continuously deformed 
by transferring cyclotron
energy to longitudinal energy, moving the family to higher values of 
$\beta$ for fixed total energy.  This can only
continue until $\varepsilon_L = \varepsilon_0$ and all of the energy is
longitudinal, yielding now a {\it maximum} allowed value of $\beta$,
\begin{equation}
\beta_{TO} = \beta_c [\sqrt{\gamma} + \sqrt{\gamma -1}].
\label{bt_to}
\end{equation}
We denote this value by $\beta_{\rm TO}$ because at this value the $(n,k)$
helical emitter family has collapsed to the traversing orbit 
(which exists and always reaches the emitter for $\gamma > 1$).
Thus the scenario at $\theta = 0$ is that two $(n,k)$ families are
born at the critical boundary each time $\beta$ increases through 
$\beta_c(n,k)$.  The collector family moves outwards in the SOS and
exists for all $\beta > \beta_c$, whereas the emitter family moves
inwards in the SOS and annihilates with the TO at $\beta_{TO}(n,k)$
(see Fig. \ref{fig_tori_dbm} ).  The consequence is that each emitter
family lives for only a finite interval, $ \beta_c < \beta < \beta_{TO}$.
By continuity all the emitter periodic orbits which evolve from
these emitter tori (in a manner similar to the SBM) will also live
in a finite interval given approximately by this inequality for small
tilt angle. To our knowledge this property of the system has not been 
demonstrated in the previous literature.  
As only the emitter orbits will play a major role in the semiclassical
theory of the tunneling spectrum (collector orbits make exponentially
small contributions), the point is of some significance.

It follows from this argument that as $\beta$ increases the collector
families evolve by transferring longitudinal energy to cyclotron
energy in the manner familiar from the SBM, whereas as $\beta$
increases the new emitter orbits {\it give up} cyclotron energy to remain in
resonance.  To understand this less familiar behavior recall
that increasing $\beta$ may be regarded as increasing $B$ with all
other parameters fixed.  As $B$ increases the cyclotron frequency 
increases and the longitudinal frequency will need to increase to
maintain the resonance condition.  As noted already, unlike 
the collector orbits, for emitter orbits the longitudinal frequency 
increases with $\varepsilon_L$.  The reason for this is that 
as $\varepsilon_L$ increases the electron traverses the fixed
distance to the emitter faster and is more rapidly returned to
the collector.  We will see below that the 
consequence of this reversal of the dependence on
$\varepsilon_L$ means that all bifurcations of emitter orbits 
in the DBM happen in the reverse direction (as a function of 
$\beta$) from the bifurcations
of the corresponding orbits in the SBM.

\subsection{Period-one orbits in the DBM}

\subsubsection{Continuity argument}

We now analyze the period-one POs of the DBM for $\theta \neq 0$.
Here we mean period-one orbits with respect to iteration of the 
Poincar\'e map defined at the collector of the DBM, i.e. the orbits
must collide with the collector only once before retracing.
For zero tilt angle these orbits will be of three types. 1) The collector
orbits corresponding to the $n=1,k=1,2,\ldots$ resonances, which do
not collide with the emitter. 2) The emitter orbits 
corresponding to the $n=1,k=1,2,\ldots$ resonances which do
reach the emitter. 3) The traversing orbit, which has zero cyclotron
energy and which hence must
reach the emitter for $\gamma > 1$. The TO has the period:
\begin{eqnarray}
T_{\rm TO} & = & \frac{\beta}{\omega_c} \left( 1 - 
\sqrt{ 1 - \frac{1}{\gamma} } \right). 
\label{t_to_dbm}
\end{eqnarray} 
As in the SBM, The helical families of orbits will 
generate pairs of POs when $\theta \neq 0$
and by continuity, for infinitesimal tilt angle,
the orbits arising from emitter families will be emitter orbits
and those arising from collector families will be collector orbits.

\begin{figure}[htbp]
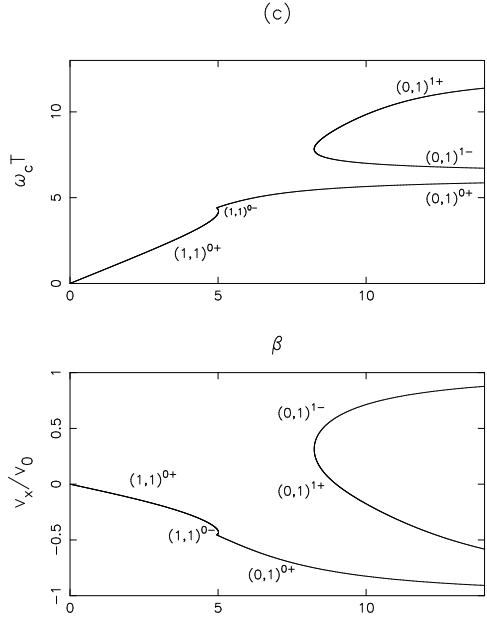

\begin{center}
\leavevmode
\epsfysize=4.2in 
\epsfbox{fig21a.epsi}
\end{center}
\vskip .7cm
\begin{center}
\leavevmode
\epsfysize=4.2in 
\epsfbox{fig21b.epsi}
\end{center}
\vskip 0.cm
\begin{center}
\leavevmode
\epsfysize=3.2in 
\epsfbox{fig21c.epsi}
\end{center}
\caption{
The scaled period $\omega_c T$ as function of $\beta$  and the 
corresponding bifurcation diagrams for the period-$1$ orbits in 
the double-barrier model at
zero tilt angle. The tilt angle (a) $\theta = 0.5^\circ$, 
(b) $ 11^\circ$ , (c) $ 20^\circ$. $\gamma = 1.17$.
The vertical axis in the bifurcation diagrams represents the 
scaled cyclotron velocity 
$\tilde{v}_c$ (a) or to the $x$
component of the scaled velocity of the electron at the point of
collision with the collector barrier.
\label{fig_t_and_bifdiag_po1_dbm}
}
\end{figure}

We must now classify periodic orbits not only by the number of
bounces with the collector, but also by the number of bounces with the
emitter.  We introduce the generalization of our earlier notation
\begin{eqnarray}
&(1,1)^{\pm k} & {\rm \ \ \ \ for \ the \ emitter \ orbits} 
\nonumber \\
&(0,1)^{\pm k} & {\rm \ \ \ \ for \  the \ collector \ orbits} 
\nonumber
\end{eqnarray}
where the first number in the parentheses denotes the number of
collisions with the emitter barrier and the second the number
with the collector barrier per period.
$k$ is the integer defining the interval as in the SBM; 
the period of an order $k$ single-bounce orbit is 
between $k T_c$ and $(k+1) T_c$. This notation
is used in Fig. \ref{fig_t_and_bifdiag_po1_dbm}. 

For infinitesimal tilt angle and $\beta < \beta_c \simeq 2 \pi$ 
there will exist only 
one single-bounce orbit, the analog of the TO, which we denote
as $(1,1)^{+0}$.  This orbit differs only 
infinitesimally from a straight line when $\beta \to 0$, but
gains more cyclotron energy as $\beta$ is increased, just as in the SBM.

As $\beta$ is increased to $\approx \beta_c$ {\it four} new 
single-bounce orbits
arise in an infinitesimal interval; these are the two non-mixing
orbits from each of the collector and emitter $n=1,k=1$ families.
Due to the breaking of the symmetry between these two orbits in
each family, they are created pairwise at slightly different $\beta$ values
and with slightly different periods.  However the corresponding 
collector and emitter orbits are still born at the same $\beta$ value
in a cusp bifurcation.  The two orbits which survive 
from the single-bounce collector orbit families are identical to those already 
discussed in the SBM, they are denoted by $(0,1)^{0+}$ and
$(0,1)^{-(1)}$, because they are now born in different intervals
(see Fig.  \ref{fig_t_and_bifdiag_po1_dbm}) of the period 
(the period of the orbit $(0,1)^{+(0)}$ is greater than $T_c$, while 
the period of $(0,1)^{-(1)}$ is less than $T_c$).
The single-bounce collector orbits must be non-mixing
by the simple argument given in discussing the SBM.  
The single-bounce emitter orbits
collide twice in each period and so it is less obvious that they
must be non-mixing in their collision with the collector barrier;
however it can be rigorously proved that this must be the case. 
Therefore, again our continuity arguments implies that only the
two emitter orbits with $v_y=0, v_x = \pm v_c$ will survive. 
The one with period
shifted slightly down from $T_c$ will
be denoted $(1,1)^{-(0)}$; the one with period shifted up
will be denoted $(1,1)^{+(1)}$.

Above $\beta_c$ in the first interval there now exist three single bounce
orbits, the $(0,1)^{+(0)}$ orbit which doesn't reach the emitter, the
$(1,1)^{-(0)}$ ``helical'' emitter orbit and the $(1,1)^{+(0)}$ ``traversing
orbit'', which has the shortest period of the three.  As in the SBM,
for $\theta \neq 0$ there is no qualitative difference between 
traversing orbits and helical orbits, since both must have non-zero
cyclotron energy.  As $\beta$ increases to $\approx \beta_{\rm TO}$
(see Eq. (\ref{bt_to}),
the helical $(1,1)^{-(0)}$ orbits loses cyclotron energy (as would
the corresponding orbits at $\theta=0$ discussed above) whereas
the $(1,1)^{0+}$ orbit gains cyclotron energy.  Eventually the
two orbits become degenerate and annihilate in a backwards tangent
bifurcation, the analog of the annihilation of the $n=1,k=1$ emitter
family at $\theta=0$ (see Fig.  \ref{fig_t_and_bifdiag_po1_dbm}).

At $\beta$ larger than the value for this TB the $(1,1)^{+(0)}$
orbit does not exist, and this is apparently in contradiction with the
behavior of the TO at $\theta=0$ which survives unscathed through
the annihilation of the helical family.  Moreover, by continuity,
for an infinitesimal tilt angle
the analog of the (normally) isolated TO must survive at all but a discrete 
set of values of $\beta$.  The resolution of this apparent paradox is
that, just as in the SBM, an orbit in the next interval, the 
$(1,1)^{+(1)}$, 
which is the partner of the $(1,1)^{-(1)}$, takes over the role of the
TO at this value of $\beta$, see Fig.  \ref{fig_t_and_bifdiag_po1_dbm}.
The same scenario repeats
then in the $k=1$ and higher intervals.  Note that in this scenario
all period-one emitter orbits only survive for a finite interval, being born
at some threshold value of $\beta$ by cusp bifurcation and disappearing
at higher $\beta$ by backwards tangent bifurcation.

\begin{figure}[htbp]
\begin{center}
\leavevmode
\epsfysize=4.in 
\epsfbox{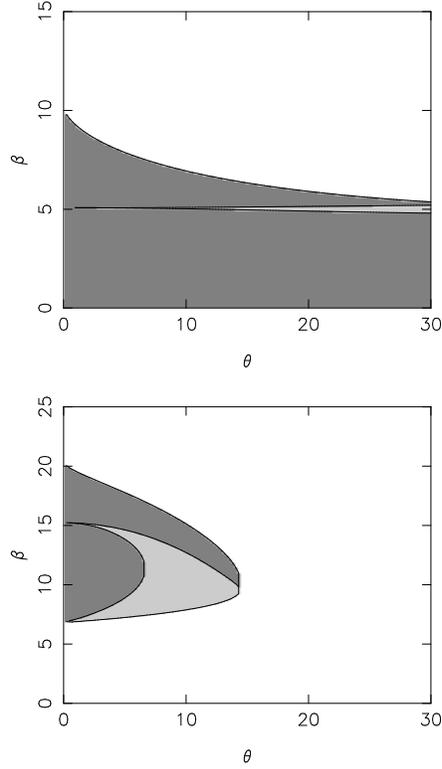}
\end{center}
\caption{
The intervals of existence of the period-$1$ ``emitter'' orbits
shown as shaded areas in the $(\theta,\beta)$ plane for (a) 
$(1,1)^{-(0)}$,
(b) $(1,1)^{+(0)}$, (c) $(1,1)^{-(1)}$, (d) $(1,1)^{+(1)}$.
Dark and light shading represent existing stable and unstable
periodic orbits respectively.  
\label{fig_exist_intervals_po1_dbm}
}
\end{figure}

The behavior of the single-bounce orbits for larger tilt angle 
differs in one important respect.  It becomes 
more and more difficult for the $(1,1)$ orbits to reach the 
emitter barrier and as a result their intervals of existence in
$\beta$ (which initially fill the entire $\beta$ axis)
shrink monotonically until they go to zero at a critical angle
which differs for each interval (see Fig. 
\ref{fig_exist_intervals_po1_dbm}b ).
The only exception is in the first interval where for sufficiently small
$\beta$ it is always possible to have a 
$(1,1)^{+(0)}$ analogous to the TO of the untilted system.
The reason the $(1,1)^{+(0)}$ orbit always exists is that we may
regards the limit $\beta \to 0$ as the limit of vanishing magnetic
field, so its tilt can have no effect on the orbit, which does have
enough energy to reach the emitter ($\gamma > 1$).  However since
all other single-bounce orbits require finite $\beta$, tilting the
field sufficiently for fixed $\gamma$ can prevent the electron
from reaching the emitter.  As these intervals shrink the scenario
also changes.  Instead of the $(1,1)^{+(k)}$ orbit being created
directly by a cusp bifurcation, it is created in a tangent bifurcation
as a $(0,1)^k$ orbit and then evolves at higher $\beta$ into
$(1,1)^{+(k)}$ orbit.  This is the first example of an orbit continuously
changing its connectivity with the emitter as a function of $\beta$;
these events also play a role in the theory of the two-bounce or
three-bounce orbits, as discussed below.

Now we discuss the stability of the single-bounce orbits.  Clearly, the
collector $(0,1)^\pm$ orbits have identical stability properties as
their SBM counterparts.  As for the emitter orbits, their stability can 
also be understood using qualitative arguments similar to the ones 
we applied in our SBM analysis. Just as in the SBM,
in the DBM for zero tilt angle the traversing orbit is stable for any 
$\beta$ and $\gamma$ except when it's period is either an integer
or a half integer multiple of the cyclotron period $T_c$, 
when it is marginally stable. When the period takes the values
$T=k T_c$ the corresponding value of $\beta$ is $\beta = \beta_{\rm TO}(1,k)$; 
when $T = (k + \frac{1}{2}) T_c$ the corresponding $\beta$ values are
and $\beta = \beta_{\rm ms}(k) \equiv 
(1 + \frac{1}{2 k}) \beta_{\rm TO}(1,k)$.
Therefore, for a small tilt angle the single-bounce orbit which
evolved from the TO of the untilted system, can become unstable
only near $\beta_{TO}$ and $\beta_{\rm ms}$. 
In particular, the $(1,1)^{+(0)}$ orbit is stable for small
$\beta$, but goes unstable and soon restabilizes near 
$\beta_{\rm ms}(0) \equiv \pi / (1 - \sqrt{1 - 1/\gamma})$.
As in the SBM, this instability for period $\approx T_c/2$ locates
the bifurcations involving the important period-two orbits.

Whereas in the SBM the $(1)^{0+}$ orbit simply evolves into
a helical orbit when $\beta \gg 2 \pi$, its analog, the 
$(1,1)^{0+}$ annihilates with the $(1,1)^{-(0)}$ orbit near
$\beta_{\rm TO}$.  Due to the general 
properties of tangent bifurcations, one of these orbits must be
stable, while the other must be unstable. Since the $(1,1)^{+(0)}$
orbit is a deformation of the stable TO it is the stable one just
before the TB, while the orbit $(1,1)^{-(0)}$ is unstable.  
This is illustrated by
the plot of the monodromy matrix for these orbits (Fig. 
\ref{fig_trM_po1_dbm}).

This $(1,1)^{-(0)}$ is worth further consideration because it
appears at the critical 
boundary near $\beta = \beta_c$ in a cusp bifurcation 
together with the collector orbit $(0,1)^{+(0)}$. 
A detailed analysis of cusp bifurcations is given in section 
IV.B.3 below.  Here we simply note that due to the singularity in the
Poincar\'e map at the critical boundary the monodromy matrices
defining the stability of the new orbits cannot be uniquely defined.
We will show that therefore the two orbits need not be born as
unstable-stable pairs as in tangent bifurcations (this is why
we have introduced the new term cusp bifurcation (CB)).
Moreover, one can show that of the two orbits born in a CB, the
one with the greater number of collisions with the emitter barrier 
is necessarily unstable.  It follows that the orbit 
$(1,1)^{-(0)}$ is unstable immediately after it is born, and turns
out to be unstable over its entire interval of existence until
it vanishes in the TB with $(1,1)^{+(0)}$.

These principles allow us to understand the behavior in the next
interval as well.  The emitter orbit $(1,1)^{+(1)}$ is also born in
a cusp bifurcation with the $(0,1)^{-(1)}$ collector orbit and hence is
born unstable.  Initially it plays the role of the ``other'' 
emitter helical orbit. However, near $\beta = \beta_{\rm TO}$ the orbit
$(1,1)^{+(1)}$ loses almost all
it's cyclotron energy (see Fig. \ref{fig_t_and_bifdiag_po1_dbm}) and 
becomes a recognizable
deformation of the TO of the untilted system. By continuity, since
away from $\beta_{\rm TO}$ the TO was stable, the $(1,1)^{+(1)}$ 
periodic orbit must restabilize near 
$\beta_{\rm TO}$. Its further evolution is similar to that of the
first interval orbit $(1,1)^{+(0)}$ just discussed.  It will bifurcate and
then restabilize near $\beta_{\rm ms}(1)$ and later annihilate with the
unstable orbit $(1,1)^{-(1)}$ in a tangent bifurcation - see Fig.
\ref{fig_trM_po1_dbm}.  This scenario is repeated in higher intervals
although the first interval of stability (below $\beta_{\rm ms}(1)$)
may disappear.  We note however, that as long as a
$(1,1)^{+(k)}$ orbit exists in each interval, it must have a region
of stability just before it annihilates with the $(1,1)^{-(k)}$
orbit (which is always unstable), although these intervals will
shrink with increasing tilt angle and $k$.

\subsubsection{Exact analysis}

The derivation of the periods of the period-one emitter orbits 
in the DBM can
be performed using a technique similar to the one employed for the
description of period-two non-mixing orbits in the SBM, since both the
emitter and collector bounces are non-mixing. The calculation 
is given in Appendix G and yields the following equation :
\begin{eqnarray}
\beta^2 & = & \left(\frac{\omega_C T}{2}\right)^2 
\left( 1 + \frac{\beta^2}{\gamma \left(\omega_c T\right)^2}
\frac{1 - f\left(\omega_c T\right)}
{1 - \cos^2\theta f\left(\omega_c T\right)} \right)^2 
+ 4 \sin^2\theta f^2\left(\omega_c T\right)
\left( 1 + \frac{\beta^2}{16 \gamma} 
\frac{1}{f\left(\omega_c T \right) 
\left(1 - \cos^2\theta f\left( \omega_c T \right) \right)}
\right)^2 ,
\label{t_po1_dbm}
\end{eqnarray}
where $\tilde{v}^-_e$ is the scaled velocity immediately {\it before}
the collision of the electron with the emitter barrier and
\begin{equation}
f(x) = 1 - \frac{x}{4}\cot\left(\frac{x}{4}\right) \nonumber
\end{equation}

This is a quadratic equation for $\beta$ for given $T$; it should be
solved along with condition (\ref{vz_gt_0_dbm}), that $v_z$ just
before the collision with the emitter is positive, to determine
the physically meaningful roots.
Solving Eq. (\ref{t_po1_dbm}) together with the condition
(\ref{vz_gt_0_dbm}), one can 
obtain the dependence $\beta(T)$, which was plotted in 
Fig. \ref{fig_t_and_bifdiag_po1_dbm} 
and used to obtain the corresponding bifurcation diagrams.
The equation Eq. (\ref{fig_trM_po1_dbm}) and the
condition (\ref{vz_gt_0_dbm})
imply that $\beta(T)$ is not monotonic in each interval
$[(k-1)T_c < T < k T_c$, but always has a single maximum.  
Therefore it describes {\it two} different $(1,1)$ orbits, 
which we already identified as the $(1,1)^{\pm}$ orbits.

Using Eqs. (\ref{t_po1_dbm}) and (\ref{vz_gt_0_dbm}), one can show,
that, as for the period-one orbits in the SBM, for a nonzero tilt
angle the period of the $(1,1)$ orbits cannot be equal to integer 
multiples of the cyclotron period $k T_c$. Moreover, the period also 
can not take values too close to $k T_c$. The width of each of these
``forbidden'' regions in each interval increases (from zero at 
$\theta = 0$) with the increase of tilt angle, so that at some
critical angle (which depends on the interval number $k$) the 
``forbidden'' regions originating from $T =  (k-1) T_c$ and $T = k
T_c$ merge and as already noted, it becomes impossible for the
period-one orbits to reach the emitter in this interval of period.
When period-one emitter orbits exist in an interval, we can calculate
their interval of existence in $\beta$ from Eqs. 
(\ref{t_po1_dbm}), (\ref{vz_gt_0_dbm}).  The results
for the $(1,1)^{+(0)}$ and $(1,1)^{+(1)}$ orbits are shown in 
Fig. \ref{fig_exist_intervals_po1_dbm}.

One can also calculate the stability properties of the $(1,1)$ orbits
as outlined in Appendix H.  The results for the trace of the monodromy
matrix for different $(1,1)$ orbits are shown in 
Fig. \ref{fig_trM_po1_dbm}.   The qualitative behavior is as discussed
above.  The key new feature that emerges is an analytic understanding
of the cusp bifurcations at the birth of the $(1,1)^-$ and $(0,1)^+$
orbits.  

\subsubsection{Cusp Bifurcations and Connectivity Transitions}

\begin{figure}[htbp]
\begin{center}
\leavevmode
\epsfysize=3.2in 
\epsfbox{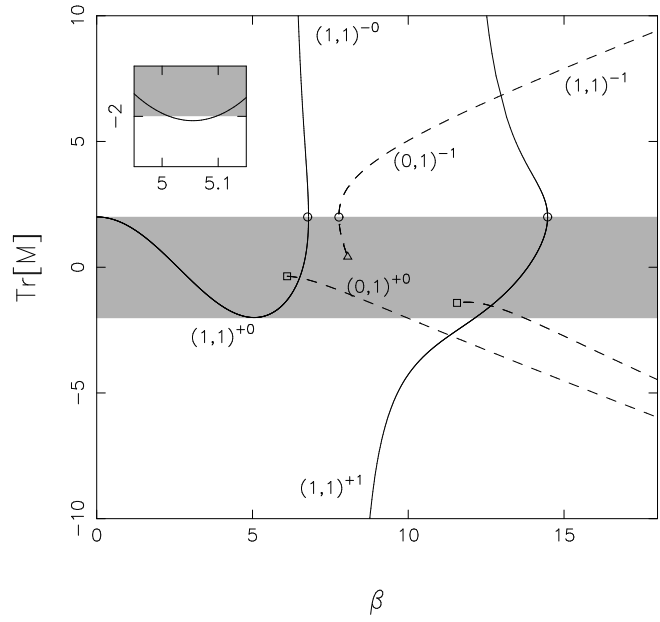}
\end{center}
\caption{
Trace of monodromy matrix of the period-$1$ orbits of the first two
intervals at $\theta = 11^\circ$, $\gamma = 1.17$. The tangent 
bifurcations, cusp bifurcations and connectivity transitions are 
labeled by open circles,
open squares and open triangles respectively.
Shaded area corresponds to the stable region.
\label{fig_trM_po1_dbm}
}
\end{figure}

First, we note again that {\it all} relevant emitter orbits
are born in cusp bifurcations at the low $\beta$ side of 
their existence interval.  As shown in Appendix H, the monodromy matrix
for the emitter orbit born in a CB involves terms proportional to
the inverse of the velocity at the emitter barrier. Since at the 
cusp bifurcation the emitter velocity goes to zero, the trace of the
monodromy matrix of the corresponding orbit will diverge (see 
Fig. \ref{fig_trM_po1_dbm}).  
Therefore {\it all} emitter orbits are extremely unstable just
after their appearance in a CB (unless both orbits born in a cusp
bifurcation are emitter orbits, in which case the one with greater
number of collision with the emitter barrier will be extremely unstable).
On the other hand, their companion collector orbits, for $\beta$ just
above the CB no longer ``feel'' the emitter barrier and must have
stability properties as in the SBM, where there is no such divergence
for any values of $\beta$.  Therefore the monodromy matrix for this
orbit as $\beta$ is reduced to the CB value does not tend to infinity
but tends toward a finite value (see Fig. \ref{fig_trM_po1_dbm}).
Whether this value is in the stable region or not depends on the
value of the tilt angle and of $\gamma$.  For large tilt angle
the companion collector orbit is typically unstable just above the CB
and {\it two} unstable orbits are born at the CB, in contrast to
the generic behavior at tangent bifurcations.

There is an interesting and important variant on the concept of
cusp bifurcation.  It is possible that
orbits may be born as collector orbits in a TB, and lose
cyclotron energy with increasing $\beta$ 
until at some higher $\beta$ they reach the emitter and
evolve into emitter orbits.  We will refer to these events as
{\it connectivity transitions} since the orbit changes its
connectivity to the emitter.  However in this case no new orbit is created
at the value of $\beta$ at which the emitter is reached, so this
is not a bifurcation point in any sense.  Nonetheless, the behavior of
the monodromy matrix of this one orbit in the neighborhood of the
connectivity transition is similar to that near a CB.  
The $Tr[M]$ tends to a finite value on the low $\beta$ side, whereas
it diverges at the high $\beta$ side. For a not too small tilt angle
this behavior occurs for the $(0,1)^{+(1)}$ and $(1,1)^{+(1)}$ orbits 
(see Fig. \ref{fig_trM_po1_dbm}).  Interestingly enough, the dynamics
does not seem to favor these connectivity changes although they are
allowed.  For tilt angles larger than a few degrees they are typically
replaced by a tangent bifurcation and a new cusp bifurcation which
ultimately results in the appearance of an orbit with higher 
connectivity and the disappearance of one with lower connectivity.

\subsection{ Period-two orbits}
As in the SBM, the most important set of period-two orbits, for small
tilt angles, are those associated with the period-doubling bifurcations
of the (deformed) traversing orbit $(1,1)^{+0}$ 
which occurs near $T \approx T_c/2$ (so that
the relevant period-two orbits have $T \approx T_c$).
The scenario for their creation and evolution is in many respects
similar to the behavior of the helical period-one orbits just
described.  For $\theta=0$ a pair of emitter and collector families
are created at the critical boundary at the threshold 
$\beta_c(n=2,k=1)= \pi \sqrt{\gamma}$.  The emitter family loses 
cyclotron energy 
with increasing $\beta$, moves inward in the SOS and annihilates with the TO
at $\beta_{TO}(2,1)= \pi \left(\gamma + \sqrt{\gamma^2 - \gamma}\right)$.
The collector family gains cyclotron energy with increasing $\beta$,
moves outward, and exists for all $\beta$.

When $\theta \neq 0$ two orbits survive from each of the collector and
emitter families.  These four orbits are born pairwise in two cusp 
bifurcations involving degenerate collector and emitter orbits, which
occur at slightly different values of $\beta$.  The two collector
orbits involved are identical to the mixing $(2)^{+}$ orbit
of the SBM and the non-mixing $(2)^{-}$ orbit. According to our 
notation, these collector orbits are denoted as $(0,2)^\pm$. 
The emitter orbit created in a CB with the non-mixing orbit $(0,2)^-$,
which will be referred to as the $(2,2)^-$ orbit 
(see Fig. \ref{fig_po2_YZ_dbm}), has the simplest 
qualitative behavior and we will discuss it first.

\subsubsection{$(2,2)^-$ orbits}

The period-$2$ emitter orbit, which appears
together with the $(0,2)^-$ orbit, at the cusp bifurcation is
degenerate with $(0,2)^-$ and has therefore the same shape. However, 
as the parameter $\beta$ is increased, it begins striking the 
emitter wall with a nonzero velocity. Since at the point of this 
collision the angle in the $(y,z)$ plane between the electron velocity
and the normal to the barrier is {\it not $90^\circ$}, it is a {\it 
mixing collision}. In fact it can be shown that any orbit in either 
the SBM or DBM with more than two total collisions must be partially 
mixing.

As a result of the mixing collision with the emitter barrier this
emitter orbit acquires a cusp at the emitter.
Although this $(2,2)^-$ orbit is mixing in a strict sense, it remains 
non-mixing at the ``collector'' barrier. Since the magnitude of the 
velocity is very low 
at the emitter collision the mixing for this orbit remains very weak.  

Whereas the $(0,2)^-$
orbit moves away from the emitter with increasing $\beta$ in the usual
manner, the $(2,2)^-$ orbit transfers more and more energy to
longitudinal motion until its ``two legs'' come together and it
becomes degenerate with the $(1,1)^+$  traversing orbit.  It is then
absorbed in a backwards period-doubling bifurcation, causing a
change in the stability of the $(1,1)^+$ orbit.

\begin{figure}[htbp]
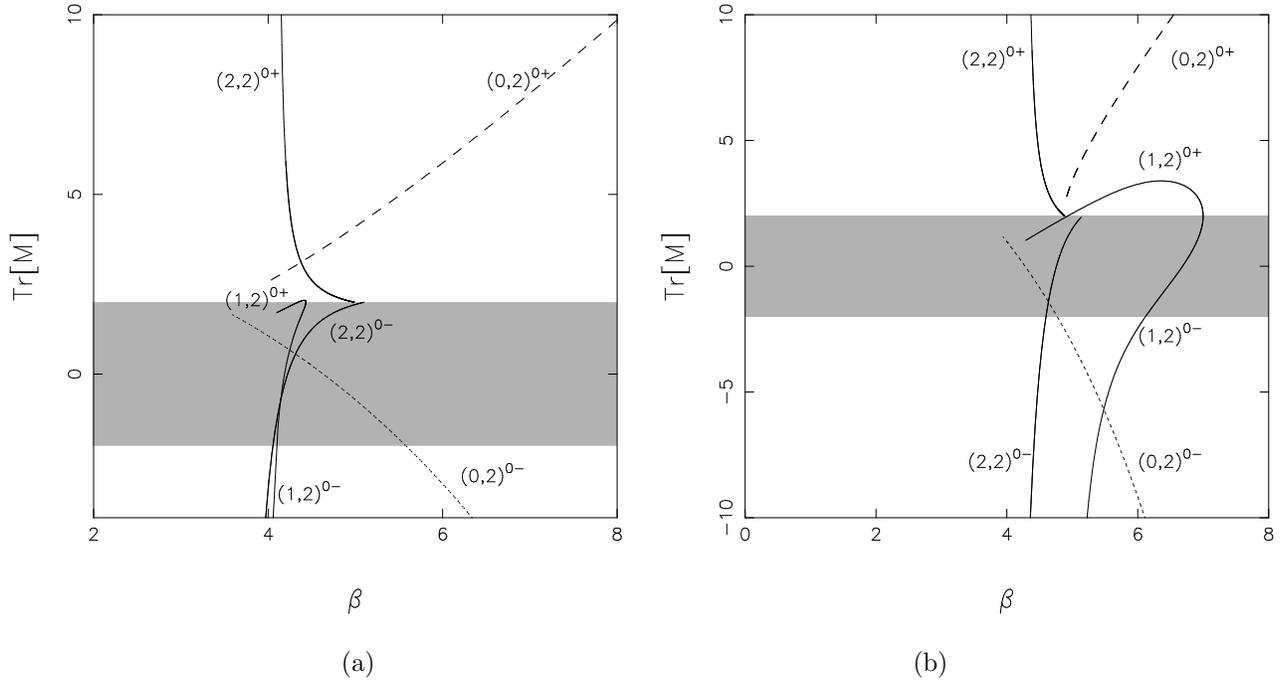

\begin{center}
\leavevmode
\epsfysize=3.2in 
\epsfbox{fig27a.epsi}
\hskip 0.5 truecm
\leavevmode
\epsfysize=3.2in 
\epsfbox{fig27b.epsi}
\end{center}
\begin{center}
\ \ \ \ \ \ \ \ \ \ \ \ \ \ \ 
(a)
\ \ \ \ \ \ \ \ \ \ \ \ \ \ \ 
\ \ \ \ \ \ \ \ \ \ \ \ \ \ \
\ \ \ \ \ \ \ \ \ \ \ \ \ \ \
\ \ \ \ \ \ \ \ \ \ \ \ \ \ \ 
(b)
\ \ \ \ \ \ \ \ \ \ \ \ \ \ \ 
\end{center}
\caption{
The trace of monodromy matrix for different period-$2$ orbits of the 
first interval at (a) $\theta = 17^\circ$ and (b) $\theta = 28^\circ$. 
\label{fig_trM_po2_dbm}
}
\end{figure}

We have already shown by continuity that the $(1,1)^+$ orbit must 
destabilize and restabilize in a short interval when its period
is $\approx T_c/2$.  And we have argued that all its bifurcations
must be backwards, since in the DBM orbits are born at lower $\beta$
in cusp bifurcations.  Therefore this backwards PDB of the emitter
$(2,2)^-$ orbit
corresponds to one of these stability changes.  To decide which one,
we note that although the $(2,2)^-$ orbit
must be born unstable because
it is the more connected partner in a cusp bifurcation, it should
typically be more stable than other period-two orbits which are
mixing at the collector, when the velocity is large.
Thus, we expect it to restabilize at higher $\beta$ and 
therefore to restabilize the $(1,1)^+$ orbit when the $(2,2)^-$ orbit
is absorbed as a stable period-two orbit in the backwards
PDB (see Fig. \ref{fig_bifdiag_po2nm_dbm}).  The exact 
calculation of the monodromy matrix (see Appendix H for the details)
confirms this scenario - see Fig. \ref{fig_trM_po2_dbm}.
Furthermore, increasing the tilt angle does not change the scenario
for the $(2,2)^-$ orbit, it only reduces its 
interval of existence. This orbit is relevant in the first
peak-doubling region observed at small tilt angles in the data of
Muller et al. \cite{Muller}.

\subsubsection{$(1,2)$ and $(2,2)^+$ orbits}

As just noted above, a collector orbit identical to the mixing 
$(2)^+$ orbit of the SBM (the $(0,2)^+$  orbit) is also 
created in a cusp bifurcation with
an emitter orbit which must have similar morphology.  The simplest
scenario would have this emitter orbit evolving exactly as did
the $(2,2)^-$ orbit, losing cyclotron energy until it is absorbed
by the $(1,1)^+$ in the other backwards PDB.  However we can immediately
see that this simplest scenario is impossible.  
The mixing collector orbit $(0,2)^+$ with zero emitter collisions 
per period and an emitter orbit $(2,2)^+$ with {\it two} emitter collisions
per period can never be created in a {\it single} cusp bifurcation.

If it were possible, than at the cusp bifurcation these two orbits
would have zero $z$ {\it and} $y$ components of the velocity at
two {\it different} points of collision with the emitter barrier
\cite{explain_why_vy=0}. Since the total
kinetic energy of the electron must be the same at any collision with
the emitter barrier, this means that the velocities at each of the
collisions with the emitter wall will differ only by the sign of $v_x$.
That is possible only for a zero tilt angle, when the system possesses
reflection symmetry.

\begin{figure}[htbp]
\begin{center}
\leavevmode
\epsfysize=3.5in 
\epsfbox{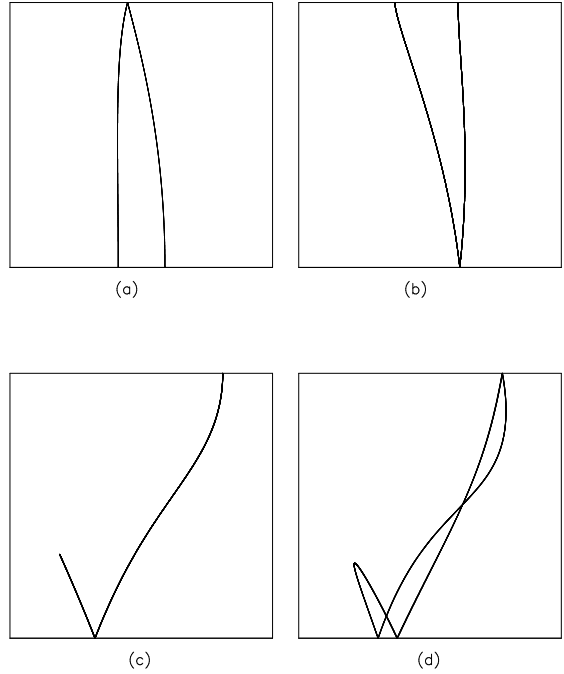}
\end{center}
\caption{
Examples of the different types of period-$2$ orbits in the DBM, 
projected onto
$(y,z)$ planes : (a) a $(2,2)^-$ orbit, 
(b) a $(2,2)^+$ orbit, (c) a self-retracing $(1,2)$ orbit,
(d) a non-self-retracing $(1,2)$ orbit.
\label{fig_po2_YZ_dbm}.
}
\end{figure}

What must happen instead is that the $(0,2)^+$ is born in a cusp
bifurcation with an orbit of the type $(1,2)^+$ 
(see Fig. \ref{fig_po2_YZ_dbm}), which infinitesimally
above the CB is connected to the emitter at one point and not two.
For small tilt angle the reflection symmetry is only weakly broken
and the other leg of this orbit will be quite close to the emitter,
but it may not touch.  Eventually, the creation of this orbit leads
to the creation of a $(2,2)^+$ orbit (see Fig. 
\ref{fig_po2_YZ_dbm}), which is absorbed by the 
$(1,1)^+$ in a backwards inverse PDB.  However the qualitative scenario
changes several times with increasing tilt angle and may be quite subtle,
with no less than four regimes which are relevant to the recent experiments.
Since the orbits involved control much of the peak-doubling behavior
at larger tilt angles, we will describe these scenarios in some detail
here.  In the next paper \cite{sc} we will make specific connections
to the data of Muller et al. \cite{Muller}

{\bf Regime One} ($\theta < \hat{\theta}_1$): 
This regime is described completely by continuity arguments once it
is understood that the mixing $(0,2)^+$ collector 
orbit must pair with a $(1,2)_1$ orbit.  
As $\beta$ increases above the threshold 
$\beta_c \approx \pi \sqrt{\gamma} (1 + 2 k)$ (where $k = 0,1,\ldots$ is
the interval number) 
the $(0,2)^+$ and $(1,2)_1$ orbit are created in a CB.  In 
a very small interval of $\beta$ this $(1,2)_1$ orbit attaches its other leg
to the emitter and becomes a $(2,2)^+$ orbit in a connectivity
transition of the type described in section IV.B.3 above.
The $(1,2)_1$ orbit must have been
born unstable at the CB and since the $(0,2)^+$ orbit it creates is 
mixing at the collector we
expect it to remain unstable as it loses cyclotron energy until it
is absorbed in a backwards inverse PDB with the $(1,1)^+$ orbit.
The $(1,1)^+$ then becomes unstable and is shortly after restabilized by
its  backwards PDB with the $(2,2)^-$ orbit.  All steps are consistent with
the continuity argument from $\theta=0$.  The bifurcation diagram 
in Figs. \ref{fig_bifdiag_po2m_1_dbm}
illustrates the behavior in this regime.

The $(1,1)^+$ continues its evolution until it vanishes in the
backwards tangent bifurcation described above and neither creates nor
destroys any further period-two emitter orbits.  However there is
a new period-two orbit created by the $(0,1)^+$ collector orbit.
It behaves just as in the SBM and goes unstable creating a $(0,2)^*$ 
orbit which is the exact analog of the $(2)^*$ orbit of the SBM.
However this only occurs at large $\beta$ values and the orbit 
never reaches the emitter once it is created, so it is not relevant
to the experiments at small tilt angle.  We mention it
because it will become very relevant at large tilt angles.

{\bf Regime Two} ( $\hat{\theta}_1 < \theta < \hat{\theta}_2$):
The behavior in this regime is as follows.  As $\beta$ increases,
as before, the first event is the creation of the $(0,2)^+$ collector orbit 
and the $(1,2)_1$ orbit via CB.  This $(1,2)_1$ orbit evolves for some
interval in $\beta$ without becoming a $(2,2)^+$ and in this interval
a second CB occurs in which a distinct orbit $(1,2)_2$ 
and a $(2,2)^+$ are created - see Fig. \ref{fig_bifdiag_po2m_dbm}a)
(this can happen because their connectivity only differs by one).
At slightly higher $\beta$ still the two orbits $(1,2)_1,(1,2)_2$ 
annihilate in a backwards TB and a yet higher $\beta$ the $(2,2)^+$ 
orbit 
is absorbed by the traversing orbit in the now-familiar PDB.  The net
effect of the creation of this second orbit $(1,2)_2$ is to eliminate the
connectivity transition directly from $(1,2)_1$ to $(2,2)^+$.  The
dynamics seems to rapidly eliminate these transitions
even though they are not strictly forbidden; preferring to replace
one connectivity transition with a CB and TB which results in the
same final state.  The total number of $(1,2)$ orbits is increased to two
by this change.

{\bf Regime Three} ( $\hat{\theta}_2 < \theta < \hat{\theta}_3$):
As already mentioned, a further period-two orbit, $(0,2)^*$ is
created by the PDB of the $(0,1)^+$ collector orbit, exactly as
the $(2)^*$ orbit is created in the SBM.  As tilt angle is increased
this PDB moves to lower and lower $\beta$ until at the value
$\hat{\theta}_2$, it coincides with the cusp bifurcation which
creates the $(1,1)^-$ and $(0,1)^+$ orbits.  For larger $\theta$ 
a period-two emitter orbit of type $\nu$ is created at this CB.
Thus in a somewhat mysterious manner this CB is a ``point of accumulation''
for the creation of higher period orbits (a similar thing happens 
for period-three here as well).  We may call this orbit $(1,2)^*_1$ since
it is similar in many ways to the $(2)^*$ orbit of the SBM.  For example
it has no analog in the untilted system.
Just above the critical angle $\hat{\theta}_2$ this $(1,2)^*_1$ 
orbit is barely
reaching the emitter and it rapidly detaches for higher $\beta$ and 
becomes a collector orbit.  As $\theta$ is increased, very quickly this
connectivity transition is again replaced by a combination of CB and
TB, where in this case the CB involves the $(0,2)^*$
collector orbit and
a second $(1,2)^*$ orbit,$(1,2)^*_2$.  The orbits $(1,2)^*_1,
(1,2)^*_2$ then
annihilate at higher $\beta$ in tangent bifurcation
(see Fig. \ref{fig_bifdiag_po2m_dbm}b,c).
So except for very near the critical 
angle $\hat{\theta_2}$, there are now a total of four $\nu$ orbits
associated with the first interval.  These are the two $\nu^*$ orbits 
just mentioned, which are connected with the cusp bifurcation of
the $(0,1)^+$, $(1,1)^-$ orbits, and the two $(1,2)^+$ orbits which can
be associated with the destabilizing PDB of the $(1,1)^+$ traversing
orbit.  Therefore, although the scenario is substantially more
complicated than in the SBM, the bifurcations of the period-one orbits
in the first interval determine all the relevant period-two orbits.

For most of this interval the two $(1,2)^+$ orbits
exist at lower $\beta$ than the two $(1,2)^*$ orbits.
However as the next critical 
angle $\hat{\theta_3}$ is approached the intervals of existence of
these pairs of orbits begin to overlap and their associated fixed
point move together(see  Fig.  \ref{fig_bifdiag_po2m_dbm}c).
The final act is about to take place.

{\bf Regime Four} ( $\theta > \hat{\theta}_3$):  Recall that in the
SBM the different branches of the $(2)^*$ and $(2)^+$ orbits
linked up above the critical angle $\theta^{\dag}$.  In that case
the link was established by the merging of the PDBs at which these
orbits were created from the traversing orbit.  In the DBM a similar
connection now occurs for the $(1,2)^*$ and $(1,2)^+$ orbits via
an ``exchange of partners'' bifurcation (note, that we already 
encountered this bifurcation in the SBM - see the description of
three-bounce orbits). The $(1,2)_1$
and $(1,2)^*_2$ orbits are both created at cusp bifurcations with 
collector orbits (which are identical to the $(2)^+$, $(2)^*$ 
orbits of the
SBM) and are annihilated at tangent bifurcations with their partners
$(1,2)_2$, $(1,2)^*_1$.  At a critical angle $\hat{\theta}_3$ the 
$(1,2)_2$
and $(1,2)^*_1$ orbits exchange partners.  Above this angle, the
$(1,2)_1$ orbit born in CB with the $(0,2)^+$ annihilates in a TB with
the $(1,2)^*_2$ orbit born in a CB with the $(0,2)^*$; whereas the
$(1,2)_2$ orbit born in a CB with the (one and only) $(2,2)^+$ orbit
now annihilates with the $(1,2)^*_1$ orbit born at the CB of the period-one
orbits - see Fig. \ref{fig_bifdiag_po2m_dbm}d.

After the ``exchange of parters'' transition the $(1,2)_1$ orbits 
exists for a very large interval of $\beta$ and has relatively
low cyclotron energy.  Thus it plays a dominant role in the tunneling
spectrum in this interval of $\beta$.  The importance of this orbit
has been emphasized in work of Fromhold \cite{Fromhold}.

In contrast, the other pair of orbits, $(1,2)^+_2, (1,2)^*_1$, decrease
their interval of existence because the PDB and CB to which they
are connected move together.  

In Fig. \ref{fig_trM_po2_dbm} we show the behavior of the trace of the
monodromy matrix for different period-$2$ orbits. Note, that the orbit
$(1,2)^+$ does not become highly unstable in the whole interval of it's
existence and is therefore expected to produce strong scars in the
quantum-mechanical wavefunctions \cite{heller,bogomolny,berry}.

To summarize the complicated story of the period-two orbits :  For
small tilt angles the important orbits are the $(2,2)$ orbits
we have denoted as $(2,2)^-$  orbits.  As tilt angle increases
the importance of $(1,2)$ orbits increases and eventually they
become the dominant period-two orbits in the first interval.  Since
higher intervals correspond to greater chaoticity, they become 
important more quickly in the second interval.  These $(1,2)$ orbits are
created in a complicated bifurcation tree which connects to
a period-doubling bifurcation of the period-one traversing orbit, as well
as cusp bifurcations with various period-one and period-two collector orbits. 
It is very difficult to discern these relationships from simple
observations of the SOS as many of these orbits are born highly 
unstable in cusp bifurcations and certain of the transitions 
described occur over very small angular intervals.  

\subsection{ Period-$3$ orbits}

All of the qualitative differences between the periodic orbit
theory of the SBM and that of the DBM already
have entered into the description of the period-one and two orbits.
However, peak-tripling regions have been clearly observed in experimental
tunneling spectra, indicating that the behavior of period-three orbits
is relevant to these experiments. Moreover there has been a recent 
Comment questioning the interpretation proposed for these 
peak-tripling regions \cite{Fromcomment,Reply}
in ref. \cite{Muller}, where they were attributed to trifurcations of
the traversing orbit.  Since we are able to reach a complete
understanding of these orbits based on the principles used in discussing
the period-one and two orbits, we will briefly summarize their properties.

As for the period-$2$ orbits, for small tilt angles the main
period-$3$ orbits are those related to the resonances of the 
traversing orbit. When the tilt angle is exactly zero, the 
traversing orbit has 
{\it two} $1:3$ resonances in each interval, when its period is 
equal to $2 \pi (k + 1) T_c /3$ and $4 \pi (k + 1) T_c /3$
respectively. The behavior near each of these resonances is
essentially the same for small tilt angles, so we just consider the 
first one.  First, an emitter and collector family
is created at the critical boundary at $\beta_{c1} < \beta_1$.
The emitter family moves inwards in the SOS and 
collapses to the TO at resonance.  When the field is tilted
only two period-three orbits survive from each emitter family and they are
now created in cusp bifurcations with the corresponding collector
families at slightly different values of $\beta$.

As with the period-two orbits in the DBM, these emitter orbits
will move inwards in the SOS until they annihilate.  The one
difference in their behavior has already been noted in the discussion of
of the SBM (see section III.D).
Because period-three orbits generically are not
born or absorbed in bifurcations with a period-one orbit, these
two orbits cannot disappear precisely on resonance with the TO.
Instead one of them (the unstable one) passes through the fixed point
associated with the $(1,1)^{k+}$ traversing orbit in a 
touch-and-go bifurcation and then annihilates
with the other in a backward tangent bifurcation.  For all tilt
angles the interval between the TAG bifurcation and the TB is
negligibly small, and so practically speaking it is as if these
two orbits vanish in a ``backwards trifurcation''.

Again, as with the period-two orbits, for finite tilt angle the emitter
orbits cannot be created as $(3,3)$ orbits at the initial cusp
bifurcation.  Therefore the two emitter orbits just described
are created in the form of a $(1,3)$ and a $(2,3)$ orbit.
These orbits are the analogs of the
period-two $(1,2)$ orbits, but now there are two different types of 
orbits with less than the maximum $(3,3)$ connectivity to the emitter.
In $y-z$ projection the $(3,3)$ orbits each
have a mixing collision point (where two collisions occur) and a
non-mixing collision point (where only one collision occurs, 
see Fig. \ref{fig_po3_YZ_dbm}.).
The $(1,3)$ orbits correspond to detaching the orbit at the mixing
collision point, the $(2,3)$ orbits correspond to detaching it at 
the non-mixing collision point. As noted, both occur for each resonance.

\begin{figure}[htbp]
\begin{center}
\leavevmode
\epsfysize=3.2in 
\epsfbox{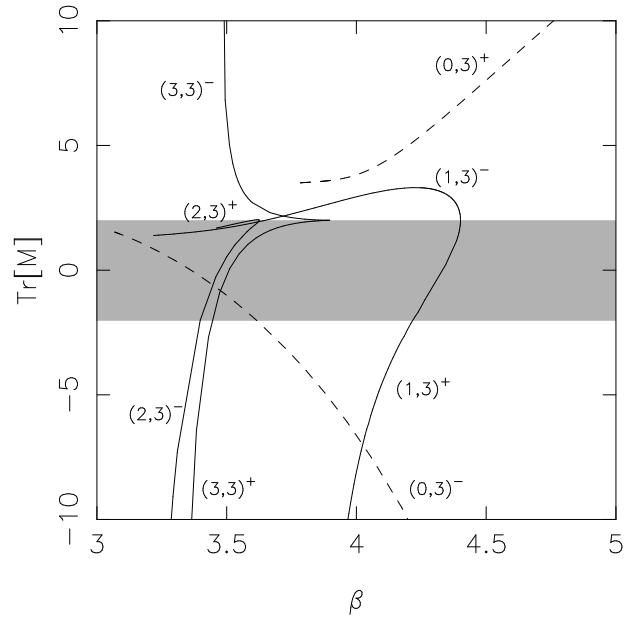}
\end{center}
\caption{
The trace of monodromy matrix for different period-$3$ orbits 
realted to the first $1:3$ resonance of the traversing
orbit $(1,1)^{+(0)}$ at $\theta = 17^\circ$.
\label{fig_trM_po3_dbm}
} 
\end{figure}

For small tilt angles the $(1,3)$ and $(2,3)$ orbits created at
these cusp bifurcations evolve by connectivity transitions into
the stable and unstable $(3,3)$ orbits which participate in the
TAG/TB behavior already described.  At higher tilt angles, as for
the period-two orbits, the connectivity transitions are replaced by
the appearance of a new $(1,3)$ and $(2,3)$ orbit which through
a combination of CB and TB leads to the same final state.
In the regime of small tilt angle there are six period-three orbits 
created in the neighborhood of
each resonance: two collector orbits, a $(1,3)$, a $(2,3)$ and 
two $(3,3)$ orbits.  For large tilt angles there are {\it eight} period-three
orbits due to the new $(1,3)$ and $(2,3)$ orbits which arise
to replace the connectivity transitions (see Fig. \ref{fig_bifdiag_po3_dbm}). 
The bifurcation
diagrams of Fig.  \ref{fig_bifdiag_po3_dbm} summarize the behavior 
of the family of 
period-three orbits related to the first resonance; qualitatively the
same behavior is observed at the second resonance as well.
In Fig. \ref{fig_trM_po3_dbm} we show the behavior of the trace of the
monodromy matrix for these orbits. Note, that as for the period-$2$ 
orbits, there is one orbit which, although exists in a substantial interval,
does not become too unstable (the orbit $(1,3)^-$) and is therefore 
expected to produce strong scars. 

The $(1,3)$ and $(2,3)$ orbits in each family 
appear at lower magnetic field than the resonance value, and
evolve either directly or indirectly into the $(3,3)$ orbits.
One of these orbits has been identified previously by Fromhold et al.
\cite{Fromprb,Fromcomment} in connection with peak-tripling.  We will analyze
the relation of the entire family to the experimental observations in
the following paper.  We simply point out here that each family of
eight period-three orbits is connected to a period-three resonance
through bifurcation processes, and in the scheme presented in this 
paper they arise as a natural consequence of that resonance.

As noted, for small tilt angles both resonances between the period-three
and period-one orbits in the first interval are similar, with the
creation of six or eight period-three orbits, four of which are
related by continuity to tori of the unperturbed system.
As with the period-two orbits, there is another resonance corresponding
to $T=3T_c$ which occurs in the first interval, but initially for
very high $\beta$.  This resonances will give rise to $(1,3)$ and
$(2,3)$ orbits analogous to the $(1,2)^*$ period-two orbits.
For small tilt angles they are created near the $(0,1)$ collector
orbit and do not reach the emitter, as happened also for the $(1,2)^*$.
Just as for that case, as tilt angle is increased the resonance
moves ``down'' to the period-one cusp bifurcation and now gives
rise to emitter orbits.  These emitter orbits then evolve similarly
to the $(1,2)^*$ orbits with exchange of partner bifurcations, etcetera.
However,
the periods of these orbits ($T > 2T_c$) apparently are too long
for them to be resolved as resonance peaks 
in the experimental data of ref. \cite{Muller}.

Higher period orbits also appear in families in connected bifurcation
sequences which begin with collector orbits and end with fully connected
emitter orbits which are annihilated at resonances with the TO.  
The principles and analytic relations we have derived can 
be used to develop a quantitative theory of such orbits, but we
have focused here on those which are experimentally-relevant and 
will defer any such analysis to other work. The relevant orbits
at the DBM are summarized in Table II.

\section{Summary and Conclusions}

We have developed a complete qualitative and quantitative theory
of the periodic orbits relevant to the magnetotunneling spectra
of quantum wells in tilted magnetic field.  

First we introduced two model hamiltonians and showed how to scale
the variables so that the only one or two dimensionless parameters
$\beta,\gamma$ describe the classical dynamics at fixed $\theta$.
As $\gamma=eV/\epsilon_0$ is approximately constant in experiments,
the dependences on magnetic field, voltage and injection energy
are all summarized by the behavior of the Poincar\'e velocity map
as a function of the variables $\beta,\theta$. 

The periodic orbit theory was first developed for the single-barrier model 
which elucidates many of the qualitative features of the system.  
In particular, the SBM
describes a standard KAM transition to chaos as a function of
tilt angle.  The period-one orbit with the smallest cyclotron
energy (the traversing orbit) plays a fundamental role in the
transition, with the relevant periodic orbits appearing through
the bifurcations of this orbit.  These bifurcations follow the
known bifurcation rules for generic (2D) conservative maps.
However the detailed scenario for the bifurcations evolves
with tilt angle in a complicated manner, which nonetheless can be understood
using continuity arguments.  Exact analytic expressions for the period
and stability of most of the relevant orbits were obtained for all
parameter values, something which has not been possible for other
experimentally-studied chaotic quantum systems.   We note again that
the SBM could be realized in a practical double-barrier structure
in which the band profiles were chosen to reduce the emitter energy
appropriately. 

In generalizing the theory to the double-barrier model which is
relevant to the present generation of experiments we uncovered several
new features of the dynamics.  Perhaps most interesting was the
discovery that {\it all} relevant orbits (except the traversing orbit)
are created in a new kind of bifurcation, called a cusp bifurcation,
which can violate generic bifurcation rules due to the discontinuity
in the Poincar\'e map on the curve separating initial conditions which
reach the emitter from those which do not.  These orbits are created in
families near, but below, the value of $\beta$ at which resonances with the
traversing orbit occur.  They only exist
for a finite interval of $\beta$ (or magnetic field)
and then annihilate in backwards
bifurcations with the traversing orbit or in tangent bifurcations.
In a given family of period-$n$ orbits ($n$ collisions with the collector
per period) there will exist orbits with $0,1, \ldots n$ emitter
collisions, connected together by one or more bifurcation ``trees''.
Typically, several orbits in a given family will be relevant for
understanding the magnetotunneling spectra, with their relative
importance changing as a function of tilt angle.  

Having determined
the periods and stability of all the orbits which are short
enough to resolve in the experimental tunneling spectra, we can
now calculate the tunnel current semiclassically using Equation 
(\ref{tun_rate})
quoted above.  In the companion paper to this work we will derive 
this equation and compare its predictions qualitatively and quantitatively
to the data of Muller et al. \cite{Muller}.  The complicated evolution
of the observed spectra with increasing tilt angle finds a natural
explanation in this approach.  The ability to develop a semiclassical
theory in essentially analytic form makes this system unique among the 
few quantum
systems which have been studied experimentally in the transition regime
to chaos.

\section{Acknowledgements}
The authors wish to thank G. Boebinger, 
M. Fromhold, H. Mathur, T. Monteiro and D. Shepelyansky
for helpful discussions.  We particulary thank Monteiro for pointing
out to us the importance of the $(1,2)$ orbits even at tilt angles
as small as $11^{\circ}$, and for explaining that the period-three
bifurcations follow the touch-and-go scenario.
The work was partially supported by NSF grant no. DMR-9215065.
We also acknowledge the hospitality of the Aspen Center for Physics
where some of this work was done.

\appendix
\section{The monodromy matrix for the single-bounce
periodic orbits}

In this Appendix we derive the expressions for the components and
the trace of the monodromy matrix for the period - one orbits in the 
single barrier model. 

By definition, the monodromy matrix $M^* = (m^*_{ij})$ of a
period-one orbit is the matrix, which represents the linerized 
Poincar\'e map, calculated at the position of the single-bounce
periodic orbit $(\tilde{v}_x^*,\tilde{v}_x^*)$ in the Poincar\'e 
surface of section :

\begin{eqnarray}
\Phi_x\left(\tilde{v}_x^* + \delta \tilde{v}_x,
\tilde{v}_x^* + \delta \tilde{v}_x\right) 
& = & \tilde{v}_x^* + m_{11}^* \tilde{v}_x^*
+  m_{12}^* \tilde{v}_y^* + O\left(
\left(\delta\tilde{v}_x\right)_0^2, 
\left(\delta\tilde{v}_y\right)_0^2, 
\left(\delta\tilde{v}_x\right)_0
\left(\delta\tilde{v}_y\right)_0
\right) \nonumber \\
\Phi_y\left(\tilde{v}_x^* + \delta \tilde{v}_x,
\tilde{v}_x^* + \delta \tilde{v}_x\right) 
& = & \tilde{v}_y^* + m_{21}^* \tilde{v}_x^*
+  m_{22}^* \tilde{v}_y^* + O\left(
\left(\delta\tilde{v}_x\right)_0^2, 
\left(\delta\tilde{v}_y\right)_0^2, 
\left(\delta\tilde{v}_x\right)_0
\left(\delta\tilde{v}_y\right)_0
\right) \label{PmapA}
\end{eqnarray}

The monodromy matrix $m_{ij}^*$ therefore relates to each other 
the deviation $\delta\tilde{\bf v}$ from the location of the periodic 
orbit after one iteration of the Poincar\'e map to the initial 
deviation  $\delta\tilde{\bf v}_0$ in the limit  
$|\delta\tilde{\bf v}| \rightarrow 0$ :  
 
\begin{eqnarray} 
\left( \begin{array}{c} 
\delta \tilde{v}_x \\ 
\delta \tilde{v}_y 
\end{array} 
\right) 
& = &  
\left( 
\begin{array}{cc} 
m_{11}^* & m_{12}^* \\ 
m_{21}^* & m_{22}^*  
\end{array} 
\right) 
\left( 
\begin{array}{c} 
\left(\delta\tilde{v}_x\right)_0 \\ 
\left(\delta\tilde{v}_y\right)_0 
\end{array} 
\right) 
+ O\left(\delta\tilde{v}^2\right) 
\end{eqnarray} 
 
Expanding the Poincar\'e map (\ref{PmapVSBM}) in  
$\delta\tilde{\bf v}$, we obtain : 
 
\begin{eqnarray} 
\Phi_x\left(\tilde{v}_x^* + \delta \tilde{v}_x, 
\tilde{v}_x^* + \delta \tilde{v}_x\right)  
& = & \tilde{v}_x^*  
\nonumber \\ 
& + &  
\delta T  
\left(   
\frac{ \sin\theta \omega_c T^* }{\beta} 
\cos\left(\omega_c T^* \right) 
-  
\left( \tilde{v}_x^* - \frac{2 \sin\theta}{\beta} 
\right)  
\sin\left(\omega_c T^* \right) \right) 
\nonumber \\ 
& + &  
\left(\delta\tilde{v}_x\right)_0  
\left(  
\cos\left(\omega_c T^* \right) - 
\frac{\sin\theta \beta \tilde{v}_x^* }{\omega_c T^*} 
\sin\left(\omega_c T^* \right)  
\right) 
\nonumber \\ 
& - &  
\left(\delta\tilde{v}_y\right)_0  
\cos\theta  
\sin\left(\omega_c T^* \right) 
\nonumber \\ 
& + &  
 O\left( 
\left(\delta\tilde{v}_x\right)_0^2,  
\left(\delta\tilde{v}_y\right)_0^2,  
\left(\delta\tilde{v}_x\right)_0 
\left(\delta\tilde{v}_y\right)_0 
\right) 
\label{phixA} 
\\ 
\nonumber \\ 
\Phi_y\left(\tilde{v}_x^* + \delta \tilde{v}_x, 
\tilde{v}_x^* + \delta \tilde{v}_x\right)  
& = & \tilde{v}_y^* 
\nonumber \\  
& + & 
\delta T   
\cos\theta  
\left( 
\left( \tilde{v}_x^* - \frac{2 \sin\theta}{\beta} \right)  
\cos\left(\omega_c T^* \right) 
+  
\sin\theta \sqrt{1 -  \left(\tilde{v}_x^*\right)^2 } 
\sin\left(\omega_c T^* \right) 
\right.  
\nonumber \\ 
& - & \left. \frac{2 \sin\theta }{\beta}  
\right) 
\nonumber \\ 
& + &  \left(\delta\tilde{v}_x\right)_0  
\cos\theta 
\left( 
 \sin\left(\omega_c T^* \right) 
- \frac{\tilde{v}_x^* \sin\theta} 
{ \sqrt{1 -  \left(\tilde{v}_x^*\right)^2 }}  
\left( 1 - \cos\left(\omega_c T^* \right) \right) 
\right) 
\nonumber \\ 
& + &  
\left(\delta\tilde{v}_y\right)_0 
\left( \cos^2\theta  \cos\left(\omega_c T^* \right) 
 +  \sin^2\theta \right) 
\nonumber \\ 
& + &  
 O\left( 
\left(\delta\tilde{v}_x\right)_0^2,  
\left(\delta\tilde{v}_y\right)_0^2,  
\left(\delta\tilde{v}_x\right)_0 
\left(\delta\tilde{v}_y\right)_0 
\right) \label{phiyA} 
\end{eqnarray} 
where the parameter $\delta T$ is the different between the time 
interval to the next collision of the electron with the barrier  
$T(\beta, \theta; \tilde{v})$ and the period of the single-bounce 
periodic orbit $T^*$ : 
\begin{eqnarray} 
T(\beta, \theta; \tilde{v}) & = & 
T^*\left(\beta, \theta\right) +  \delta T 
\label{tdt}  
\end{eqnarray} 
 
To obtain the linearization of the Poincar\'e map in terms of  
the velocity deviations, we therefore need to calculate the  
expansion of $\delta T$ up to linear order in $(\delta\tilde{v}_x)_0$  
and $(\delta\tilde{v}_y)_0$. This result can be obtained from the  
equation (\ref{dtSBM}), which relates the scaled in-plane components 
of the velocity of the 
electron $(\tilde{v}_x,\tilde{v}_y$ at the point of collision with the 
barrier to the time interval $T$ to the next collision. Substituting  
the expression (\ref{tdt}) into the eqation (\ref{dtSBM}), we obtain : 
\begin{eqnarray}  
\delta T & = &  - \frac 
{ 
\sin\theta \left(1 - \cos\left(\omega_c T^* \right) \right) 
+ \beta \tilde{v}_x^* \left( \cos^2\theta + \sin^2\theta  
\frac{\sin\left(\omega_c T^*\right)}{\omega_c T^*} \right) 
}  
{ 
\frac{\omega_c T^*}{\beta} \left(  
\cos^2\theta - \sin^2\theta \cos\left(\omega_c T^*\right) \right) 
+ \sin\theta \sin\left(\omega_c T^*\right)  
\left( \tilde{v}_x^* + \frac{2 \sin\theta}{\beta} \right) 
} 
\left(\delta\tilde{v}_x\right)_0 
\nonumber \\ 
& + & \sin\theta\cos\theta \frac 
{ 
\omega_c T^* - \sin\left(\omega_c T\right) 
} 
{ 
\frac{\omega_c T^*}{\beta} \left(  
\cos^2\theta - \sin^2\theta \cos\left(\omega_c T^*\right) \right) 
+ \sin\theta \sin\left(\omega_c T^*\right)  
\left( \tilde{v}_x^* + \frac{2 \sin\theta}{\beta} \right) 
} 
\left(\delta \tilde{v}_y\right)_0 \nonumber \\ 
& + &  
O\left( 
\left(\delta\tilde{v}_x\right)_0^2,  
\left(\delta\tilde{v}_y\right)_0^2,  
\left(\delta\tilde{v}_x\right)_0 
\left(\delta\tilde{v}_y\right)_0 
\right) 
\label{dtv1} 
\end{eqnarray} 
 
Substituting (\ref{dtv1}) into (\ref{phixA}), (\ref{phiyA}) and  
using the expression (\ref{vxy1}) for the scaled velocity for  
the one-bounce periodic orbit, we obtain the following 
result for the components of the monodromy matrix : 
 
\begin{eqnarray} 
m_{11}^* & = &  \sin^2\theta +  
\cos^2\theta \cos\left(\omega_c T^*\right) 
- 2 \sin^2\theta \cos^2\theta \left( 1 -  
\frac{\omega_c T^*}  
{\tan\left(\omega_c T^*\right)} 
 \right) \left(1 -  
\frac{\sin\left(\omega_c T^*\right)}{\omega_c T^*} \right) 
\nonumber \\ 
m_{12}^* & = & - \cos\theta \left( \sin^2\theta {\omega_c T^*} 
+ \cos^2\theta \sin\left({\omega_c T^*}\right) \right) 
\nonumber \\ 
m_{21}^* & = &  
\cos\theta \sin\left({\omega_c T^*}\right) 
+ \frac{4 \sin^2\theta \cos\theta}{\omega_c T^*} 
\left( 1 -  
\frac 
{\frac{\omega_c T^*}{2}} 
{\tan\left(\frac{\omega_c T^*}{2} \right)} 
\right) 
\nonumber \\ 
& \times & 
\left(1 - \cos\left(\omega_c T^*\right)  
-  
\left(1 - \frac{\frac{\omega_c T^*}{2}} 
{\tan\left(\frac{\omega_c T^*}{2} \right)} 
\right) 
\left(  
\cos^2\theta +  
\sin^2\theta \frac{\sin\left(\omega_c T^*\right)}{\omega_c T^*} 
\right) 
\right) 
\nonumber \\ 
m_{22}^* & = & m_{11}^* 
\end{eqnarray} 
 
and the trace of the monodromy matrix is therefore given by 
 
\begin{eqnarray} 
{\rm tr}\left(M^*\right) & = & 2 m_{11}^* 
\end{eqnarray} 
 
For the analysis of the stability of the single-bounce periodic  
orbits it is convenient to represent the expression for  
${\rm tr}\left(M^*\right)$ as a sum of $-2$ (which is the critical  
value of the trace of the monodrony matrix, when a periodic orbit bifurcates 
and loses stability), and some additional, depending on the tilt angle 
$\theta$ and other parameters, term. This can be achieved by a  
trivial rearrangement of terms giving  
 
\begin{eqnarray} 
{\rm tr}\left(M^*\right) & = & -2 +  
4\cos^4(\theta)   
\left(  \tan^2(\theta) + 
(\omega_c   T^*/2)  
\cot(\omega_c T^*/2)  
\right) \nonumber    
\\ &  \times & 
\left(\tan^2(\theta)   
+ \sin(\omega_c  T^*)/(\omega_c T^*) \right) 
\label{tr1A} 
\end{eqnarray} 
 
which is exactly the equation (\ref{trm1sbm}). 
 
\section{Period-doubling bifurcations of single-bounce orbits
and the scaling of the Poincar\'e map}

In this appendix we consider the evolution of the single-bounce 
orbits $(1)^{+(k)}$, which appear in tangent bifurcations together
with the unstable orbits $(1)^{-(k)}$. As follows from the
expression (\ref{trm1sbm}) for the trace of the monodromy
matrix and Eq. (\ref{t1sbm}), immediately after the tangent 
bifurcation all $(1)^{+(k)}$ orbits are stable 
( $-2 < {\rm tr}(M) \leq 2$ - see Fig. \ref{fig_trm_po1_sbm}). 

At $\beta = \beta_{\rm b1}^{(k)}$, where 
\begin{eqnarray}
\beta_{\rm b1}^{(k)} & = & {\cal F}\left(\sin\theta, 
\iota_k\left(-\tan^2\theta\right) \right)  
\label{bt_pdb1_sbm} 
\end{eqnarray}
the function ${\cal F}$ is defined in (\ref{define_f}) and $\iota_k(a)$
is the $k$-th positive root of the equation
\begin{eqnarray}
\frac{\iota}{\tan\iota} & = & a ,
\label{eq_tan}
\end{eqnarray}   
the trace of the corresponding monodromy matrix reaches the value 
$-2$, and the orbit $(1)^{+(k)}$ goes unstable  
via a period-doubling bifurcation. At thet moment a new stable  
two-bounce periodic orbit with the period exactly 
twice the period of $(1)^{+(k)}$ is born in the neighborhood. 
 
However, although {\it all} one-bounce periodic orbits $(1)^{+(k)}$ ($ k 
= 0, \ldots , \infty$) show the period-doubling bifurcation at  
$\beta = \beta_k^{\rm b1}$, the  further evolution of the $(1)^{+(k)}$ 
periodic orbits depends on $\theta$ and $k$ and is qualitatively  
different for $\theta < \theta_k^\dag$ and $\theta \ge \theta_k^\dag$, 
where 
\begin{eqnarray}  
\theta_k^\dag = \arctan\left(\sqrt{- \sin\left(\xi_k\right)/\xi_k}\right) 
\end{eqnarray} 
and $\xi_k$ is the $(k+1)$-th positive root of the equation $\tan(\xi) = 
\xi$. 
 
Note, that since critical angle $\theta_k^\dag$ is a monotonically  
decreasing function of $k$ for a fixed value of the tilt angle 
$\theta$ the inequality $\theta < \theta_k^\dag$ is equivalent to the  
condition $k <k_{\rm min}(\theta)$, where the integer  
$k_{\rm min}(\theta)$ is the  
smallest integer value of $k$, for which the inequality 
$\theta_k^\dag < \theta$ 
still holds.  $k_{\rm min}(\theta)$ is a 
decreasing function of $\theta$, it diverges as ${\rm 
integer}(1/\theta)$ at $\theta \rightarrow 0$, and $k_{\rm 
min}(\theta) = 0$ for $\theta > \theta_0$. The regime $\theta < 
\theta_k^\dag$ corresponds to $k \leq k_{\rm min}(\theta)$, and the regime 
$\theta > \theta_k^\dag$ is achieved for $k \geq k_{\rm min}(\theta)$ (so 
that for arbitrary $\theta$ above at sufficiently high $\beta$ the 
system is in the regime $\theta > \theta_k^\dag$).  
 
First, we consider the case $k < k_{\rm min}$ (which is non-generic in 
a sense that it corresponds to a {\it finite} part of an {\it 
infinite} sequence $ k = 0, \ldots, \infty$). At  $\beta = 
\beta_{\rm b2}^{(k)}$, where 
\begin{eqnarray}
\beta_{\rm b2}^{(k)} & = & 
{\cal F}\left( \sin\theta, \wp\left(-\tan^2\theta\right) \right)
\label{pdb2sbm}      
\end{eqnarray} 
and $\wp_n(a)$ is the $n$-th positive root of the equation
\begin{eqnarray}
\frac{\sin\wp}{\wp} & = & a ,
\label{eq_sin}
\end{eqnarray}
the trace of the monodromy matrix of the one-bounce periodic orbit  
$(1)^{(k)}$ again passes through the value $-2$ (see Fig. 
\ref{fig_trm_po1_sbm}). At 
this point, the orbit $(1)^{+(k)}$ restabilizes via a 
period-doubling bifurcation. In this bifurcation, the period-$1$ orbit  
$(1)^{(k)}$ can either ``emit'' an unstable two-bounce orbit  
or absorb a stable two-bounce orbit. A detailed description of this
behavior will be given in section IIIB, where we analyze the properties of 
the two-bounce orbits. 
 
As follows from the equations (\ref{bt_tb_sbm}) and
(\ref{bt_pdb1_sbm}), for a fixed tilt  
angle $\theta$ the intervals of  
stability of the single bounce orbits $(1,0)_{\rm T}^{+k}$ at large $k$ 
scale as $1/k$. If we inroduce an effective ``local'' parameter  
$\beta_{\ell}$ such that  
\begin{eqnarray} 
\beta_{\ell} & = & k \left( \beta - \pi \left( 2 k + 1 \right) \right)  
\end{eqnarray} 
then in the limit $k \gg 1$ the values of this local parameter
corresponding to the bifurcations of the single-bounce orbits
do not depend on $k$. This property 
gives us a hint about the existence of a universal  
limiting behavior of the Poincar\'e map  in the regime $k \gg 1$. 
Also, using Eqs. (\ref{bt_pdb1_sbm}), (\ref{pdb2sbm}) together with
Eq. (\ref{t1sbm}), one can show that      
for $k \gg 1$ the ``nontrivial'' part of the evolution of  
the single-bounce orbit  $(1)^{(k)+}$ takes place in the 
vicinity of the origin of the surface of section,
so that the ``universality'' of the behavior of the Poincar\'e map is 
expected to show up for $\tilde{v} \ll 1$. 
 
Introducing the  rescaled velocity 
\begin{eqnarray} 
{\bf v}_{\ell} & = & \left( \frac{\tilde{v}}{k},  
\frac{\tilde{v}}{k^2} \right) 
\end{eqnarray} 
and substituting the expressions of $\beta$ and $\tilde{\bf v}$ in 
terms of the local variables $\beta_{\ell}$ and ${\bf v}_{\ell}$ 
into the exact Poincar\'e map (\ref{PmapVSBM}), in the leading order 
in $1/k$ we obtain the following mapping : 
\begin{eqnarray} 
\left({\bf v}_{\ell}\right)_{n+1} & = &  
{ \bf \Phi_\ell} \left(  
\left({\bf v}_{\ell}\right)_n ; \beta_\ell  
\right) 
\label{Pmap_bigk} 
\end{eqnarray} 
where 
\begin{eqnarray} 
\left(\Phi_\ell\right)_x & = &  
 a_{00}  
 + a_{10} \left( v_\ell \right)_x  
 + a_{10} \left( v_\ell \right)_x  
 + a_{20} \left( v_\ell \right)_x^2  
 + a_{01} \left( v_\ell \right)_y  
 + O\left(\frac{1}{k}\right) \nonumber \\ 
\left( \Phi_\ell \right)_y & = &  
 b_{00}  
 + b_{10} \left( v_\ell \right)_x  
 + b_{10} \left( v_\ell \right)_x  
 + b_{20} \left( v_\ell \right)_x^2  
 + b_{30} \left( v_\ell \right)_x^3 
 + b_{40} \left( v_\ell \right)_x^4  
\nonumber \\ 
& + & b_{01} \left( v_\ell \right)_y  
 + b_{02} \left( v_\ell \right)_y^2  
 + b_{11} \left( v_\ell \right)_x  \left( v_\ell \right)_y  
 + b_{21} \left( v_\ell \right)_x^2  \left( v_\ell \right)_y 
 + O\left(\frac{1}{k}\right) 
\end{eqnarray} 
 
and  
 
\begin{eqnarray} 
a_{00} & = & - \cos^2\theta \sin\theta  
\left( \beta_\ell + \frac{2}{\pi} \right) \nonumber \\ 
a_{10} & = & - \cos^22\theta \nonumber \\ 
a_{20} & = & \pi \cos^2\theta \sin\theta \nonumber \\ 
a_{01} & = & - 2 \pi \sin^2\theta \cos\theta 
\nonumber \\ 
b_{00} & = & \sin2\theta \frac{1 - \cos^4\theta}{2 \pi} 
-  \frac{\sin2\theta \cos^2\theta}{\pi^2} 
- \beta_\ell^2 \frac{2 \sin2\theta \cos^2\theta}{4} 
\nonumber \\ 
b_{10} & = & \frac{2}{\pi} \cos\theta \sin^2\theta  
\left( 3 - 2 \sin^2 \theta \right) - 
\beta_\ell \cos2\theta \cos^3\theta 
\nonumber \\ 
b_{20} & = & \sin2\theta  
\left( \cos^4\theta - \sin^2\theta + \frac{1}{2} \right) 
\nonumber \\ 
b_{30} & = & \pi \cos^3\theta \cos2\theta  
\nonumber \\ 
b_{40} & = & \frac{\pi^2 \cos^4\theta \sin2\theta}{4} 
\nonumber \\ 
b_{01} & = & - \cos2\theta - \frac{ \pi \beta_\ell}{2} 
\sin^22\theta \cos^2\theta 
\nonumber \\ 
b_{02} & = & - \frac{\pi^2}{4} \sin^32\theta 
\nonumber \\ 
b_{11} & = & - \frac{\pi}{2} \sin4\theta \cos\theta 
\nonumber \\ 
b_{21} & = & \frac{\pi^2}{2} \sin^22\theta \cos^2\theta    
\nonumber  
\end{eqnarray} 
 
\begin{figure}[htbp]
\begin{center}
\leavevmode
\epsfysize=4.5in 
\epsfbox{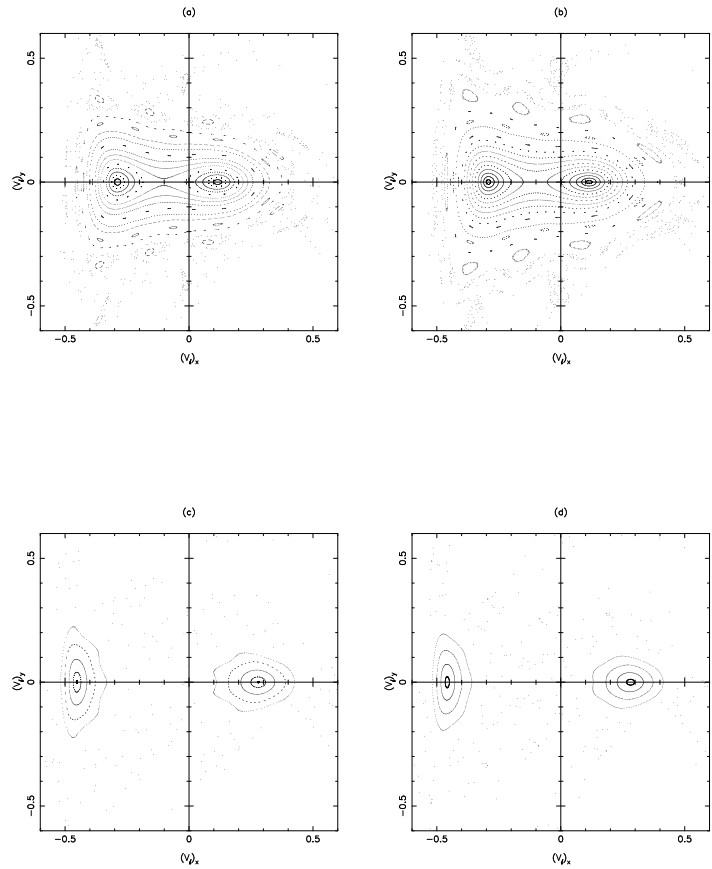}
\end{center}
\caption{
Comparison of the SOS for the limiting mapping (\protect\ref{Pmap_bigk})
(b,d) with the ones of the exact Poincar\'e map (a,c). The tilt
angle $\theta = 15^\circ$, 
$\beta_\protect{\protect\rm local\protect} = 0.2$  
(a,b)  and  $0.5$ (c,d). The SOS of the exact map is 
obtained for $k = 20$.
\label{fig_ps_compare}
} 
\end{figure}

In Fig. \ref{fig_ps_compare} we compare the Poincar\'e 
surfaces of section of the  
mapping (\ref{Pmap_bigk}) with  Poincare Surfaces of section of 
the exact map (\ref{PmapVSBM}) for different values of index $k$.  
An excellent agreement is found even for relatively small values of 
the index. 
 
\section{The monodromy matrix for a many - bounce orbit in SBM} 
 
To obtain the monodromy matrix for the period-one orbits, we  
essentially used the non-mixing property of the single-boounce 
periodic orbits. Therefore, it may seem, that an analycal expression  
for the trace of the monodromy matix may be obtained only for the 
simplest non-mixing orbits. However, it is not the case in the tilted well.  
The non-mixing property of a periodic orbit substantially simplify the  
calculation of the corresponding monodromy matrix, but it is not necessary 
to get an analyticall deccription of the stability, as it will be shown in 
the present Appendix. 
 
Consider a general (mixing) periodic orbit with $n$ collisions with the  
barrier per period. Let 
$\tilde{\bf v}_k \equiv ( (\tilde{v}_x)_k, (\tilde{v}_y)_k,
(\tilde{v}_z)_k)$ and $t_k$ be the  
scaled velocity immediately {\it after} the $k$-th collision and the time  
interval from $k$-th to $(k+1)$-th collision respectively. Once the values  
of $\tilde{\bf v}_k$ and $t_k$ are known, one can linearize the Poincar\'e  
map near the point $((\tilde{v}_x)_k, (\tilde{v}_y)_k)$ : 
 
\begin{eqnarray} 
\left(\delta\tilde{v}_x\right)_{k+1} & = &  
\left(M_k\right)_{11} \left(\delta\tilde{v}_x\right)_{k}
+
\left(M_k\right)_{12} \left(\delta\tilde{v}_y\right)_{k}
\nonumber \\ 
\left(\delta\tilde{v}_y\right)_{k+1} & = &  
\left(M_k\right)_{21} \left(\delta\tilde{v}_x\right)_{k}
+
\left(M_k\right)_{22} \left(\delta\tilde{v}_y\right)_{k}
\label{pmap_mk_sbm}  
\end{eqnarray} 
where $\delta\tilde{\bf v}_k$ and $\delta\tilde{\bf v}_{k+1}$ are the 
deviations of the velocity from  
$\tilde{\bf v}_k$ and from $\tilde{\bf v}_{k+1}$ respectively,  
and the matrix  $M_k$ is defined as follows 
\begin{eqnarray} 
M_k & = & 
\left(
\begin{array}{cc}
\left. 
\frac{\partial \Phi_x\left(\tilde{v}_x, \tilde{v}_y \right) }
{\partial \tilde{v}_x}
\right|_{ \tilde{\bf v} = \tilde{\bf v}_k }
&
\left. 
\frac{\partial \Phi_x\left(\tilde{v}_x, \tilde{v}_y  \right)}
{\partial \tilde{v}_y}
\right|_{ \tilde{\bf v} = \tilde{\bf v}_k }
\\
\left. 
\frac{\partial \Phi_y\left(\tilde{v}_x, \tilde{v}_y \right)}
{\partial \tilde{v}_x}
\right|_{ \tilde{\bf v} = \tilde{\bf v}_k }
&
\left. 
\frac{\partial \Phi_y\left(\tilde{v}_x, \tilde{v}_y \right)}
{\partial \tilde{v}_y}
\right|_{ \tilde{\bf v} = \tilde{\bf v}_k }
\end{array}
\right)
\label{define_Mk}
\end{eqnarray} 
 
Using the definition of the functions $\Phi_x, \Phi_y$  
(\ref{PmapVSBM}), we obtain the following expressions for the 
components of the matrix $M_k$ : 
\begin{eqnarray} 
\left(M_k\right)_{11} & = & \cos\left(\omega_c t_k\right)
- \frac{\left(\tilde{v}_x\right)_k \sin\theta \sin\left(\omega_c t_k\right)}
{\left(\tilde{v}_z\right)_k}
+ \kappa_{1t} \kappa_{t1} \nonumber \\
\left(M_k\right)_{12} & = & - \cos\theta \sin\left(\omega_c t_k\right)
- \frac{\left(\tilde{v}_y\right)_k \sin\theta \sin\left(\omega_c t_k\right)}
{\left(\tilde{v}_z\right)_k}
+ \kappa_{1t} \kappa_{t2} \nonumber \\
\left(M_k\right)_{21} & = & \cos\theta \sin\left(\omega_c t_k\right)
- \frac{\left(\tilde{v}_y\right)_k
\sin\theta \cos\theta
\left( 1 - \cos\left(\omega_c t_k\right) \right)} 
{\left(\tilde{v}_z\right)_k}
+ \kappa_{2t} \kappa_{t1} 
\nonumber \\
\left(M_k\right)_{22} & = & 
\cos^2\theta \cos\left(\omega_c t_k\right) + \sin^2\theta 
- \frac{\left(\tilde{v}_y\right)_k \sin\theta \cos\theta
\left( 1 - \cos\left(\omega_c t_k\right) \right)} 
{\left(\tilde{v}_z\right)_k}
+ \kappa_{2t} \kappa_{t2} 
\label{mk_analyt_sbm} 
\end{eqnarray} 
where
\begin{eqnarray}
\kappa_{1t} & = & 
\left(\tilde{v}_z\right)_k
\sin\theta  \cos\left(\omega_c t_k\right)
- \frac{2 \sin\theta \sin\left(\omega_c t_k\right)}{\beta}
\nonumber \\
& - & \left(\tilde{v}_x\right)_k \sin\left(\omega_c t_k\right)
- \left(\tilde{v}_y\right)_k \cos\theta \cos\left(\omega_c t_k\right)
\nonumber \\
\kappa_{2t} & = & 
\left(\tilde{v}_z\right)_k
\sin\theta \cos\theta 
 \sin\left(\omega_c t_k\right)
- \frac{2 \sin\theta \cos\theta
\left(1 - \cos \left(\omega_c t_k\right)\right) }{\beta}
\nonumber \\
& + & \left(\tilde{v}_x\right)_k \cos\theta 
\cos\left(\omega_c t_k\right)
- \left(\tilde{v}_y\right)_k \cos^2\theta 
\sin\left(\omega_c t_k\right)
\nonumber \\
\kappa_{t1} & = & - \left(
\sin\theta \left(1 - \cos \left(\omega_c t_k\right)\right)
+  
\frac
{
\left(
\tilde{v}_x\right)_k 
\left( \cos^2\theta + 
\sin^2\theta  \sin\left(\omega_c t_k\right) \right)
}
{
\left(\tilde{v}_z\right)_k
} \right) \varsigma_k^{-1}
\nonumber \\
\kappa_{t2} & = & \left(
\sin\theta \cos\theta \left(t_k - \sin \left(\omega_c t_k\right)\right)
-  
\frac
{
\left(
\tilde{v}_y\right)_k 
\left( \cos^2\theta + 
\sin^2\theta  \sin\left(\omega_c t_k\right) \right)
}
{
\left(\tilde{v}_z\right)_k
} \right) \varsigma_k^{-1}
\nonumber \\     
\varsigma_k & = & \sin\theta \sin\left(\omega_c t_k\right)
\left( \left(\tilde{v}_x\right)_k + \frac{2 \sin\theta}{\beta}
\right) + 
\frac{\omega_c t_k \left(\cos^2\theta - \sin^2\theta 
\cos\left(\omega_c t_k\right)\right) }{\beta}
\nonumber
\end{eqnarray}
 
The matrix $M_k$ relates the 
deviations of the velocity from the periodic orbit after two 
successive iterations of the Poincar\'e map, and is therefore 
directly connected to the monodormy matrix. The monodromy matrix 
of a period-$n$ orbit relates the velocity deviation after the 
first collision to the velocity deviation after the $n$-th collision. 
Therefore, the monodromy matrix can be obtained as : 
\begin{eqnarray} 
M & = & \Pi_{k = 1}^n M_k \label{m_mk_sbm} 
\end{eqnarray} 
The equations (\ref{m_mk_sbm}) and (\ref{mk_analyt_sbm}) give the 
analytical 
expressions for the components of the monodormy matrix in terms  
of the properties of the periodic orbit and are the final results of  
this Appendix. 

\section{Periods of non-mixing two-bounce orbits}

As in the case of single bounce orbits, the derivation of the periods 
of the two-bounce periodic orbits is most easily performed in the 
``drifting'' coordinate system $(x'',y'',z'')$, which was defined in 
(\ref{drift_frame_define}). In this coordinate system, the electron 
moves under the action of electric and magnetic fields, which are  
{\it both} parallel to the ${z''}$ axis : ${\bf E} = E \cos\theta  
\hat{z}''$, ${\bf B} = B \hat{z}''$. An immediate consequence of this  
fact is, that in this coordinate system the kinetic energy of the  
electron at the point of collision depends on the corresponding  
value of $z''$ : 
\begin{eqnarray} 
\left. \left. \frac{m^* v^2}{2}\right|_{{z''}_1} -  
\frac{m^* v^2}{2}\right|_{{z''}_2}  
& = & - e E \cos\theta \left({z''}_1 - {z''}_2 \right)  
\label{kinetic_energy} 
\end{eqnarray} 
 
Projected onto the plane $(x'',y'')$, a 
two-bounce periodic orbit forms a repeating pattern of two  
arcs of two different circles, as shown in  
fig. \ref{fig_po2_pic_driftXY_sbm}. Each ``kink'' in the projection of 
the trajectory corresponds to a collision with the barrier, when the 
direction of the electron velocity abruptly changes. The radius of 
each circle is related to the value of the cylotron velocity :  
$R_c = v_c/\omega_c$. If the periodic orbit 
is non-mixing, then there is no energy exchange between cyclotron and 
longitudinal motion. In this case the cyclotron velocity remains  
unchanged and the circles have equal radii - see  
fig. \ref{fig_po2_pic_driftXY_sbm}(b).  
 
Another consequence of the non-mixing property is that all  
the successive collisions of 
the electron with the barrier are separated by equal time intervals,  
so that the trajectory of the electron is symmetric under mirror 
reflection around any vertical (i.e. parallel to the $y''$ ) axis, 
passing through any of the collision points. If it were not true,  
then the collisions would necessarely have to change the {\it 
absolute value} of the $y''$ components of the velocity. 
Since the $x''$ component of the velocity of the electron remains  
intact at collisions,  
this will introduce a nonzero energy exchange between cyclotron and  
longitudinal motion, which contradicts the non-mixing property of the 
periodic orbit. 
 
At the point of a  ``non-mixing'' collision the electron has zero 
$y$ component of the velocity. In the drift(ing) coordinate system 
this condition is equivalent to the following relation : 
\begin{eqnarray} 
v_{y''} & = & - v_{z''} \tan\theta \label{vzyprimeprime_po2_sbm} 
\end{eqnarray} 
If this is the case, then the collision only reverses {\it sign} of  
the velocity in the $(y'',z'')$ plane, leaving the $x''$ component  
unchanged: 
\begin{eqnarray} 
v_{x''}^+ & = & v_{x''}^- \nonumber \\ 
v_{y''}^+ & = & - v_{y''}^- \nonumber \\ 
v_{z''}^+ & = & - v_{z''}^- \label{col_nonmix_drift}  
\end{eqnarray} 
where ${\bf v}^-$ and ${\bf v}^+$ are the velocities of the electron  
immediately before and immediately after the collision respectively. 
 
Let ${\bf v}_1$ and ${\bf v}_2$ be the velocities of the electron, 
corresponding to two successful (non-mixing) collisions with the 
barrier (fig. \ref{fig_po2_pic_driftXY_sbm}b). 
As follows from (\ref{col_nonmix_drift}) and (\ref{vzprime}),  
\begin{eqnarray} 
{v^+_{z''}}_2 = -\left( {v^+_{z''}}_1 - 
\frac{e E \cos\theta T }{2 m^*} \right) 
\label{vzvz_1_po2_sbm} 
\end{eqnarray} 
where $T$ is the period of the orbit, equal to twice the time interval 
between successful collisions. Due to the conservation of the cyclotron 
energy the   equation (\ref{kinetic_energy}) reduces to : 
\begin{eqnarray} 
{ v^+_{{z''}_2} }^2 -  {v^+_{{z''}_1}}^2 & = &  
- \frac{2 e E \cos\theta}{m^*} \left( {z''}_2 - {z''}_1\right) 
\label{vzvz_2_po2_sbm} 
\end{eqnarray} 
Using (\ref{vzvz_1_po2_sbm}), we can rewrite the equation  
(\ref{vzvz_2_po2_sbm}) as 
\begin{eqnarray} 
 v^+_{{z''}_2} -  v^+_{{z''}_1} & = & \frac{4}{T}  
\left({z''}_2 - {z''}_1 \right) 
\label{vz_z_prime_po2_sbm} 
\end{eqnarray} 
 
If $\alpha$ is the phase of the cyclotron rotation immediately after 
the first collision (at the point $({x''}_1, {y''}_1, {z''}_1)$ - see 
fig. \ref{fig_po2_pic_driftXY_sbm}b), then 
\begin{eqnarray} 
{v^+_{x''}}_1 & = &  v_c \cos\alpha \nonumber \\ 
{v^+_{y''}}_2 & = &  v_c \cos\left(\alpha + \omega_c T /2\right)
\label{vxprimeprime_po2_sbm} 
\end{eqnarray} 
and 
\begin{eqnarray} 
{v^+_{y''}}_1 & = & v_c \sin\alpha \nonumber \\ 
{v^+_{y''}}_2 & = & -v_c \sin\left(\alpha + \omega_c T /2\right) 
\label{vyprimeprime_po2_sbm} 
\end{eqnarray} 
Substituting (\ref{vyprimeprime_po2_sbm}) into  
(\ref{vzyprimeprime_po2_sbm}), 
we obtain 
\begin{eqnarray} 
{v^+_{z''}}_1 & = & - v_c \sin\alpha \tan\theta \nonumber \\ 
{v^+_{z''}}_2 & = & v_c \sin\left(\alpha + \omega_c T /2\right) 
\tan\theta 
\label{vzvzprimeprime_po2_sbm} 
\end{eqnarray} 
 
The distance ${z''}_2 - {z''}_1$ can be obtained as : 
\begin{eqnarray} 
{z''}_2 - {z''}_1 = \left( {y''}_2 - {y''}_1 \right) \tan\theta 
\label{dzpp_dypp_po2_sbm} 
\end{eqnarray} 
where 
\begin{eqnarray} 
{y''}_2 - {y''}_1 & = &  
\left( {x''}_2 - {x''}_1 \right) \tan\left(\pi - \alpha - 
\frac{\omega_c T}{4} \right)  
= 2 \frac{v_c}{\omega_c} \sin\frac{\omega_c T}{4}  
\sin\left(\alpha + \frac{\omega_c T}{4} \right) 
\label{dy_primeprime_po2_sbm} 
\end{eqnarray} 
 
Substituting (\ref{vzvzprimeprime_po2_sbm}) - 
(\ref{dy_primeprime_po2_sbm}) into (\ref{vz_z_prime_po2_sbm}), we 
finally obtain : 
\begin{eqnarray} 
\frac{\omega_c T}{ 4 } \cot\frac{\omega_c T}{4} & = & - \tan^2\theta 
\label{t_po2_sbm_appndx} 
\end{eqnarray} 
The $k$-th positive root of this equation gives the value of the 
period of the $(2)^{+(k)}$ orbit. 

\section{The monodromy matrix for a two-bounce non-mixing orbit}

The trace of the corresponding monodromy matrix for a (non-mixing)
two bounce orbit can be obtained 
using the general expressions developed in the Appendix C. For the 
period-2 orbits the monodromy matrix can be represented as  
\begin{eqnarray}
M & = & M_1 M_2 \label{mmatrix_po2_nm_sbm} 
\end{eqnarray} 
where the matrix $M_k$ ($k = 1, 2$) relates the velocity deviations 
from the periodic 
orbit at two successive collisions and can be calculated using the   
relations  (\ref{define_Mk}). As the input 
information for these mashinery one needs the values of the velocity 
of the electron immediately after each collision with the barrier  
($\tilde{\bf v}_1$ and $\tilde{\bf v}_2$) and the time intervals between 
successive  collisions ($t_1$ and $t_2$).  
 
For the period-2 non-mixing  
orbits, as we have shown in Appendix D, all the collisions are separated by 
equal time intervals, so that : 
\begin{eqnarray} 
t_1 = t_2 & = & \frac{T}{2} \label{t12_nm_po2_sbm} 
\end{eqnarray} 
 
To obtain the velocity at the point of collision, we can use the  
energy conservation condition :  
\begin{eqnarray} 
\varepsilon & = & \frac{m^* }{2} 
\left( \left( v_{x''} + v_d \right)^2 + v_{y''}^2    
+ v_{z''}^2 \right)
\label{energy_nm_po2_sbm} 
\end{eqnarray} 
 
Substituting the expressions for the velocity components at the point of 
collision (\ref{vxprimeprime_po2_sbm}), (\ref{vyprimeprime_po2_sbm}) and 
(\ref{vzvzprimeprime_po2_sbm}) into (\ref{energy_nm_po2_sbm}) and using  
(\ref{vzvz_1_po2_sbm}), we obtain : 
\begin{eqnarray}
\left( \frac{\beta \cos^2\theta}{2 \sin\theta} \right)^2 & = & 
\left(
1 + \sin^2 \theta\tan^2\left( \frac{\omega_cT}{4} \right)
\right)
\left( 1 + \sin^2\theta \tan^2\phi \right)
\label{eq_tanphi_po2_sbm}
\end{eqnarray} 
where we introduced a new angle $\phi$, which is defined as 
\begin{eqnarray} 
\phi & = & \pi - \alpha - \frac{\omega_c T}{4}  
\label{define_phi_po2_sbm} 
\end{eqnarray} 
It is more convenient to use $\phi$ rather than $\alpha$. In addition  
to a clear geometrical interpretation of the angle $\phi$ 
(see fig.  \ref{fig_po2_pic_driftXY_sbm}b ), when the non-mixing  
two-bounce orbit is born in a period-doubling bifurcation of the period - 1 
orbit and is indistinguishable from it's second repetition, the value of  
$\phi$ is {\it exactly} equal to zero, which makes $\phi$ a convenient
variable. 
 
Using the equation (\ref{eq_tanphi_po2_sbm}), we obtain : 
 
\begin{eqnarray} 
\tan\phi & = & \pm \frac{1}{\tan^2\theta}
\sqrt
{
\frac
{
\left(\frac{\beta}{2}\right)^2 - 
\left(\frac{\omega_c T}{4} \right)^2 - 
\frac{\tan^2\theta}{\cos^2\theta}
}
{
1 + \sin^2\theta \tan^2\left( \frac{\omega_c T}{4} \right)
} 
}
\label{tanphi_po2_sbm} 
\end{eqnarray} 
where the two different solutions correspond to the values of $\tan\phi$  
at the two nonequivalent points of collision. 
 
As follows from (\ref{tanphi_po2_sbm}), a particular period-two  
non-mixing orbit $(2)^{+k}$ exists only above the critical value  
of $\beta$ given by : 
\begin{eqnarray}
\beta_{c_2} & = & 
\sqrt{ \left(\frac{\omega_c T}{2}\right)^2 + 
\left(\frac{\tan\theta}{\cos\theta}\right)^2}
\end{eqnarray} 
which is {\it exactly} equal to the value of  $\beta = \beta_{b1}$,
 corresponding to the first period-doubling bifurcation of the 
single-bounce orbit  $(1)^{+k}$, as expected. 
 
For the velocity components at the points  
of collision in the non-tilted "stationary" system of coordinates  
$(x,y,z)$ we therefore obtain: 
\begin{eqnarray} 
\left(\tilde{v}_x\right)_{1,2} & = & 
- \frac{2 \sin\theta}{\beta}
\left(
\frac{1}{\cos^2\theta} \pm \tan\left({\omega_c T}{2}\right)
\sqrt
{
\frac
{
\left(\frac{\beta}{2}\right)^2 - 
\left(\frac{\omega_c T}{4} \right)^2 - 
\left(\frac{\tan\theta}{\cos\theta}\right)^2
}
{
1 + \sin^2\theta \tan^2\left( \frac{\omega_c T}{4} \right)
}
}
\right) 
\nonumber \\
\left(\tilde{v}_y\right)_{1,2} & = & 0
\label{vy_fp_po2_nm_sbm} 
\end{eqnarray} 
 
The relations (\ref{vy_fp_po2_nm_sbm}) and (\ref{t_po2_sbm}) 
together with (\ref{mmatrix_po2_nm_sbm}) and (\ref{define_Mk}) 
provide the complete information  we need for the stability analysis.
Substituting  (\ref{vy_fp_po2_nm_sbm}) and (\ref{t_po2_sbm}) into
(\ref{define_Mk}),  we obtain the matrices $M_1$ and $M_2$. 
Substituting these expressions into  (\ref{mmatrix_po2_nm_sbm}), one 
can obtain the monodromy matrix $M$.

\section{Periods of the type-$1$ mixing two-bounce orbits}

\begin{figure}[htbp]
\begin{center}
\leavevmode
\epsfxsize=5.in 
\epsfbox{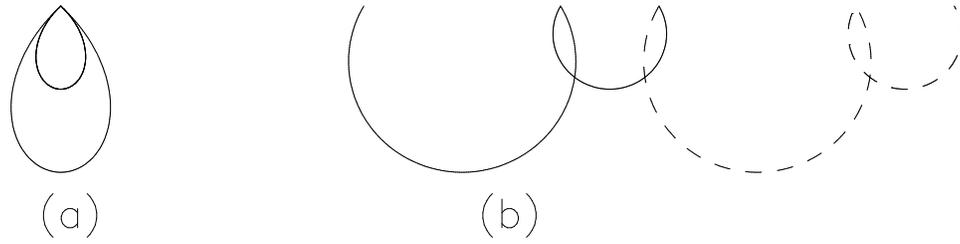}
\end{center}
\caption{
A mixing self-retracing two-bounce orbit, projected onto the 
$(x',y')$ plane of the laboratory system of coordinates (a)
and onto  $(x'',y'')$ plane of the ``drifting'' frame of reference.
\label{fig_po2m_pic_driftXY_sbm}
} 
\end{figure}

Projected onto the  plane $(x'',y'')$ of the drift(ing) frame of 
reference, a self-retracing mixing period-2 orbit forms a repeating pattern of
two portions of circles of {\it different} radii, with ``kinks'' at
the points of collision with {\it exactly} same values of $y''$ - see
fig. \ref{fig_po2m_pic_driftXY_sbm}(b).

Since the $x''$ component of the velocity is unchanged at collisions,
we obtain :
\begin{eqnarray}
v_{c_1} \cos\left(\frac{\omega_c t_1}{2}\right)
& = & 
v_{c_2} \cos\left(\frac{\omega_c t_2}{2}\right)
\label{vc1vc2_po2_m_sbm} 
\end{eqnarray}
where $v_{c}$ and $t$ are the cylotron velocity and the time interval 
between collisions respectively.

The periodicity of the orbit requires, that the distance traveled by
the electron in the drift frame of reference after two successive 
collisions 
\begin{eqnarray}
\delta {x''}_2 & = & 
\frac{2 v_{c_1}}{\omega_c} \sin\left(\frac{\omega_c t_1}{2}\right)
+
\frac{2 v_{c_2}}{\omega_c} \sin\left(\frac{\omega_c t_2}{2}\right)
\nonumber
\end{eqnarray}
is equal to the displacement of this coordinate system
\begin{eqnarray}
\delta {x''}_d = v_d \left( t_1 + t_2 \right)
\nonumber 
\end{eqnarray}
which yields
\begin{eqnarray}
{v_{c_1}} \sin\left(\frac{\omega_c t_1}{2}\right)
+
{v_{c_2}} \sin\left(\frac{\omega_c t_1}{2}\right)
& = & 
v_d \omega_c \left( t_1 + t_2 \right)
\label{vc1vc2vd_po2_m_sbm} 
\end{eqnarray}

Using (\ref{vc1vc2_po2_m_sbm}) together with
(\ref{vc1vc2vd_po2_m_sbm}), we obtain :
\begin{eqnarray}
{v_{c_1}} & = & v_d 
\frac{ \frac{\omega_c T}{2} }{ \sin\left(\frac{\omega_c T}{2}\right) }
\cos\left(\frac{\omega_c t_2}{2}\right)
\nonumber \\
{v_{c_2}} & = & v_d 
\frac{ \frac{\omega_c T}{2} }{ \sin\left(\frac{\omega_c T}{2}\right) }
\cos\left(\frac{\omega_c t_1}{2}\right)
\label{vc_po2_m_sbm}
\end{eqnarray}
where $ T \equiv t_1 + t_2$ is the period of the orbit. The
``in-plane''  components of the electron velocity 
$v_{x''}$, $v_{x'}$ and $v_{y'} \equiv v_{y''}$ are 
therefore given by :
\begin{eqnarray}
v_{x''} & = & v_d 
\frac{ \frac{\omega_c T}{2} }{ \sin\left(\frac{\omega_c T}{2}\right) }
\cos\left(\frac{\omega_c t_1}{2}\right)
\cos\left(\frac{\omega_c t_2}{2}\right)
\nonumber \\
v_{{x'}} & = & v_d \left(1 + 
\frac{ \frac{\omega_c T}{2} }{ \sin\left(\frac{\omega_c T}{2}\right) }
\cos\left(\frac{\omega_c t_1}{2}\right)
\cos\left(\frac{\omega_c t_2}{2}\right)
\right)
\nonumber \\
v_{{y''}_{1,2}} \equiv v_{{y'}_{1,2}}
& = & 
v_d 
\frac{ \frac{\omega_c T}{2} }{ \sin\left(\frac{\omega_c T}{2}\right) }
\cos\left(\frac{\omega_c t_{2,1}}{2}\right)
\sin\left(\frac{\omega_c t_{1,2}}{2}\right)
\label{v_inplane_po2_m_sbm}
\end{eqnarray}

Since the $y''$ coordinate is the same at each bounce, 
the longitudinal energy immediately after one collision
is equal to the longitudinal energy immediately before
the next collision, and the longitudinal velocities
$v^+_{{z'}_1} \equiv v^+_{{z''}_2}$ immediately after two successive
collisions  the time 
imtervals $t_1$ and $t_2$ between successive collisions must 
satisfy the relations
\begin{eqnarray}
{v_{z'}}_{1,2} & = & \frac{eE \cos\theta t_{1,2}}{ 2 m^*}
\label{vzprime_po2_m_sbm}
\end{eqnarray} 

Substituting (\ref{v_inplane_po2_m_sbm}) and (\ref{vzprime_po2_m_sbm})
into (\ref{eqns_reflect}) and using the conservation of the total energy 
\begin{eqnarray}
\varepsilon & = & \frac{m^*}{2} 
\left( v_{x'}^2 + v_{y'}^2 + v_{z'}^2 \right)
\nonumber
\end{eqnarray}
we obtain :
\begin{eqnarray}
\left\{
\begin{array}{ccl}
\frac
{\textstyle \sin\left(\frac{ \textstyle \omega_c T}{2}\right)}
{\frac{\textstyle \omega_c T}{ \textstyle 2}}
& = & - \tan^2\theta
\frac
{\textstyle \sin\left(\frac{\textstyle \omega_c \delta T}{2}\right)}
{\frac{\textstyle \omega_c \delta T}{ \textstyle 2}} 
\\
\left(\frac{\beta}{2}\right)^2
& = & 
\sin^2\theta \left( 1 - 
\frac{
\frac{\omega_c T}{2} \left( \cos\left(\frac{\omega_c T}{2}\right)
+  \cos\left(\frac{\omega_c \delta T}{2}\right) \right)
}
{
2 \sin\left(\frac{\omega_c T}{2}\right)
}
\right)^2 
+
\left( \frac{\omega_c T}{4}\right)^2
+ 
\cot^2\theta \left( \frac{\omega_c \delta T}{4}\right)^2
\end{array}
\right.
\label{t_po2_m_sbm_appndx}
\end{eqnarray}
where $\delta T \equiv | t_2 - t_1 |$. This system of two equations
defines the periods of all of the type-1 mixing period-2 orbits as 
functions of $\beta$ and the tilt angle.

\section{ Double Barrier Model : periods of $(1,1)$ orbits }

\begin{figure}[htbp]
\begin{center}
\leavevmode
\epsfxsize=5.in 
\epsfbox{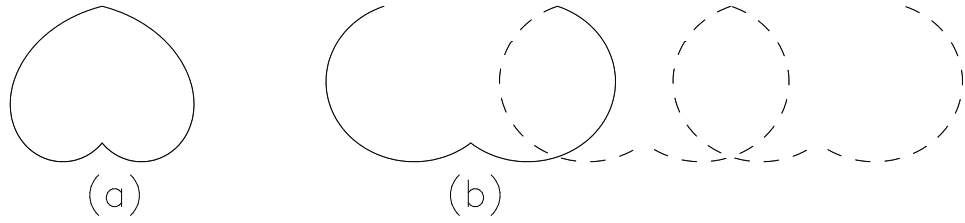}
\end{center}
\caption{
A non-mixing two-bounce orbit, projected onto the 
$(x',y')$ plane of the laboratory system of coordinates (a)
and onto  $(x'',y'')$ plane of the ``drifting'' frame of reference.
\label{fig_po2_pic_driftXY_sbm}
} 
\end{figure}

In this appendix we derive the equation (\ref{t_po1_dbm}).
We perform the derivation ``drifting'' coordinate system 
$(x'',y'',z'')$, which was defined in 
(\ref{drift_frame_define}). In this coordinate system, the electron 
moves under the action of electric and magnetic fields, which are  
{\it both} parallel to the ${z''}$ axis : ${\bf E} = E \cos\theta  
\hat{z}''$, ${\bf B} = B \hat{z}''$. 
Since the $(1,1)$ orbit is non-mixing, the 
cyclotron velocity $v_c$ is conserved and the cyclotron radius $R_c 
\equiv v_c/\omega_c$ is the same for each part of the trajectory.
Therefore, the $(x'',y'')$ projection of the $(1,1)$ orbit
produces a pattern  of two  
arcs of two different circles with {\it equal} radii
and it looks exactly
like the $(x'',y'')$ projection of a two-bounce non-mixing orbit 
$(2)^-$ in the single-barrier model (see fig. 
\ref{fig_po2_pic_driftXY_sbm}). Hovewer, the ``kink'' at
$(x_2^{''},y_2^{''})$ is due to collision at the {\it emitter}
barrier (Fig.  \ref{fig_po2_pic_driftXY_sbm} b), so that the periods of 
the $(1,1)$ orbits are different from the ones of $(2)^-$.

In the drifting coordinate system the kinetic energy of the  
electron at the point of collision depends on the corresponding  
value of $z''$, so that (cf. (\ref{kinetic_energy}) ): 
\begin{eqnarray} 
\frac{m^* v^2_2}{2} -  \frac{m^* v_1^2}{2}  
& = & - e E \cos\theta \left( \frac{d}{\cos\theta} +
\left({y''}_2 - {y''}_1\right) \right)  
\label{kinetic_energy_dbm} 
\end{eqnarray} 
 
As for the non-mixing two-bounce orbits $(2)^-$ in the single
berrier model, the successive collisions of the $(1,1)$ with 
different barriers are are separated by equal time intervals,  
so that the trajectory of the electron is symmetric under mirror 
reflection around any vertical (i.e. parallel to the $y''$ ) axis, 
passing through any of the collision points.

At the point of a  ``non-mixing'' collision with both the emitter
and the collector barriers the electron has zero 
$y$ component of the velocity, therefore at each collision of the 
$(1,1)$ orbits the corresponding $y''$ and $z''$
components of the electron velocity are realted to each other by 
(\ref{vzyprimeprime_po2_sbm}), while the velocity immediately before 
the collision ${\bf v}^-$ and the velocity immediately after the
collision  ${\bf v}^+$ satisfy the relations (\ref{col_nonmix_drift}).
 
Let ${\bf v}_1$ and ${\bf v}_2$ be the velocities of the electron, 
corresponding to two successful (non-mixing) collisions with the 
collector and emitter barrier respectively.
As follows from (\ref{vzprime}),  
\begin{eqnarray} 
{v^-_{z''}}_2 = \left( {v^+_{z''}}_1 - 
\frac{e E \cos\theta T }{2 m^*} \right) 
\label{vzvz_1_po2_dbm} 
\end{eqnarray} 
where $T$ is the period of the orbit, equal to twice the time interval 
between successful collisions. Substituting (\ref{vzvz_1_po2_dbm} )
into (\ref{kinetic_energy_dbm}) and using the conservation of the
cyclotron velocity, we obtain :
\begin{eqnarray} 
 v^+_{{z''}_1} +  v^-_{{z''}_2} & = & \frac{4}{T}  
\left(
\frac{d}{\cos\theta} + 
\left({y''}_2 - {y''}_1 \right) 
\tan\theta \right)
\label{vz_y_prime_po1_dbm} 
\end{eqnarray} 
 
If $\alpha$ is the phase of the cyclotron rotation immediately after 
the collision with the collector wall (at the point 
$({x''}_1, {y''}_1, {z''}_1)$ - see 
Fig. \ref{fig_po2_pic_driftXY_sbm}b), then 
\begin{eqnarray} 
{v^+_{x''}}_1 & = & v_c \cos\alpha \nonumber \\
{v^+_{y''}}_1 & = & v_c \sin\alpha \nonumber \\ 
{v^-_{x''}}_2 & = & v_c \cos\left(\alpha + \omega_c T /2\right)
\nonumber \\
{v^-_{y''}}_2 & = & v_c \sin\left(\alpha + \omega_c T /2\right) 
\label{vxyprimeprime_po1_dbm} 
\end{eqnarray} 
and (see (\ref{vzyprimeprime_po2_sbm}) )
we obtain 
\begin{eqnarray} 
{v^+_{z''}}_1 & = & - v_c \sin\alpha \cot\theta \nonumber \\ 
{v^-_{z''}}_2 & = & - v_c \sin\left(\alpha + \omega_c T /2\right) 
\cot\theta 
\label{vzvzprimeprime_po1_dbm} 
\end{eqnarray} 
 
The distance ${y''}_2 - {y''}_1$ can be obtained as (see  
Fig.  \ref{fig_po2_pic_driftXY_sbm}b and cf. (\ref{dy_primeprime_po2_sbm}) : 
\begin{eqnarray} 
{y''}_2 - {y''}_1 & = &  
 2 \frac{v_c}{\omega_c} \sin\frac{\omega_c T}{4}  
\sin\left(\alpha + \frac{\omega_c T}{4} \right) 
\label{dy_primeprime_po1_dbm} 
\end{eqnarray} 
 
Substituting (\ref{vzvzprimeprime_po1_dbm}) and 
(\ref{dy_primeprime_po1_dbm}) into (\ref{vz_y_prime_po1_dbm}), we 
obtain : 
\begin{eqnarray} 
v_c \sin\left(\alpha + \frac{\omega_c T}{4} \right) 
\cos\left( \frac{\omega_c T}{4} \right)
& = & 
- \frac{d \omega_c \tan\theta}{2 \cos\theta} 
\frac
{\cot\left( \frac{\omega_c T}{4}\right)}
{\tan^2\theta + \frac{\omega_c T}{4} \cot\left({\omega_c T}{4}\right)}
\label{vc_sin_po1_dbm}
\end{eqnarray}

The periodicity of the orbit requires, that the distance traveled by
the electron in the drift frame of reference between two successive 
collisions with the collector barrier ${x''}_2 - {x''}_1$
(see Fig. \ref{fig_po2_pic_driftXY_sbm}b )
is equal to the displacement of this coordinate system $v_d T$,
which yields 
\begin{eqnarray}
v_c \cos\left( \alpha + \frac{\omega_c T}{4}
\right) \sin\left( \frac{\omega_c T}{4}\right)
& = & v_d
\frac{\omega_c T}{4}
\label{vc_cos_po1_dbm}
\end{eqnarray}

Using Eqs. (\ref{vc_sin_po1_dbm}) and (\ref{vc_cos_po1_dbm}), one 
can easily obtain
\begin{eqnarray}
v^+_{y'}{}_1 & = & v_c \sin\alpha =
-  v_d \frac{\omega_c T}{4}
- \frac{d \omega_c \tan\theta}{2 \cos\theta}
\frac{\cot\left( \frac{\omega_c T}{4}\right)}{\tan^2\theta + 
 \frac{\omega_c T}{4} \cot\left( \frac{\omega_c T}{4}\right) }
\nonumber \\
v^+_{x'}{}_1 & = & - v_d + v_c \sin\alpha =
-  v_d \left( 1 - 
\frac{\omega_c T}{4}\cot\left( \frac{\omega_c T}{4}\right)
\right)
- \frac{d \omega_c \tan\theta}{2 \cos\theta}
\frac{ 1 }{\tan^2\theta + 
\frac{\omega_c T}{4} \cot\left( \frac{\omega_c T}{4}\right) }
\label{vprime_po1_dbm}
\end{eqnarray}
Substituting (\ref{vprime_po1_dbm}) into the equation for energy
conservation
\begin{eqnarray}
v_{x'}^2 + v_{y'}^2 + v_{z'}^2 = 1 \nonumber
\end{eqnarray}
and using (\ref{vzyprimeprime_po2_sbm}), we finally obtain :
\begin{eqnarray}
\beta^2 & = & \left(\frac{\omega_C T}{2}\right)^2 
\left( 1 + \frac{\beta^2}{\gamma \left(\omega_c T\right)^2}
\frac{1 - f\left(\omega_c T\right)}
{1 - \cos^2\theta f\left(\omega_c T\right)} \right)^2 
+ 4 \sin^2\theta f^2\left(\omega_c T\right)
\left( 1 + \frac{\beta^2}{16 \gamma} 
\frac{1}{f\left(\omega_c T \right) 
\left(1 - \cos^2\theta f\left( \omega_c T \right) \right)}
\right)^2
\end{eqnarray}
which is exactly the Eq. (\ref{t_po1_dbm}). To obtain the 
period of the period-$1$ orbits from thsi equation, one has solve it
together with the condition 
\begin{eqnarray}
\left(\tilde{v}_e\right)^-_z & = &
\frac{2 \cos\theta}{\beta} \frac{\omega_c T}{4}
+ \frac{ \beta \cos\theta}{2 \gamma \omega_c T}
\frac{1 - f\left({\omega_c T}\right)}
{1 - \cos^2\theta f\left( \omega_c T \right)}
> 0
\label{vz_gt_0_dbm}
\end{eqnarray}
whcih ensures  that $v_z$ just
before the collision with the emitter is positive and allows to select
the physically meaningful roots.

\section
{ 
The monodromy matrix for a general periodic orbit in the DBM
}

In this Appendix we consider the monodromy (stability) matrix for a
general orbit in the double-barrier model. As in our stability
analysis for the periodic orbits in the SBM, the 
velocity at each collision with the barriers and the time interval
between successive collisions for the periodic orbit are 
considered already known . 

By definition, the monodromy matrix is the  linearizitaion
the Poincar'e map around the periodic orbits. It is straightforward to
show, that since the evolution of the electron velocity {\it between}
successive collisions is exactly the same in both SBM and DBM, and
any collsion only reverses the sign of $z$-component of the velocity,
the monodromy matrix will still be given by Eqs. (\ref{m_mk_sbm}) and
(\ref{mk_analyt_sbm}), where the index $k$ now labels all successive
collisions of the electron (with {\it both} emitter and collector
barriers.

Note, that the components of the matrices $M_k$ contain terms
proportional to $1/\left(\tilde{v}_z)\right)$. Therefore, if at any of
the collisions with the emitter barrier the $z$ component of the 
velocity goes to zero (as it happens in a cusp bifurcation), the
components of the matrix $M_k$ diverge, which leads to the divergence
of the trace of the monodromy matrix. An additional consequence of
this behavoiur is that by continutiy any orbit with sufficiently small
$v_z$ at at least one of the collisions of the emitter barrier per
period {\it must} be unstable.

\newpage

\begin{table}[htbp]
\caption{Period-$1$ and period-$2$ orbits in the SBM}
\begin{center}
\leavevmode
\epsfysize=7.in 
\epsfbox{table1.epsi}
\end{center}
\end{table}
 
\begin{table}[htbp]
\begin{center}
\caption{Relevant peridic orbits in the DBM}
\leavevmode
\epsfysize=7.in 
\epsfbox{table2.epsi}
\end{center}
\end{table}

\begin{figure}[htbp]
\begin{center}
\leavevmode
\epsfysize=4.5in 
\epsfbox{fig10.epsi}
\end{center}
\caption{
Examples of the different types of period-$2$ orbits, projected onto
$(x,z)$ and $(y,z)$ planes : a non-mixing orbit (a), 
a self-retracing mixing orbit (b) 
and a non-self-retracing mixing orbit (c).
\label{fig_po2_pic_YZ_sbm}
}
\end{figure}

\begin{figure}[htbp]  
\begin{center}
\leavevmode
\epsfysize=3.5in 
\epsfbox{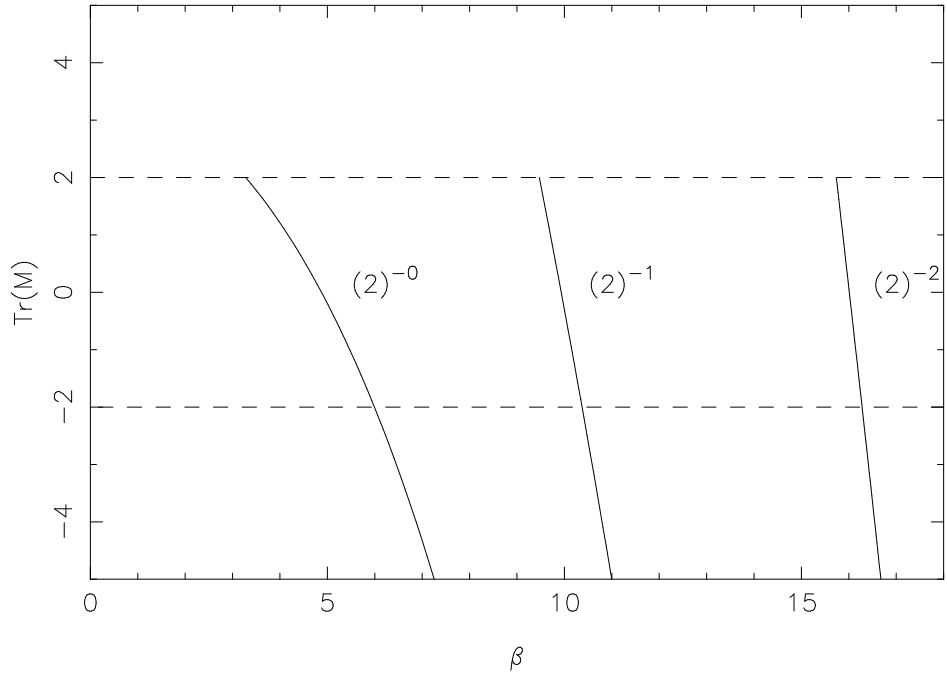}
\end{center}
\caption{
Trace of monodromy matrix as a function of $\beta$ for different
non-mixing two-bounce periodic orbits : $(2)^{+0}$, $(2)^{+1}$, 
$(2)^{+2}$ for $\theta = 15^\circ$.
\label{pic_trM_po2_nm_sbm}
}
\end{figure}

\begin{figure}[htbp]
\begin{center}
\leavevmode
\epsfysize=3.in 
\epsfbox{fig14.epsi}
\end{center}
\caption{
Trace of the monodromy matrix as a function of $\beta$ for
mixing two-bounce orbits (a) $(2)^{+(0)}$
and $(2)^{*(0)}$, $(2)^{+(1)}$ and $(2)^{*(1)}$.
The tilt angle is $\theta = 15^\circ$.
\label{pic_trM_po2_m_sbm}
}
\end{figure}

\begin{figure}[htbp]
\begin{center}
\leavevmode
\epsfysize=4.5in 
\epsfbox{fig15.epsi}
\end{center}
\caption{
Surface of Section near the one-bounce periodic orbit $(1)^{+0}$
close to it's $1:3$ resonance and the corresponding touch-and-go
bifurcation of the orbits $(3)_1^{(1)}$ : (a) just before and (b) soon after 
the touch-and-go bifurcation.
\label{fig_touch_and_go_po3_sbm}
}
\end{figure}

\begin{figure}[htbp]
\begin{center}
\leavevmode
\epsfysize=3.in 
\epsfbox{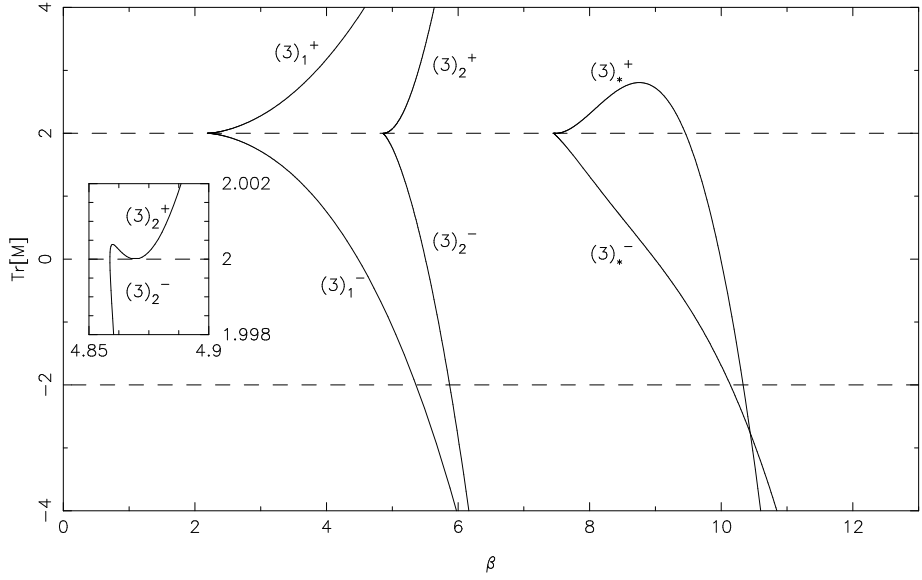}
\end{center}
\caption{
Trace of the monodromy matrix as a function of $\beta$ for 
self-retracing 
three-bounce orbits. The inset shows the behavior of ${\rm Tr}[M]$
near the ``touch-and-go'' bifurcation. 
\label{fig_trM_po3_sbm}
}
\end{figure}

\begin{figure}[htbp]
\begin{center}
\leavevmode
\epsfysize=2.5in 
\epsfbox{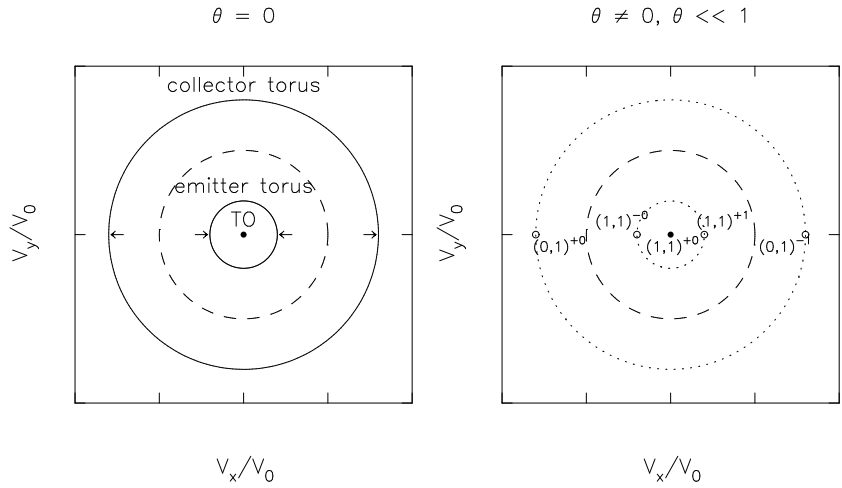}
\end{center}
\caption{
A schematic representation of (a) the two resonant tori of the 
period-$1$ orbits
at $\theta = 0^\circ$ and (b) the surviving orbits at 
$\theta \ll 1$.
\label{fig_po1_schematics_po1_dbm}
}
\end{figure}

\begin{figure}[htbp]
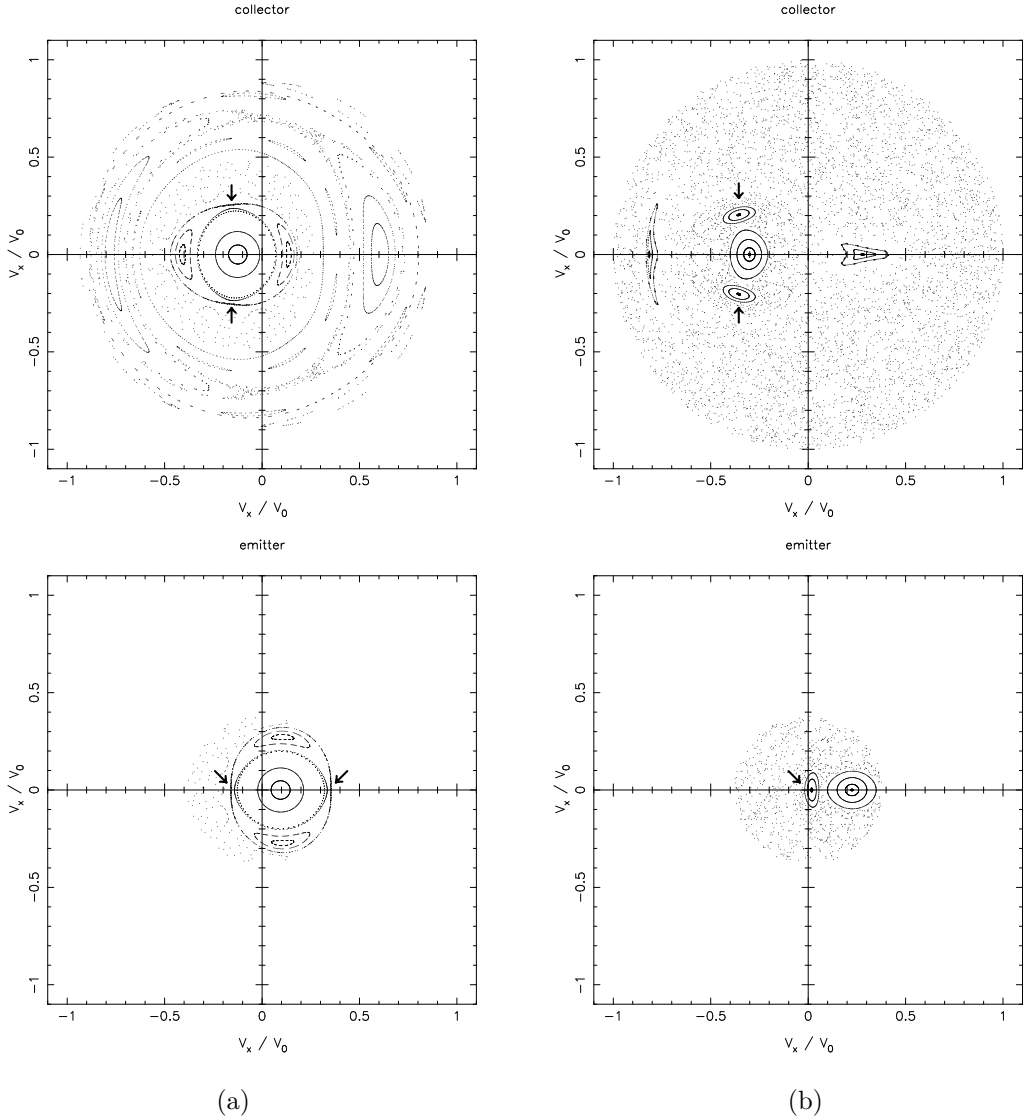

\begin{center}
\leavevmode
\epsfysize=5.5in 
\epsfbox{fig25a.epsi}
\hskip 1. truecm
\leavevmode
\epsfysize=5.5in 
\epsfbox{fig25b.epsi}
\end{center}
\begin{center}
\ \ \ \ \ \ \ \ \
(a)
\ \ \ \ \ \ \ \ \ \ \ \ \ \ \ 
\ \ \ \ \ \ \ \ \ \ \ \ \ \ \
\ \ \ \ \ \ \ \ \ \ \ \ \ \ \
\ \ \ \ \ \ \ \ \ \ \ \ \ \ \  
(b)
\ \ \ \ \ \ \ \ \  
\end{center}
\caption{
Surfaces of section, showing the fixed points 
of $(2,2)^-$, $(0,2)^-$,
$(2,2)^+$
and $(1,2)$  orbits for $\beta = 4.5$, $\gamma = 1.17$ 
and the tilt angle $\theta = $ (a) $11^\circ$,  (b) $= 28^\circ$.
The top and bottom panels correspond to the surfaces of section 
at the collector and the emitter barriers respectively.
{\protect\bf a :} one can clearly see one big stable island of the 
period-$1$ orbit $(1,1)^+$, and  stable islands of the $(2,2)^-$
and $(0,2)^-$ orbits. The stable islands of the $(0,2)^-$ orbit
lie at the $\tilde{v}_x \equiv v_x/v_0$ axis 
at the periphery of the collector surface of section, they 
are absent at the emitter SOS. This $(2,2)^-$ orbit produces
two islands centered on the $\tilde{v}_x $ axis 
at the collector barrier and
two islands at the emitter barrier. To show the $(0,2)^-$ and
$(2,2)^-$ orbits in a single bifurcation diagram it is therefore 
natural to represent these orbits  by their  values of the 
$x$ component of the scaled velocity at the collector barrier.
The fixed points of the generally
unstable orbit $(2,2)^+$ are not so easy to see by an 
(untrained) eye and
pointed out by the arrows. Both fixed 
points of $(2,2)^+$ have zero $\tilde{v}_y$ at the emitter barrier 
and nonzero $\tilde{v}_y$ at the collector barrier. Note, that at the
collector barrier the $(2,2)^+$ orbit has the same values of the $x$
component of the scaled velocity (since $\tilde{v}_x \sim y$ and
the $(2,2)^+$ orbit strikes the collector wall at the same point). 
Therefore,
this value is a convenient representation for the $(2,2)^+$ orbits 
in the
bifurcation diagrams. 
{\protect\bf b :} one can see a relatively large 
stable islang of the $(1,1)^+$
orbit, two islands of the $\Omega$ orbit (in collector barrier SOS
only) and stable islands of the $\nu$ orbit (two islands at the 
collector barrier surface of section and one island at the emitter 
barrier SOS). Just as for the $(2,2)^+$ and $(0,2)^+$ orbits, the 
fixed points of the 
$(1,2)$ orbits at the emitter wall have exactly the same values of 
$\tilde{v}_x$, which can therefore be used as their representation
in the bifurcation diagrams. 
\label{fig_psos_po2_dbm}. 
}
\end{figure}

\begin{figure}[htbp]
\begin{center}
\leavevmode
\epsfysize=3.2in 
\epsfbox{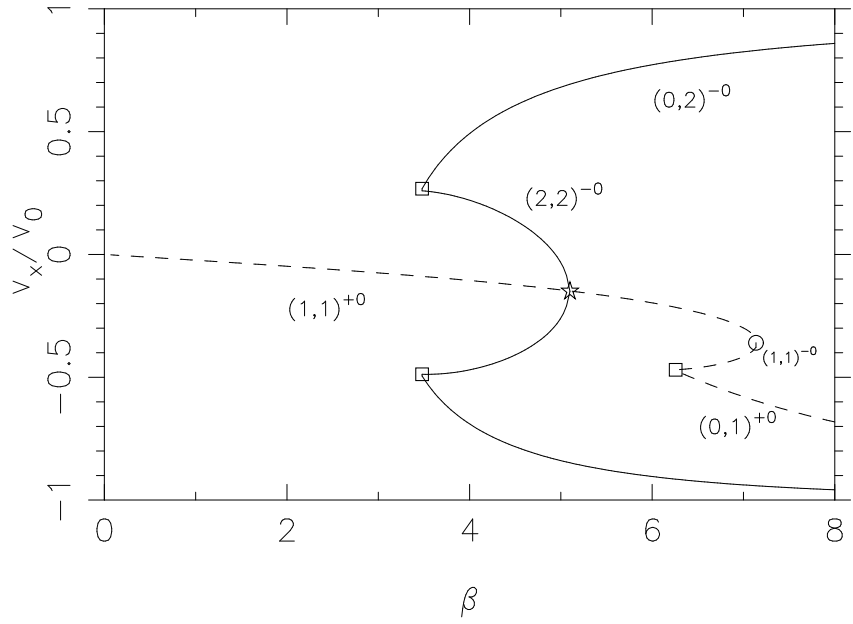}
\end{center}
\caption{
The bifurcation diagram of the $(2,2)^-$ and $(0,2)^-$ orbits in the 
DBM. The vertical axis represents $x$
component of the scaled velocity of the electron at the point of
collision with the collector barrier (see also Fig. 
\protect\ref{fig_psos_po2_dbm}a ). The tilt angle $\theta =
15^\circ$, and $\gamma = 1.17$.
\label{fig_bifdiag_po2nm_dbm}
}
\end{figure}

\begin{figure}[htbp]
\begin{center}
\leavevmode
\epsfysize=6.2in 
\epsfbox{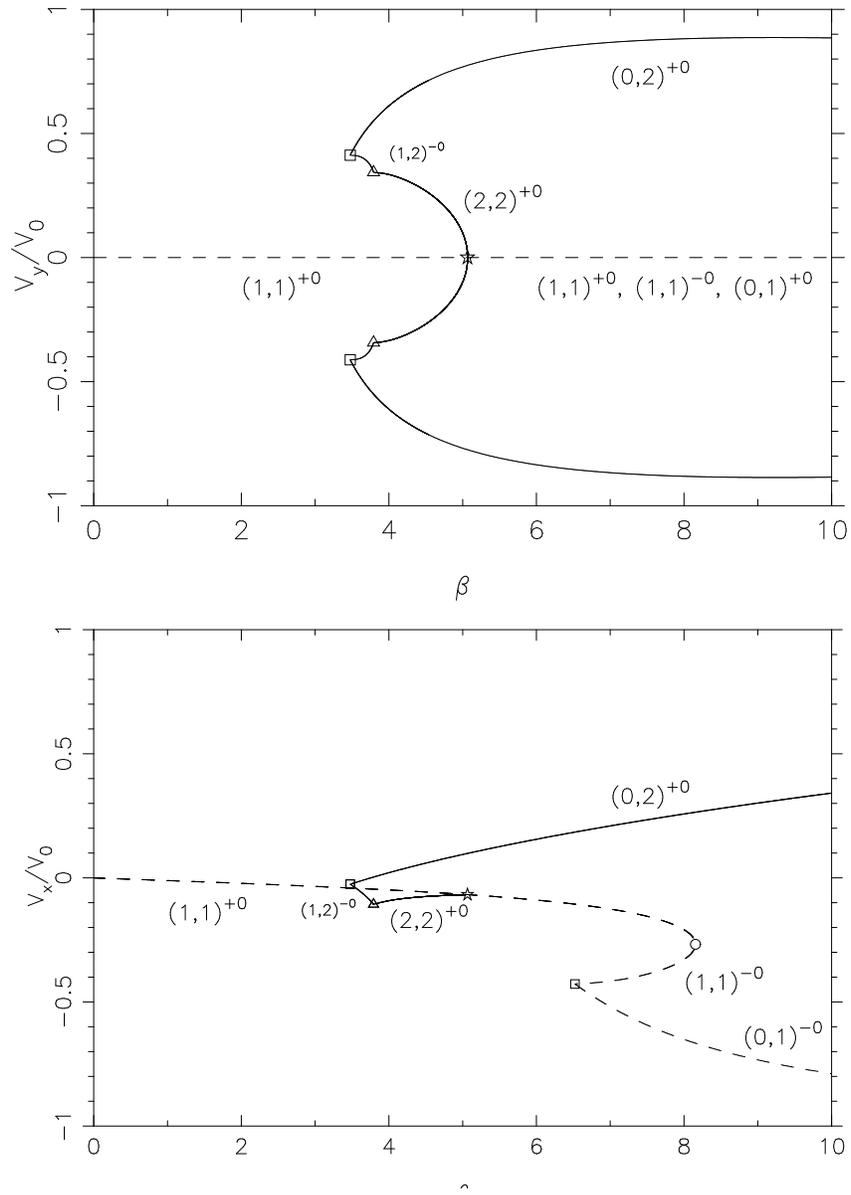}
\end{center}
\caption{
The bifurcation diagram of the $(2,2)^+$, $(1,2)$  and $(0,2)^-$ 
orbits in the DBM in ``regime one''.
The vertical axis represents $y$ (top panel) and $x$ (bottom panels)
components of the scaled velocity of the electron at the point of
collision with the collector barrier (see Fig.
\protect\ref{fig_psos_po2_dbm}b) ; $\gamma = 1.17$; the tilt 
angle $\theta = 5^\circ$.
\label{fig_bifdiag_po2m_1_dbm}
}
\end{figure}

\begin{figure}[htbp]
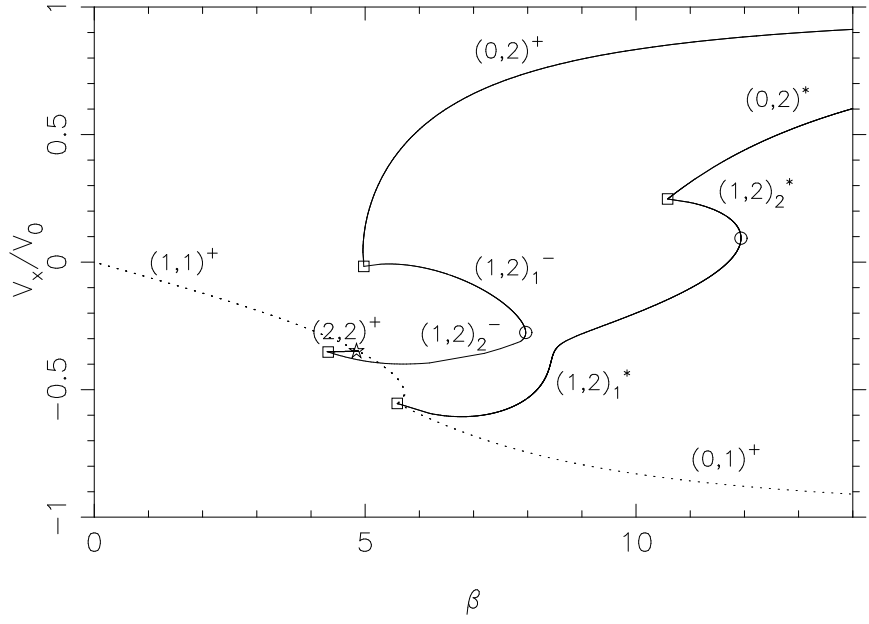
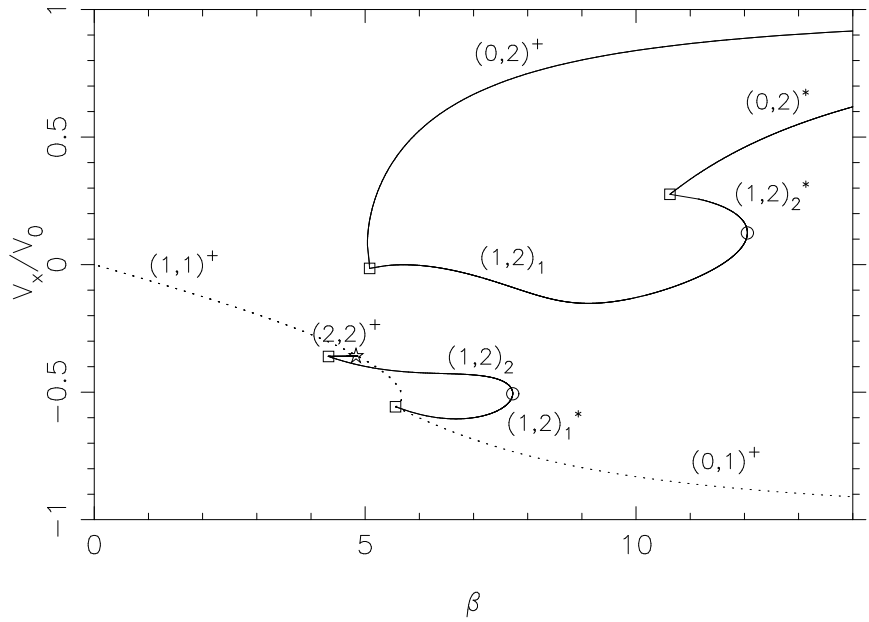

\begin{center}
\leavevmode
\epsfysize=3.2in 
\epsfbox{fig29a.epsi}
\end{center}
\begin{center}
(a)
\end{center}
\vskip 1.4 cm
\begin{center}
\leavevmode
\epsfysize=3.2in 
\epsfbox{fig29b.epsi}
\end{center}
\begin{center}
(b)
\end{center}
\vskip 1.4 cm
\begin{center}
\leavevmode
\epsfysize=3.2in 
\epsfbox{fig29c.epsi}
\end{center}
\begin{center}
(c)
\end{center}
\vskip 1.4 cm
\begin{center}
\leavevmode
\epsfysize=3.2in 
\epsfbox{fig29d.epsi}
\end{center}
\begin{center}
(d)
\end{center}
\caption{
The bifurcation diagram of the $(2,2)^+$, $(1,2)$  and $(0,2)^-$ 
orbits in the DBM in $(\beta, v_x/v_0)$ coordinates
(see Fig. \protect\ref{fig_psos_po2_dbm}b) in regimes 
(a) two, (b),(c)  three, and (c) four;
$\gamma = 1.17$; the tilt 
angle $\theta = $ (a) $20^\circ$, (b) $27^\circ$,
(c) $29^\circ$, and (d) $30^\circ$.
\label{fig_bifdiag_po2m_dbm}.
}
\end{figure}

\begin{figure}[htbp]
\begin{center}
\leavevmode
\epsfysize=3.in 
\epsfbox{fig30.epsi}
\end{center}
\caption{
Examples of the different types of period-$3$ orbits in the DBM, 
projected onto
$(y,z)$ planes : a $(3,3)^-$ orbit (a), 
a $(1,3)^+$ orbit (b), a $(2,3)$ orbit (c).
\label{fig_po3_YZ_dbm}
}
\end{figure}

\begin{figure}[htbp]
\begin{center}
\leavevmode
\epsfysize=5.2in 
\epsfbox{fig31.epsi}
\end{center}
\caption{
The bifurcation diagrams of the period-$3$ orbits in the DBM, at
$\gamma = 1.17$, $\theta = $ (a) $11^\circ$, (b) $38^\circ$. 
\label{fig_bifdiag_po3_dbm}
}
\end{figure}

\begin{references}
\bibitem{caveat} Actually we shall see that the current experimental
system, which we describe by the double-barrier model introduced
below, is not a standard KAM system due to a discontinuity in its 
classical 
mechanics which is important for the periodic orbit theory and is
addressed in detail in the text.
\bibitem{explain_kicked-top}
The kicked - top map in the standard form in variables
$(X_n, Y_n, Z_n;\alpha, \lambda)$ \cite{Haake} 
is identical to the approximate SBM mapping, when 
$ X_n \equiv (\tilde{v}_{x'})_n, Y_n \equiv 
(\tilde{v}_{y'})_n, 
Z_n \equiv (\tilde{v}_{z'})_n, \alpha  \equiv  2 \theta, 
\lambda \equiv  \frac{\beta}{\cos\theta}$
\bibitem{SBMrealizable} With a slightly different doping profile
the bottom of the well could be raised above the emitter state at 
zero bias allowing an experimantal study of the pure DBM.
\bibitem{ExplainNoLimit} The unperturbed system ($\theta = 0$) 
represents the limiting behavior of the tilted well only if the 
chaos parameter  $K \sim \beta \theta \rightarrow 0$. 
The parameter $\beta_{b3}$, corresponding to the value of $\beta$  
at the appearance of the $(2)_s^{+k}$ orbit, diverges at $\theta 
\rightarrow 0$, which resolves what seems to be a contradiction.
\bibitem{expalin_qualitatively_po1_dbm}. The following simple 
qialitative argument can illustrate the origin of this property.
When the total energy is just a litte above the
potential drop between the barriers $eV$ ($\gamma - 1 \ll 1$), then
the electron returns to the collectror barrier with very small kinetic
energy. Therefore, the energy exchange between cyclotron and
longitudunal degrees of freedom at the emitter wall, if takes place,
results in a substantially smaller energy transfer between
cyclotron and longitudinal degrees of freedom, than the energy
exchange at the collector barrier for a similar
angle of incidence. Therefore, it is intuitively clear, that the energy 
exchange at the emitter barrier alone can not compensate for the energy 
exchange at the collector wall, and for  the
period-$1$ emitter orbits are expected to  be non-mixing.
\bibitem{explain_why_vy=0} A nonzero  $y$ component of the velocity 
at the point of collision 
necessarily leads to a nonzero energy exchange between cyclotron and 
longitudinal motion at the emitter wall, while for the $(0,2)^+$ 
orbit (which at the cups bifurcation is degenerate with an emitter orbit)
the cyclotron velocity must not be changed between successive collisions
with the collector barrier.
\end{references}
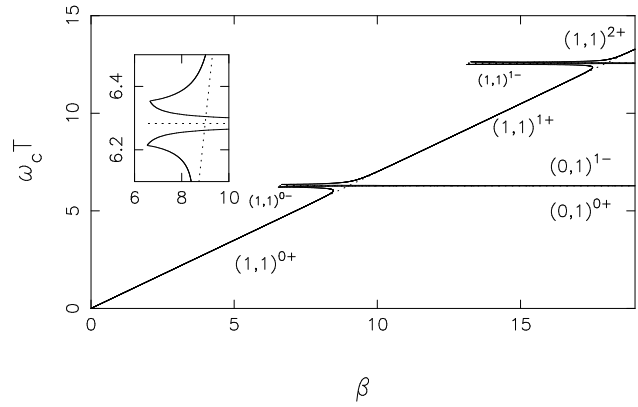
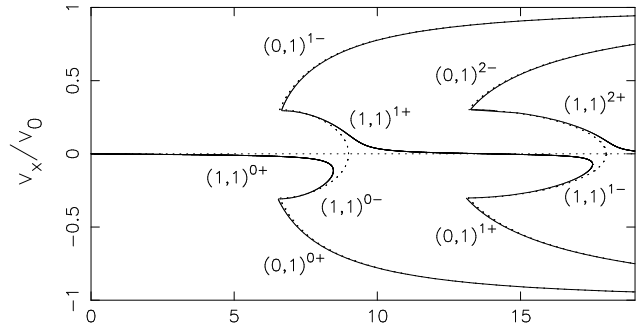
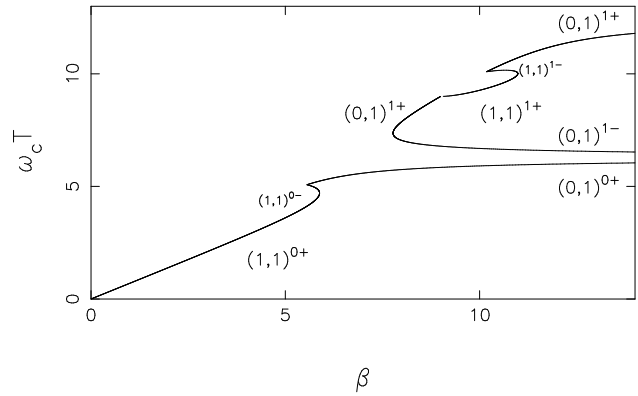
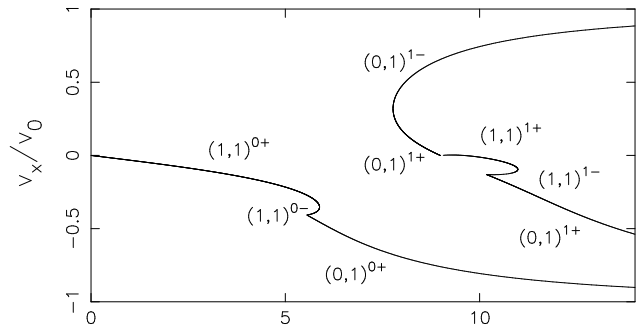
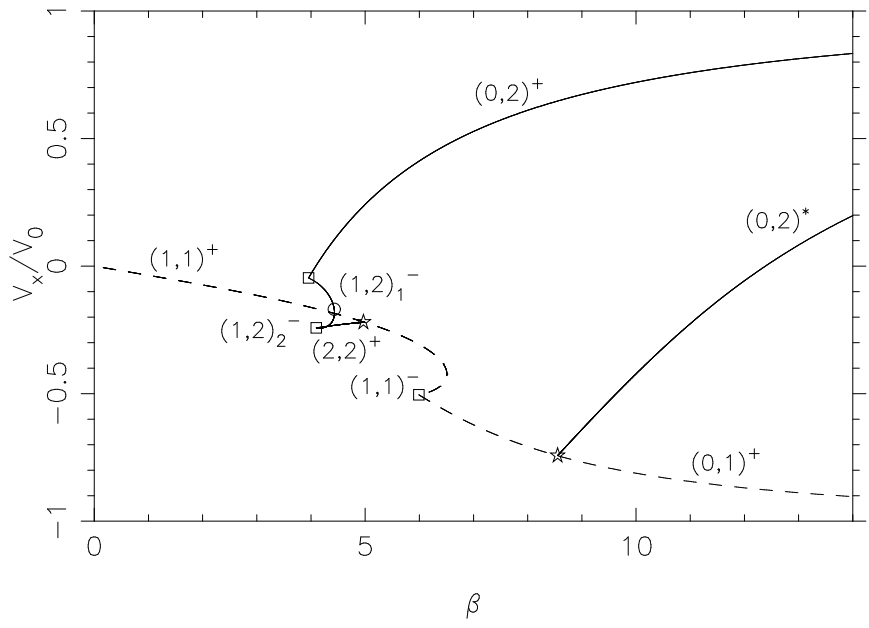
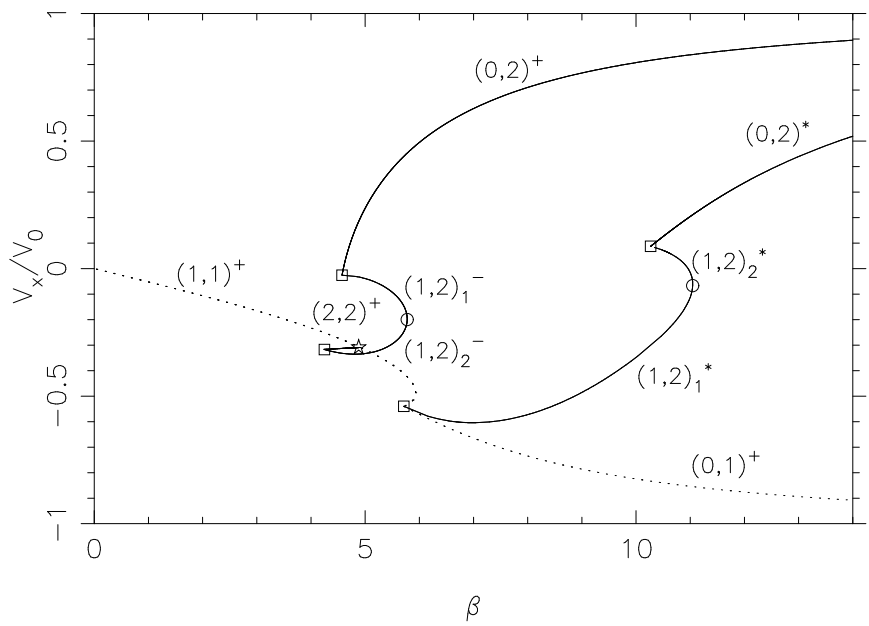
\end{document}